# Topology optimization in the framework of the linear Boltzmann equation – a method for designing optimal nuclear equipment and particle optics

Sébastien Chabod
Univ. Grenoble Alpes, CNRS, Grenoble INP, LPSC-IN2P3, 38000 Grenoble, France
Postal address: LPSC, 53 avenue des Martyrs, 38026 Grenoble, France
Email address: sebastien.chabod@lpsc.in2p3.fr
ORCID ID: https://orcid.org/0000-0003-2154-2012

**Abstract.** In this study, we describe a procedure of topology optimization in the framework of the linear Boltzmann equation, implemented using the reference Monte-Carlo particle transport code MCNP. This procedure can design complex structures that optimize the transport of particles, according to preset objectives and constraints. Although simple and perfectible, this procedure has been successfully tested on a series of difficult problems, with results outperforming human capabilities. Improved, it could be used to assist or automate the design of particle optics or nuclear devices.

0. Introduction

*Aim of the study.* Suppose we have (i) a source of particles, positioned at a point $r_s$ in space, (ii) a target or detector, positioned at a point $r_d$, (iii) and a material, which propagates the particles. This material may be subject to constraints, e.g. of cost or weight. How to organize this matter so that the flux of particles at the point $r_d$ be maximal, or minimal? Can we machine this material so that the energy spectrum of particles at the point $r_d$ be a Gaussian, or a reference spectrum, such as the neutron spectrum in a fast nuclear reactor? These problems, whose short statements hide their resolution difficulty, are topology optimization problems. The purpose of a topology optimization procedure is to give to an object the shape best suited for its use. From tools once fashioned in wood up to modern architecture and engineering, this discipline is an old and common asset of our cultures. Like others nevertheless, this field has experienced a substantial acceleration with the development of computers, which allowed the calculation of increasingly complex structures, and the resolution of more and more subtle problems: design of materials with negative coefficients of thermal expansion [1], 3D printing [2], lightering of aeronautical components [3], EM or seismic cloaking devices [4], to name just a few recent spectacular examples. The applications are numerous and varied [5], the theoretical bases of this discipline are hence now firmly established [6]. Our study therefore does not pretend to bring fundamental mathematical developments to this thematic. Its interest lies rather in its application to the optimization of particle transport in the framework of the linear Boltzmann equation.
The linear Boltzmann equation models the transport of particles in matter provided their density remains low enough so that they do not interact with each other. The involved disciplines hence include e.g. radiological protection, nuclear engineering and medicine, or particle detection and emission. Despite the economic and societal importance of these disciplines, while topology optimization in the framework of the Boltzmann equation is the subject of an intensive research in fluid dynamics [7, 8], there is not much application of this method to the transport of particles such as neutrons or gamma-rays, presumably because of the once large computing cost required to solve these problems. The closest example found in the literature is a study by Asbury et al. on the design of radiation shields for spatial applications [9]. Current optimization studies in these areas are therefore still essentially parametric, see e.g. [10-12], or use the approximation of diffusion, see e.g. [13, 14] [1]. We hence propose in this manuscript a topology optimization procedure that can be used directly with a reference particle transport code, the MCNP code in its recent versions 5, 6 or X [15]. We then propose examples of application of this tool to the design of nuclear equipment and particle optics.

*The challenges to overcome.* The Boltzmann equation can be written for particles that do not interact with each other:

$$B_{\underline{r}E\underline{\Omega}}\varphi(\underline{r},E,\underline{\Omega}) = Q(\underline{r},E,\underline{\Omega}) \quad (0.1)$$

where the Boltzmann operator B is given by:

$$B_{\underline{r}E\underline{\Omega}}\varphi(\underline{r},E,\underline{\Omega}) = \underline{\Omega}.\nabla\varphi(\underline{r},E,\underline{\Omega}) + \left(\sigma_t(E)n(\underline{r}) + \Sigma_{fixed,t}(\underline{r},E)\right)\varphi(\underline{r},E,\underline{\Omega})$$
$$- \int_{E'=0}^{+\infty} \int_{\underline{\Omega}'\in 4\pi} \left( \begin{array}{c} n(\underline{r})\sigma_s(E')f(E'\to E,\underline{\Omega}'\to\underline{\Omega}) \\ + \Sigma_{fixed,s}(\underline{r},E')f_{fixed}(\underline{r},E'\to E,\underline{\Omega}'\to\underline{\Omega}) \end{array} \right) \varphi(\underline{r},E',\underline{\Omega}')dE'd\underline{\Omega}' \quad (0.2)$$

Function $\varphi(\underline{r},E,\underline{\Omega})$ is the angular flux of particles, defined by the product $vn(\underline{r},E,\underline{\Omega})$ of the speed $v$ and the density $n(\underline{r},E,\underline{\Omega})$ of particles. The quantity $n(\underline{r},E,\underline{\Omega})d\underline{r}dEd\underline{\Omega}$ is the number of particles: (i) contained in an infinitesimal volume $d\underline{r} = dxdydz$ centered at the point $\underline{r}$ of coordinates $(x,y,z)$, (ii) of energies ranging from $E$ to $E+dE$, $dE$ being infinitesimal, (iii) of directions $\underline{\Omega}$ included in the infinitesimal portion of solid angle $d\underline{\Omega}$. Functions $\sigma_s$ and $\sigma_t$ are respectively the scattering and total microscopic cross-sections of the propagating medium to be optimized. Functions $\Sigma_{fixed,s}$ and $\Sigma_{fixed,t}$ are the scattering and total macroscopic cross-sections of the materials not covered by the optimization procedure, for example the ground, roof and walls of a work area. Functions $f(\underline{r}, E'\to E, \underline{\Omega}'\to\underline{\Omega})$ are the probability densities that a particle that propagates with an energy $E'$ in the direction $\underline{\Omega}'$ ends, after a collision with an atom of the material at the point $\underline{r}$, with an energy $E$ and a direction $\underline{\Omega}$. Function $Q(\underline{r},E,\underline{\Omega})$ models the particle source. By definition, $Q(\underline{r},E,\underline{\Omega})d\underline{r}dEd\underline{\Omega}$ is the number of source particles injected per unit of time in the volume $d\underline{r} = dxdydz$ centered at the point $\underline{r}$, with energies comprised between $E$ and $E+dE$, and with directions $\underline{\Omega}$ comprised in the portion of solid angle $d\underline{\Omega}$. Finally, function $n(\underline{r})$ is the atomic density at the point $\underline{r}$ of the material to be optimized. One of the objectives of this study will be to find the density $n(\underline{r})$ that optimizes some desired characteristics of the population $\varphi(\underline{r},E,\underline{\Omega})$ of particles in a region of space, taking into account constraints.

To do this, the first challenge to overcome is the non-linearity of $\varphi(\underline{r},E,\underline{\Omega})$ with respect to $n(\underline{r})$. Suppose we have a cube of matter, of volume $dV_1$, placed at a point $\underline{r}_1$. The particles emitted by the source will interact with this cube, be deviated or absorbed therein, and induce a flux $d\varphi_1$ at a point $\underline{r}_d$. Let us remove this cube, and put another, $dV_2$, at a point $\underline{r}_2$. The particles that pass through $dV_2$ will induce a flux $d\varphi_2$ at $\underline{r}_d$. Let us now put both cubes, $dV_1+dV_2$, at the same time: one observes that the flux thus induced, $d\varphi_{1+2}$, is not equal to the sum of the fluxes $d\varphi_1$ and $d\varphi_2$. Indeed, some particles passing through $dV_1$ may interact with the cube $dV_2$ before reaching the detector, and vice-versa. This non-linearity of the flux, $d\varphi_{1+2} \neq d\varphi_1 + d\varphi_2$, with respect to $n(\underline{r})$ (see Supplemental Material A) poses a major challenge. It shows that modifying the material at a location $dV_1$ to optimize the desired characteristics of the particle population at $\underline{r}_d$, then modify it at another location $dV_2$, restart the optimization, and so on, will give in the end a wrong result. Since all volumes $dV_i$ of the material play intricate roles in the particle transport, it is necessary to modify them all at the same time, in the right direction.

---

[1] We will show in section 2.2 that the approximation of diffusion can lead to very suboptimal configurations.

The second challenge to overcome is the impossibility to solve this problem by brute force, by testing all possible configurations of the material to be optimized. To illustrate this point, let us cut out the volume of the available material into pieces. If one wishes that the final shape of the object to be optimized is not too coarse, too "pixelated", these pieces must not be too big: their size should not reasonably exceed ~5% of the size of the object. We must therefore divide each dimension of the object into ~1/5% = 20 units, i.e. divide the total volume into 20×20×20 = 8000 cells. The number of configurations to explore, even in the simple case where the choice of material per cell is binary (material or void), is thus $2^{8000} \approx 10^{2408}$. To fully explore such a large number of configurations is impossible nowadays. Even in the simpler case where the problem has a symmetry, e.g. axial, we must divide the volume into ~20×10 = 200 cells, which gives $2^{200} \approx 10^{60}$ configurations to examine, ~1000 times more than the number of atoms in the Sun: a systematic exploration is still impossible. To solve a topology optimization problem in the framework of the linear Boltzmann equation, one must therefore find an algorithm capable of (i) filtering the huge number of possible configurations, (ii) in a human-compatible time, (iii) while managing the strong non-linearity of the problem (in particular by finding the global optimum, not just a local one). A priori, it is a tall order. However, the solution proposed in this study, although simple, already meets these criteria: it has passed with relative ease a varied set of difficult tests detailed below.

*Organization of the study.* This study is divided into six sections. In section 1, we describe the mathematical bases of the proposed topology optimization procedure. In sections 2 to 5, we test the strengths and weaknesses of this procedure on four categories of problems, chosen to have practical applications. For each problem, we compare the results of the optimization procedure with analytical or intuitive solutions. In section 2, we calculate shapes of neutron propagators that maximize or minimize the flux of these particles in a region of space. In section 3, we optimize the structure of an object designed so that the neutron energy spectrum in an area of space be as close as possible to an objective spectrum. In section 4, we use the proposed optimization procedure to minimize the critical mass of fuel in a schematic nuclear reactor. In section 5, we optimize the shape of a gamma-ray collimator and study the impact that the choice of the objective to achieve can have on the performances of the calculated optimal structure. Finally, in section 6, we conclude this study by summarizing its successes, but also its limits, its areas for improvement and its perspectives [2].

*Computing resources used.* The computations of this study were performed on machines dating from the early 2010s, operating each 16 CPU used in parallel [3]. With the exception of the case discussed in Supplemental Material G.1, each problem dealt used one machine at a time. With these (modest) computing resources, for problems with ~1000 cells, the durations of an iteration of the proposed optimization algorithm varied between about an hour for the simplest cases to one day for the most complex, with a convergence reached in ~50-100 iterations in most cases.

1. Mathematical framework of the optimization procedure

Let us introduce a functional $O$, called the objective functional, which acts on the flux $\varphi(\underline{r},E,\underline{\Omega})$ of particles. The term $O\varphi$ is the quantity we want to optimize. It can be for example the total flux of particles in a region of space. For the applications studied in section 2, this functional can be written in the following way:

---

[2] Eleven videos illustrate the results obtained in this study. Before starting its reading, they can be downloaded at https://lpsc.in2p3.fr/chabod/videos.rar.
[3] The CPU used were 16 Intel Xeon E5-2670 at 2.6 GHz.

$$O_{rE\Omega}\varphi(\underline{r},E,\underline{\Omega}) = \int_{\underline{\Omega}\in 4\pi}\int_{\underline{r}\in\Re^3}\int_{E=0}^{+\infty} f(\underline{r})g(E)h(\underline{\Omega})\varphi(\underline{r},E,\underline{\Omega})dEd\underline{r}d\underline{\Omega} \quad (1.1)$$

In the formula (1.1), the functions $f(\underline{r})$, $g(E)$ and $h(\underline{\Omega})$ are defined on all or part of the domains of variation of the variables $\underline{r}$, $E$, $\underline{\Omega}$ of the problem. For example, if one wishes to minimize or maximize the total flux at the point $\underline{r}_d$, one can take $f(\underline{r}) = \delta(\underline{r}-\underline{r}_d)$, $g(E) = 1$ and $h(\underline{\Omega}) = 1$. If one wishes to minimize or maximize the dose delivered in a volume $V$ of space, one can take $f(\underline{r}) = \Theta[\underline{r}\in V]$, where $\Theta$ is a function that returns 1 if the condition between brackets is verified and 0 otherwise, $g(E) = Q(E) =$ quality factor of the radiation at energy $E$, and $h(\underline{\Omega}) = 1$. If one wishes to optimize the dpa in a material in a volume $V$, one can take $f(\underline{r}) = \Theta[\underline{r}\in V]$, $g(E) = D(E) =$ dpa induced at energy $E$, and $h(\underline{\Omega}) = 1$.

Let us introduce now a vector of constraints, $\underline{C}(\rho,V,\chi,\varphi)$, whose components are functions of the physicochemical properties of the material to be optimized, e.g. its shape $\Theta[\underline{r}\in V]$, its volume density $\rho(\underline{r})$ and/or its isotopic composition $\chi(\underline{r})$. The constraints $\underline{C}$ can also depend on the flux $\varphi(\underline{r},E,\underline{\Omega})$ for some applications, see section 4. For usual weight or volume constraints, they can be written for example:

$$C(\rho) = \int_{\underline{r}\in\Re^3}\rho(\underline{r})d\underline{r} - P_{max} \quad \text{or} \quad C(V) = \int_{\underline{r}\in\Re^3}\Theta[\underline{r}\in V]d\underline{r} - V_{max} \quad (1.2)$$

where $P_{max}$ or $V_{max}$ are the weight or the volume of available material.

We wish to optimize the objective $O\varphi$ under constraints $\underline{C}$, i.e. to solve a problem of constrained optimization which is written for example for an optimization of the volume density $\rho(\underline{r})$:

$$\text{Find } \rho_{opt}(\underline{r}) \text{ such as}$$
$$O\varphi(\rho_{opt}) = \min_{\rho}/\max O\varphi(\rho) \quad (1.3)$$
$$\text{subject to } B(\rho)\varphi(\underline{r},E,\underline{\Omega},\rho) = Q(\underline{r},E,\underline{\Omega})$$
$$\underline{C}(\rho,\varphi(\rho)) = \underline{0}$$

where $B\varphi = Q$ is the Boltzmann equation and $\rho_{opt}(\underline{r})$ is the optimal density profile sought.

To solve this class of problems, a possible solution is to use a Lagrangian, which is written for the problem (1.3):

$$L(\rho,\underline{\lambda}) = O\varphi - \underline{\lambda}.\underline{C}(\varphi(\rho),\rho) \quad (1.4)$$

where $\underline{\lambda}$ is a vector of Lagrange multipliers. The derivatives of the Lagrangian $L$ with respect to the quantities to be optimized, here $\rho(\underline{r})$ for the problem (1.3), give the optimum of $O\varphi$. Its derivatives with respect to the multipliers $\underline{\lambda}$ give the constraints $\underline{C}$. The optimal characteristics of the particle propagator are thus solution of a system of equations, which is written:

$$\frac{\partial L}{\partial \rho} = 0, \quad \frac{\partial L}{\partial \underline{\lambda}} = -\underline{C} = \underline{0} \quad (1.5)$$

To make a practical use of this result, we can tile the available space using a union of cells $\Theta_i(\underline{r})$, e.g. $\Theta_i(\underline{r}) = \Theta[x_i \leq x \leq x_{i+1}]\Theta[y_{i'} \leq y \leq y_{i'+1}]\Theta[z_{i''} \leq z \leq z_{i''+1}]$ if parallelepipeds are chosen.

Then, let $\rho_i$ be the volume densities of the material in the cells $\Theta_i$, taken as constant. With this parametrization, the density $\rho(\underline{r})$ is rewritten:

$$\rho(\underline{r}) = \sum_i \rho_i \Theta_i(\underline{r}) = \underline{\rho}.\underline{\Theta} \quad (1.6)$$

Then taking for example a weight constraint, we obtain, cf. (1.2):

$$C(\underline{\rho}) = \sum_i \rho_i V_i - P_{max} = \underline{\rho}.\underline{V} - P_{max} \quad (1.7)$$

where $\underline{V}$ is the vector of the volumes $V_i$ of the cells $\Theta_i$. With this setting, the problem (1.3) is rewritten:

$$\min/\max_{\underline{\rho}} O\varphi(\underline{\rho})$$
$$\text{s.t.} \quad B(\underline{\rho})\varphi(\underline{r}, E, \underline{\Omega}, \underline{\rho}) = Q(\underline{r}, E, \underline{\Omega}) \quad (1.8)$$
$$\underline{\rho}.\underline{V} - P_{max} = 0$$

The Lagrangian $L$ becomes:

$$L(\underline{\rho}, \lambda) = O\varphi - \lambda(\underline{\rho}.\underline{V} - P_{max}) \quad (1.9)$$

and its derivatives according to $\underline{\rho}$ are:

$$\frac{\partial L}{\partial \rho_i} = \frac{\partial O\varphi}{\partial \rho_i} - \lambda V_i = 0 \Rightarrow C_i = \frac{1}{V_i}\frac{\partial O\varphi}{\partial \rho_i} = \lambda, \forall i \quad (1.10)$$

This gives a simple yet useful result: the density profile, $\rho_{opt}(\underline{r}) = \underline{\rho}_{opt}.\underline{\Theta}$, of the material that optimizes the objective $O\varphi$ is given by the isovalue $\lambda$ of coefficients $C_i$ that satisfies the weight constraint $\underline{\rho}.\underline{V} = P_{max}$. To solve the topology optimization problem (1.8), we must hence find a way to compute the derivatives $\partial O\varphi/\partial \rho_i$ of the objective. In conventional topology optimization problems, this task is usually performed with an adjoint problem based method (see e.g. [5], [6] pp 16-18, [8] pp 26-27) or a direct sensitivity analysis. However, for reasons detailed in Supplemental Material B [16-18], these conventional methods cannot be applied in our study. We will use instead a differential operator sampling method developed by Olhoeft, Takahashi, Hall, Rief et al. [19-23], implemented in recent distributions of the Monte-Carlo particle transport code MCNP [15], starting from its version MCNP5 (see [24] for a detailed description, or [25] chapter 2 section 12 pp 2-195 to 2-203 for a summary) [4]. This tool is commonly used in reactor physics for performing sensitivity studies, but it does not seem to have been used up to now for writing a topology optimization algorithm.

To illustrate how this procedure works, let us take for example $O\varphi = \phi$, the total flux of particles in the detector cell. In order to compute the derivatives $\partial O\varphi/\partial \rho_i$ with the MCNP code, it is first necessary to model the problem, by defining the geometry of the cells $\Theta_i$, the properties of the material to be optimized in each of these cells, the particles to be propagated, the characteristics of the particle source, as well as any models to be used or cuts to be made to propagate the

---

[4] The MCNP distribution used for all the calculations performed in this study is MCNP6.1.

particles. The MCNP user guide describes in details the procedures to follow [26]. Once this modeling work is completed, it is necessary to insert in the MCNP input file a tally card to compute the particle flux (e.g. a F4 tally), as well as PERT cards. For users with a distribution of MCNP6.1, instructions for using the PERT card can be found in the MCNP6.1 user guide, chapter 3 pp 3-156 to 3-161. For the MCNPX version, these instructions can be found online, see [26] section 5.6.22. The arguments of the PERT card to be used are, for the problem (1.8):

PERT$i$:$x$  CELL=$i$  MAT=$m$  RHO=$-(\rho_i+\delta\rho_i)$  METHOD=2

The entry $x$ specifies the type of particles propagated. The entry CELL specifies the number $i$ of the cell $\Theta_i$ whose density is to be perturbed. The entry MAT specifies the number $m$ of the card M used to define the isotopic composition of the material. The entry RHO gives the density after perturbation, $\rho_i+\delta\rho_i$, of the material in the cell $\Theta_i$. The sign − before $\rho_i+\delta\rho_i$ indicates that the density considered is a volume density, to be expressed in g/cm$^3$. In this study, we will take the following perturbation, $\delta\rho_i = \varepsilon\rho_i$, where $\varepsilon$ is a perturbative coefficient fixed at 0.05. Finally, the entry METHOD allows to specify the number of terms in the Taylor expansion of $O\varphi(\rho_i+\delta\rho_i)$ at $\delta\rho_i = 0$ to be taken into account in the calculation of the difference $O\varphi(\rho_i+\delta\rho_i) - O\varphi(\rho_i)$, as well as the format of the result written in the MCNP output files. By using the option METHOD=2, the MCNP output files contain after computation an entry giving the value of:

$$PERT_i = \frac{\partial O\varphi}{\partial \rho_i}\delta\rho_i = \varepsilon\frac{\partial O\varphi}{\partial \rho_i}\rho_i \quad (1.11)$$

as well as its statistical error. Hence, by using PERT cards, we can have access to the derivatives of the objective, $\partial O\varphi/\partial \rho_i$, sought [5]. All that remains now is to develop an algorithm that makes these derivatives converge towards the isovalue imposed by the constraint, (1.10), then validate it on a problem that has a reference solution. This work will be done in section 2.

*Comment*. Note that the computational method (1.11) has a limitation. The code MCNP returns an error message then stops when a PERT card is used on a cell whose density is too low. To avoid this, an artificial minimum bound, $\rho_{min} > 0$, on the densities $\rho_i$ has to be imposed. Additionally, a realistic material has a physical upper bound, $\rho_{max}$, on its density, for example 11.34 g/cm$^3$ for natural lead. These two bounds on $\rho_i$, one computational and the other one physical, imposes an additional constraint in the optimization problem (1.8), which is:

$\rho_i \in [\rho_{min}, \rho_{max}] \; \forall i \quad (1.12)$

---

[5] As compared to the methods described in Supplemental Material B, this method had a decisive advantage for our study. Let us take e.g. a tiling of the available space with 20×20×20 cells. To compute the 8000 associated derivatives $\partial O\varphi/\partial \rho_i$, 8000 cards PERT$i$ are to be added into the MCNP input, then we run this code, once. The gain in time and accuracy is considerable. Even if the addition of a large number of PERT cards causes a slowdown of the MCNP simulation, the associated computing times remain human-compatible, even short. For example, in section 4, we will perform computations involving a tiling with 2500 cells. These computations lasted for ~10h each on a 16-CPU machine. Despite the limited capabilities of this machine, the 40 iterations of the topology optimization algorithm took 18 days in total. By using a farm of 100-10000 CPUs of modern technology, without even anticipating the increase in computing power in the coming decades, this type of computations should already pose no insurmountable difficulties nowadays, in 2018.

2. Problem n°1: optimal neutron concentrators and shields

Suppose we have a neutron source, positioned at a point $\underline{r}_s$, and a quantity of matter. How to shape this matter so that the neutron flux at a point $\underline{r}_d$ be maximal?
Before trying to solve this problem, we should check if it is well-posed. Does it admit a trivial solution, or even multiple solutions? If we take $\underline{r}_d = \underline{r}_s$, the problem has a central symmetry. The solution should therefore be a ball of center $\underline{r}_s = \underline{r}_d$. With no upper limit on the volume of usable matter, the maximum neutron flux will be obtained for a ball of infinite radius, as any increase in radius will reduce neutron leakage at the surface of the ball. So, without a volume constraint, the solution of the problem is infinity. To avoid this unrealistic solution, we must impose the maximum volume, $V_{max}$, of usable matter. It is anyway a mandatory practical constraint, if only for evident cost or clutter reasons. By fixing this maximum volume $V_{max}$, and by denoting $S$ the surface of the material to be shaped, the aforementioned problem can be written as follows:

Find $S_{opt}$ such as
$$O\varphi(S_{opt}) = \max_{S} O\varphi(S) \qquad (2.1)$$
s.t. $\quad B(S)\varphi(\underline{r}, E, \underline{\Omega}, S) = Q(\underline{r}, E, \underline{\Omega})$
$$C(S) = V - V_{max} = \int_{\underline{r} \in \Re^3} \Theta[S(\underline{r}) \geq 0] d\underline{r} - V_{max} = 0$$
with $\quad O\varphi(S) = \phi(\underline{r}_d, S) = \int_{E=0}^{+\infty} \int_{\underline{\Omega} \in 4\pi} \varphi(\underline{r}_d, E, \underline{\Omega}, S) dE d\underline{\Omega}$
$$n(\underline{r}) = n_0 \Theta[S(\underline{r}) \geq 0]$$

where B is the Boltzmann operator, given in (0.2), and $Q(\underline{r},E,\underline{\Omega})$ is the neutron source.
In (2.1), $S_{opt}$ is the optimal surface sought, for which the total neutron flux $\phi(\underline{r}_d,S)$ at the point $\underline{r}_d$ is maximum. Function $n(r)$ is the atomic density of the material at the point $\underline{r}$, function $\Theta[]$ returns 1 when the condition between brackets is verified and 0 otherwise. The set of points $\underline{r}$ that defines the surface of the object is given by the equation $S(\underline{r}) = 0$. The equation $S(\underline{r}) \geq 0$ defines a volume, that of the object to be shaped.
In section 2.2, we will show how to solve this problem with a topology optimization algorithm. But, for the moment, we propose in the following section, 2.1, to solve it analytically in the particular case of a diffusive, monoenergetic, neutron transport. This particular solution will then serve us as a reference for testing the proposed optimization procedure.

2.1. Reference solution for a monoenergetic diffusive transport

Suppose now that the material to be shaped is made of heavy low absorbing atoms, i.e. of atoms whose nuclei have masses $m$ much greater than the neutron mass $m_n$, and whose neutron capture cross-sections are much lower than the scattering ones. Suppose also that the neutrons keep a constant energy throughout their transport, despite undertaking a large number of collisions between their emission point, $\underline{r}_s$, and their arrival point, $\underline{r}_d$. These assumptions are restrictive, but there are examples of realistic configurations that verify them. We can cite e.g. the case of a material made of lead, at a temperature $T$, with a source positioned at a point $\underline{r}_s$ sufficiently far from $\underline{r}_d$, which emits thermal neutrons of initial energies $E_0 = kT$ (in the absence of fission reactions, the neutrons will thereby keep throughout their transport an energy close to $kT$). The interest of this particular framework is that it makes it possible to formulate an explicit solution to the complex problem (2.1). Indeed, if the hypotheses listed above are satisfied, the Boltzmann equation (0.1) can be simplified using the P1 approximation (see [27] pp 99-104). The result is a diffusion equation, which is written:

$$-D\Delta_r \phi(\underline{r}_s \to \underline{r}, S) = Q(\underline{r}) \quad (2.1.1)$$

Its source term, $Q(r)$, is the density of source neutrons emitted in the material at the point $\underline{r}$. For a point source positioned at $\underline{r}_s$, it is written:

$$Q(\underline{r}) = Q_0 \delta(\underline{r} - \underline{r}_s) \quad (2.1.2)$$

where $Q_0$ is the number of source neutrons injected in the material. In this equation, function $\phi(\underline{r}_s \to \underline{r})$ is the neutron fluence generated at the point $\underline{r}$ by the point source positioned at $\underline{r}_s$. The operator $\Delta_r$ is a Laplacian acting on the coordinates $\underline{r}$. Coefficient $D$ is the diffusion coefficient of the material, function of the macroscopic scattering cross-section $\Sigma_s$ and of the ratio $A = M/m_n$ of the nuclei mass over the neutron mass. For a material made up of heavy low absorbing nuclei, its formula is $D = 1/(3\Sigma_s)$, see [27] pp 99-104. The equation (2.1.1) is valid inside the material. At the surface $S$ or outside the material, if we neglect the extrapolation distance (see [27] pp 273), the neutron flux falls to 0, which imposes the following boundary condition:

$$\phi(\underline{r}_s \to \underline{r}, S) = 0 \text{ if } \underline{r} \in S \quad (2.1.3)$$

For a monoenergetic diffusive transport, the problem (2.1) is thus rewritten:

$$\begin{aligned} &\max_S \phi(\underline{r}_s \to \underline{r}_d, S) \\ &\text{s.t. } -D\Delta_r \phi(\underline{r}_s \to \underline{r}, S) = Q_0 \delta(\underline{r} - \underline{r}_s), \forall \underline{r}/S(\underline{r}) > 0 \\ &\phi(\underline{r}_s \to \underline{r}, S) = 0, \forall \underline{r}/S(\underline{r}) = 0 \\ &C(S) = 0 \end{aligned} \quad (2.1.4)$$

In Supplemental Material C [28-30], we solve this problem: (i) by combining its equations into a single integral equation, making use of Green's second identity and an adapted Green's function (a description of these mathematical tools can be found in [28]); (ii) then by performing on it a calculus of variations. At the end of section C.2, we thereby show that the surface $S_{opt}$, solution of (2.1.4), obeys a short implicit equation, which is written [6]:

$$\underline{J}(\underline{r}_s \to \underline{r}, S_{opt}) \cdot \underline{J}(\underline{r}_d \to \underline{r}, S_{opt}) = \lambda Q_0 D, \forall \underline{r} \in S_{opt} \quad (2.1.5)$$

where $\lambda$ is a Lagrange multiplier, set by the constraint $V = V_{max}$, and $\underline{J}(\underline{r}' \to \underline{r})$ is the current of neutrons generated at the point $\underline{r}$ by a point source positioned at $\underline{r}'$. In section C.3, by using: (i) a result in electromagnetism, see [30] pp 54-55; (ii) the symmetries of the problem (2.1.4); (iii) a spherical coordinate system, $\underline{r} = (r, \theta, \varphi)$, whose origin $\underline{0}$ is the middle of the segment SD, with S being the point of coordinates $\underline{r}_s = (H/2, \pi, 0)$ and D the point of coordinates $\underline{r}_d = (H/2, 0, 0)$, whose axis $z$ is the line passing by $\underline{r}_s$ and $\underline{r}_d$, whose coordinate $r$ is the distance $\|\underline{r}\|$ between $\underline{0}$ and $\underline{r}$, and whose angle $\theta$ is the angle between $\underline{r}$ and the axis $z$, we solve (2.1.5) and show that the equation of the surface $S_{opt}$ is:

$$\frac{r}{R} = F(\theta) \quad (2.1.6)$$

where the function $F(\theta)$ obeys a system of three equations, given below:

$$\frac{1}{\sqrt{F(\theta)^2 + F(\theta)\beta\cos\theta + (\beta/2)^2}}$$
$$= \int_{\theta'=0}^{\pi} \int_{\varphi'=0}^{2\pi} \frac{u(\theta')F(\theta')\sin\theta'\sqrt{F(\theta')^2 + (dF/d\theta')^2}\,d\theta'\,d\varphi'}{\sqrt{F(\theta)^2 + F(\theta')^2 - 2F(\theta)F(\theta')(\cos\theta\cos\theta' + \sin\theta\sin\theta'\cos\varphi')}}, \forall\theta \in \left[0, \frac{\pi}{2}\right] \quad (2.1.7)$$
$$u(\theta)u(\pi-\theta) = \alpha, \forall\theta \in [0, \pi/2]$$
$$\int_{\theta=0}^{\pi} F(\theta)^3 \sin\theta\,d\theta = 2$$

---

[6] As the diffusion equation (2.1.1) is equivalent to Laplace's steady state heat equation, the problem (2.1.1)-(2.1.3) also gives the temperature field in a conductor containing a point heat source placed at $\underline{r}_s$ and having a fixed temperature $T_0$ at its surface. Thus, interestingly, the result (2.1.5) also gives the shape of such a conductor that maximizes the temperature at the point $\underline{r}_d$, with $\underline{J}$ being in this case the heat current.

In (2.1.7), $\alpha$ is an unspecified constant and $u(\theta)$ is an unspecified function, both constrained by the system (2.1.7). In (2.1.6)-(2.1.7), the parameters $R$ and $\beta$ are given by:

$$R = \left(\frac{3V_{max}}{4\pi}\right)^{1/3}, \quad \beta = \frac{H}{R} \quad (2.1.8)$$

with $H$ being the distance between the source $\underline{r}_s$ and the detector $\underline{r}_d$. The numerical resolution of the system (2.1.7) is described in Supplemental Material D [31]. Fig. 1 shows the results obtained for $\beta$ varying from 0.2 to 2 in steps of 0.2, by increasing the distance $H$ while keeping the volume $V_{max}$ constant. The corresponding surfaces $S_{opt}$ are generated by revolving the curves presented in this figure around the axis SD, which is the line connecting the symbols.

*Comment* 1. Modeling the transport using a diffusion equation imposes additional constraints on the optimal shape of the material: (1) the volume $V$ of the object must contain the source $\underline{r}_s$. Otherwise, the neutron flux will be zero everywhere; (2) the optimal propagating medium must be continuous, made up of a union of volumes in contact containing the points $\underline{r}_s$ and $\underline{r}_d$. Indeed, according to point (1), any discontinuous part that does not contain the source will have a zero flux inside. Since the total volume $V_{max}$ of material is fixed, these discontinuous topologies will thus be suboptimal, because they will not bring any positive contribution to the neutron flux at the point $\underline{r}_d$ while consuming useful volume; (3) the optimal propagating medium obtainable in the framework of a diffusive transport cannot contain holes. Indeed, a hole induces neutron losses, as the flux must vanish at its surfaces. Obviously, the constraints (1)-(3) are incompatible with the actual behavior of neutrons in empty areas, e.g. the decay of $\phi(\underline{r}_s \to \underline{r})$ as $1/\|\underline{r}-\underline{r}_s\|^2$ in vacuum far from the source, which can be correctly obtained with the Boltzmann equation by applying to it a 3D Fourier transform acting on the coordinates $\underline{r}$. Identifying these fundamental discrepancies between a diffusive transport and an actual Boltzmann transport will be useful in interpreting the results obtained in section 2.2.

*Comment* 2. By using the resolution procedure described in Supplemental Material C, the result (2.1.5) can be easily generalized. For example, we demonstrate in Supplemental Material E that this result holds for a monoenergetic Boltzmann transport in the case of a heavy low-absorbing material. We then obtain a more general formula of $S_{opt}$ using an expansion of the neutron angular flux on the spherical harmonics basis [32]. In Supplemental Material F, we propose an equation of $S_{opt}$ valid in the framework of the Boltzmann equation, with no restriction on the physicochemical properties of the material nor on the neutron energies. The conclusion of section F is the starting point of the topology optimization procedure developed in this study.

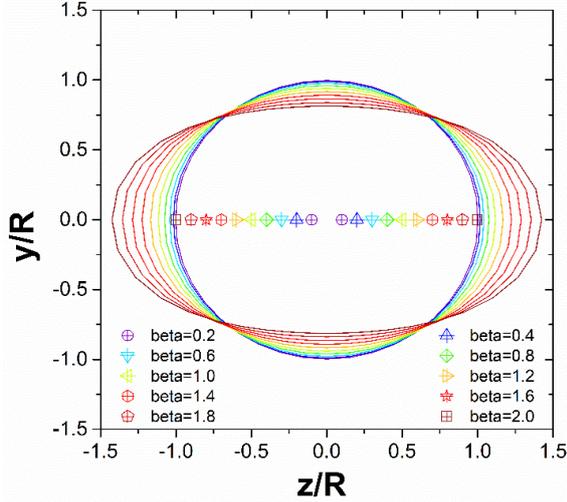

**Figure 1.** Cuts according to the plane $x = 0$ of the surfaces $S_{opt}$ of a propagator that maximizes the total neutron flux at the point $\underline{r}_d$, in the case of a monoenergetic diffusive transport. These surfaces are obtained for $\beta$ ranging from 0.2 (most circle-like surface) to 2 (most elongated surface) in steps of 0.2. The symbols indicate the locations of the points $\underline{r}_s$ and $\underline{r}_d$, which are symmetric with respect to the plane $z = 0$.

2.2. Resolution of the problem using a topology optimization algorithm

In Supplemental Material G, we describe the prototype of an algorithm capable of solving the problem (2.1). This prototype, called A0, was an iterative procedure, which took an arbitrary initial density configuration as a seed, then built from it a more optimal solution, iteration after iteration. Algorithm A0 was based on a result obtained at the end of Supplemental Material F. We noted there that, in the framework of the Boltzmann equation, the surface $S_{opt}$ should be the interface between two regions of space, $\partial L/\partial \rho > 0$ and $\partial L/\partial \rho < 0$. Therefore, to solve (2.1), one just had to find a way to correctly position the available matter in one of these regions, according to the objective and constraints. The first results obtained with algorithm A0 were promising, but their convergence was not satisfactory. In order to improve it, we understood at the end of section G.1 that the matter to be positioned should not flow at an excessive speed from one cell to another. To introduce some viscosity in these matter transfers, the densities $\rho_i$ in the cells $\Theta_i$ had to vary by a small amount between two iterations of the algorithm. A way for implementing this restriction is to impose a quantization of the densities $\rho_i$, which may be written:
(i) for a linear quantization:

$$\rho_i^{(k)} = \rho_{\min} + k\delta\rho, \quad \delta\rho = (\rho_{\max} - \rho_{\min})/M, \quad k \in [0, M] \quad (2.2.1)$$

(ii) for a logarithmic quantization:

$$\rho_i^{(k)} = \rho_{\min}\varepsilon^k, \quad \varepsilon = (\rho_{\max}/\rho_{\min})^{1/M}, \quad k \in [0, M] \quad (2.2.2)$$

where $\rho_{min}$ and $\rho_{max}$ are the lower and upper bounds on the densities $\rho_i$, cf. equation (1.12).
In (2.2.1)-(2.2.2), $k$ numbers a density level $\rho_i^{(k)}$ in the cell $\Theta_i$, and $M+1$ is the total number of density levels. The quantization to be chosen, linear or logarithmic, depends on the problem to be solved, and on the computing resources available. When the number *NPS* of source particles sampled in MCNP is small, e.g. for reducing the computing time, choosing a linear quantization can induce noise or vibrations on the calculated structure. Indeed, the statistical fluctuations on the values $C_i$ can make the density level of a normally empty cell, $\rho_{min}$ (often $10^{-5}$ g/cm$^3$), pass at the higher density level $\rho_{min}+\delta\rho$, which is a value $\gg \rho_{min}$ if $M$ is small, with as a result the creation a false substructure. Conversely, the logarithmic quantization is very stable: it manages the statistical noise well, because a statistical fluctuation on $C_i$ makes at worse the density of a normally empty cell passes from $\rho_{min}$ to $\varepsilon\rho_{min}, \approx \rho_{min}$ even for small values of $M$. There is thus less chances of suboptimal solid structures appearing due to the statistical noise. The stability of the logarithmic quantization comes however at the cost of a sometimes excessive rigidity of the algorithm, which can for instance prevent potentially interesting structures from emerging from initially empty areas. In use, its linear counterpart appears more reactive, more "creative", allowing the fast emergence or suppression of structures in a small number of iterations.
Building on the principle of algorithm A0, the implementation of this last reasoning step leads to the development of an improved topology optimization, A1, described below. Starting from an arbitrary vector of initial densities $\underline{\rho}(0) = (\rho_1(0), \ldots, \rho_N(0))$, where $N$ is the number of cells $\Theta_i$ usable in the optimization procedure, an algorithm capable to solve a general optimization problem of the type (1.8) is:

Algorithm A1 (valid in the case one wishes to maximize $O\varphi$, [7])
For $n$ from 1 to $n_{max}$
  1) Write the MCNP input by taking for densities $\rho_i$ in the cells $\Theta_i$ the densities $\rho_i(n-1)$ obtained at the previous iteration of the algorithm. Add the PERT$i$ cards described section 1. Modify the seed used for the random draws.
  2) Run MCNP, read its output file, extract the derivatives $\partial O\varphi/\partial \rho_i$, calculate the coefficients $C_i$ given in (1.10).
  3) Create $N$ vectors $\underline{F}_i = (C_i, V_i, \rho_i(n-1))$, one per cell $\Theta_i$ usable in the optimization procedure. Reorder them in descending order of $C_i$. Renumber the vectors $\underline{F}_j$ thus classified from $j = 1$ to $j = N$.
  4) Count and select the $N_{lim}$ configurations that contribute positively to $O\varphi$:
$N_{lim} = 0$. For $j$ from 1 to $N$ if $C_j > 0$ then $N_{lim} = N_{lim}+1$ endif endfor
  5) Increment the densities using the quantization (2.2.1) or (2.2.2), taking into account the constraint (1.12), and calculate the weights $P$ of the structures:
For $m$ from 1 to $N_{lim}$
    $P(n,m) = 0$
    For $j$ from 1 to $N$
      If $j \leq m$ then
        For the linear quantization (2.2.1), $temp = \rho_j(n-1)+\delta\rho$
        For the logarithmic quantization (2.2.2), $temp = \rho_j(n-1)\times\varepsilon$
        If $temp \leq \rho_{max}$ then $r(j,m) = temp$ else $r(j,m) = \rho_{max}$ endif
        $P(n,m) = P(n,m) + r(j,m)\times V_j$
      Else
        For the linear quantization (2.2.1), $temp = \rho_j(n-1)-\delta\rho$
        For the logarithmic quantization (2.2.2), $temp = \rho_j(n-1)/\varepsilon$
        If $temp \geq \rho_{min}$ then $r(j,m) = temp$ else $r(j,m) = \rho_{min}$ endif
        $P(n,m) = P(n,m) + r(j,m)\times V_j$
This step can be refined by specifically treating the case $N_{lim} = 0$.
  6) Find the integer $m_{opt}$ for which the weight $P(n,m_{opt})$ is the closest to the limit $P_{max}$ while verifying the constraint $P(n,m_{opt}) \leq P_{max} = \rho_{max}V_{max}$. Note that this instruction cannot penalize the objective: as $m_{opt} \leq N_{lim}$, all of these tested configurations contribute positively to $O\varphi$. The new densities $\rho_i(n)$ are then given by $r(i,m_{opt})$.

---

[7] In the case one wishes to minimize $O\varphi$, the corresponding algorithm is obtained: (i) at step 3, by classifying the vectors $\underline{F}_i$ in increasing order of $C_i$; (ii) at step 4, by taking $C_j < 0$ instead of $C_j > 0$.

In algorithm A1, the index $n$ is the iteration number, and $n_{max}$ is a preset upper bound. For the problems addressed in this study, we indeed advise against stopping the iterations automatically using a convergence criterion. Indeed: (i) in some cases, the calculations seem to have converged, with a slope of the objective that is almost zero over a sometimes long interval, until the algorithm ends hitting a more subtle and more optimal solution, which would have been missed otherwise; (ii) the statistical fluctuations on the objective can be large enough to mislead an automatic stopping criterion.

The step 5 of A1 distributes the available material among the $N$ cells of the tiling, according to the objective and constraints. It also makes it possible to solve the optimization problem by replacing the strict constraint $V = V_{max}$ by a more flexible one, $P = \rho.V \leq P_{max} = \rho_{max} V_{max}$. Indeed: (i) unless we specifically choose the union of cells $\Theta$ so that the volume of any subset of it can be equal to $V_{max}$, it is normally impossible to have $V = V_{max}$ exactly with a union of cells of arbitrary shapes in finite number; (ii) it can be counterproductive to try to verify at all cost the constraint $V = V_{max}$. Suppose for instance one wishes to minimize $O\varphi = \phi(\underline{r}_d)$, as in the counterexample given in Supplemental Material I.2. One may end up in a situation where a part of the volume of the computed shape is added only to verify the constraint $V = V_{max}$, although the addition of this volume increases $\phi(\underline{r}_d)$. In other words, in order to verify the strict constraint $V = V_{max}$, one would end increasing the cost of the device while decreasing its efficiency, which is obviously not desirable.

Suppose now that algorithm A1 converges (which was not guaranteed). If it converges, to validate the proposed topology optimization procedure, we must verify that the result obtained is correct. In section 2.1, we obtained the optimal shape $S_{opt}$ of the propagator in the framework of a monoenergetic diffusive transport. We hence propose there to use this solution as a reference. To verify the assumptions behind the calculations performed in section 2.1, we must use in the MCNP computations a material that is heavy, non-absorbing, diffusive, and in which the energy of the neutrons remains constant throughout their transport. An eligible material is lead. Indeed, lead isotopes have masses ~200 times larger than that of a neutron, which allows a good isotropy of the neutron scatterings in the laboratory frame as well as a low variation of their energies after scattering. They are also little absorbing: for example, the $^{207}$Pb (n,$\gamma$) capture cross-section evaluated in ENDF/B-VIII is 0.7 b at 2.53 $10^{-8}$ MeV. In this section, we will therefore take a material made of 100% of $^{207}$Pb, with a maximum density $\rho_{max} = 11.34$ g/cm$^3$. We now have to find a way to simulate a monoenergetic transport. As mentioned in section 2.1, a simple solution is to take for energy $E_0$ of the source neutrons the thermal energy $kT$, where $T$ is the temperature of the material. In this section, we took the usual ambient temperature, $T = 293$ K. Finally, we need to take a distance $H = ||\underline{r}_d - \underline{r}_s||$ between the source and the detector that is: (i) large enough to have a transport as diffusive as possible, i.e. so that the number of collisions made by the neutrons be large; (ii) but not too large either, otherwise the neutron flux in cell D will be small, and therefore plagued by a large statistical error. Taking into account the natural density of lead $\rho_{max}$ and its elastic scattering cross-section, ~11 b at 2.53 $10^{-8}$ MeV, we took in this section $H = 20.0$ cm and $V_{max} = 10^4$ cm$^3$, so $P_{max} = \rho_{max} V_{max} = 113.4$ kg.

The line passing by S and D being an axis of symmetry of the problem, we can use cylindrical coordinates, $\underline{r} = (r, \theta, z)$, taking for origin $\underline{r} = \underline{0}$ the middle of the segment SD, for axis Oz the line SD, and for $r$ the distance from a point $\underline{r}$ to the axis Oz. In this reference system, the points S and D have the coordinates $\underline{r}_s = (0, 0, -H/2)$ and $\underline{r}_d = (0, 0, H/2)$. By exploiting this axial symmetry to reduce the number of cells, thereby the computing time, we have chosen for cells $\Theta_i$ cylindrical rings of volumes $V_i$ parametrized as follows:

$$\Theta_i(\underline{r}) = \Theta[R_{j-1} \leq r < R_j] \times \Theta[X_k \leq z < X_{k+1}]$$

$$X_k = H\left(\frac{2k-1}{NX} - 1\right), \quad R_j = \begin{cases} 0 & \text{if } j = 0 \\ H(j-1/2)/NR & \text{if } j > 0 \end{cases} \quad (2.2.3)$$

$$i = (NX+1)j + k - NX, \quad k \in [0, NX], \quad j \in [1, NR+1]$$

In (2.2.3), integers $NX$ and $NR$ are used to adjust the pixelisation of the shape to be calculated. The cells $\Theta_{1+NX/4}$ (source S) and $\Theta_{1+3NX/4}$ (detector D) are left empty, and are not usable in the optimization procedure. The weight $P(n)$ of the structure at iteration $n$ is therefore given by the sum of the products $\rho_i(n)V_i$ for all cells $\Theta_i$, with the exception of the source and detector cells:

$$P(n) = \sum_{\substack{i \neq 1+NX/4 \\ i \neq 1+3NX/4}} \rho_i(n) V_i \quad (2.2.4)$$

The neutron propagators designed by algorithm A1 to solve the optimization problem (2.1) are shown in fig. 2. The density configurations, $\rho(\underline{r}, n) = \rho(n).\Theta(\underline{r})$, of these devices are shown in a plane containing the symmetry axis SD. The full 3D structures are generated by revolving these density maps around SD. In fig. 2, the gray scale gives the density values in g/cm$^3$, from $\rho_{min} = 10^{-5}$ g/cm$^3$ (white) to $\rho_{max} = 11.34$ g/cm$^3$ (black). The red cells are the source (left) and the detector (right) cells. The red line indicates the surface $S_{opt}$ of the propagator obtained section 2.1 equations (2.1.6)-(2.1.8) in the approximation of a diffusive transport, for $H = 20.0$ cm and $V_{max} = 10^4$ cm$^3$, i.e. for $\beta = 1.496$ (for reminder, this propagator is a plain solid of density $\rho_{max}$, with no hole). In order to have a good spatial resolution and convergence, we took $NX = 48$ and $NR = 20$ in (2.2.3), and used the linear quantization (2.2.1) with $M = 50$. For each MCNP calculation, we sampled $NPS = 10^8$ source neutrons. The calculations turned on a single 16-CPU machine, lasting ~2 hours per iteration. The initial densities $\rho_i(0)$ were taken uniform, $\rho_i(0) = \rho(0) \; \forall i$, with $\rho(0)$ chosen so that the weight (2.2.4) of the initial structure, $P(0)$, be as close as possible to the maximum weight $P_{max}$, while keeping the condition $P(0) \leq P_{max}$. For the linear quantization (2.2.1), we obtain in Supplemental Material G.2:

$$\rho(0) = \rho_{min} + floor\left(\frac{1}{\delta\rho}\left(\frac{P_{max}}{V_a} - \rho_{min}\right)\right)\delta\rho, \quad V_a = \sum_{\substack{i \neq 1+NX/4 \\ i \neq 1+3NX/4}} V_i \quad (2.2.5)$$

and, for the logarithmic quantization (2.2.2), we find:

$$\rho(0) = \rho_{min} \exp\left[\ln(\varepsilon) floor\left(\frac{1}{\ln(\varepsilon)} \ln\left(\frac{P_{max}}{V_a \rho_{min}}\right)\right)\right] \quad (2.2.6)$$

where the function $floor(x)$ returns the largest integer smaller than $x$. The coefficients $\delta\rho$ and $\varepsilon$ are given in (2.2.1) and (2.2.2). In (2.2.5)-(2.2.6), $V_a$ is the total volume available for the optimization procedure, i.e. the sum of the volumes $V_i$ of the cells $\Theta_i$, minus the volumes of the source and detector cells, $\Theta_{1+NX/4}$ and $\Theta_{1+3NX/4}$.

The evolution of the total neutron flux, $\phi(n)$, in the detector, whose maximization is the goal of these calculations, is given in fig. 3 on the left side as a function of the iteration number $n$. The fluxes $\phi(n)$ are expressed in neutrons/cm$^2$ per source neutron. They were calculated with a tally F4, they are thus averaged fluxes on the volume of $\Theta_{1+3NX/4}$. The fluxes $\phi(n)$ are compared to the flux $\phi_{diff}$ calculated with MCNP for the shape (2.1.6) of the propagator. The MCNP statistical errors on the fluxes $\phi(n)$ being lower than one percent, their error bars are not plotted. Finally, the right-hand side of fig. 3 gives the evolution of the weight, $P(n)$, of the structure in kg with the iteration number $n$. The maximum permissible weight $P_{max}$ is indicated by the red line. Finally, to better visualize how algorithm A1 builds the optimal propagator, we add in Supplemental Material a video, phimax_therm_A1.mp4, that shows the sequence of the density configurations $\rho(n)$ obtained for each $n$ ranging from 0 to 100.

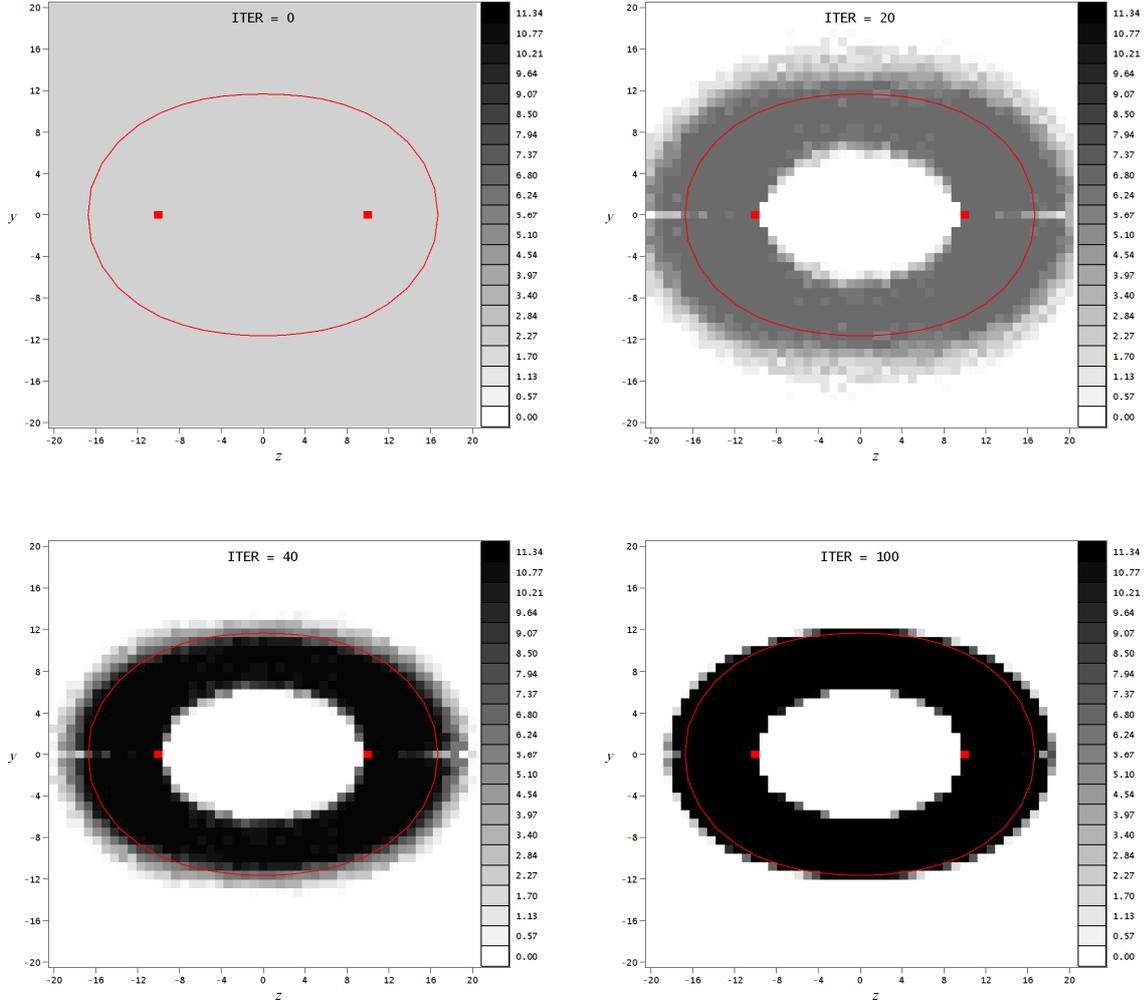

**Figure 2.** Density maps obtained at iterations 0 to 100 of algorithm A1, using the linear quantization (2.2.1) with $M = 50$. The units of axes $y = \pm r$ and $z$ are cm. The gray scale gives the values of the densities $\rho_i(n)$ in g/cm$^3$.

The results obtained are excellent: algorithm A1 correctly accretes the matter available to build a structure close to that of fig. 1. However, it does better than that:
(i) in fig. 2, we note that algorithm A1 designs a hole in the propagator between the source and the detector. This hole[8] allows source neutrons emitted with directions $\underline{\Omega} \propto \underline{r}_d - \underline{r}_s$ to reach the detector in ballistic flight, or after a small number of collisions, without suffering losses by attenuation in the material. This solution could not have been foreseen with a diffusion model, as the latter implicitly assumes that the neutrons undertake a large number of collisions during their transport, and because its boundary condition imposes the nullification of the flux at a surface or in an empty area, hence no hole in the structure.

---

[8] The large apparent size of this hole is an illusion, due to the 2D representation of the structure. As the squares shown in fig. 2 are in fact cross-sections of cylindrical rings, whose volumes increase linearly with their radii, the volume of the hole is actually small compared to that of the structure. This explains why the structures calculated with A1 do not greatly exceed the surface (2.1.6) materialized by the red line.

(ii) As a result, the flux $\phi(n)$ generated by the propagator designed by A1 exceeds the flux $\phi_{diff}$ generated by the propagator designed with the diffusive model, for all $n$. At $n = 0$ in A1, the available space is filled with lead at density $\rho(0)$, which reduces neutron leakage at the surface $S_{opt}$, thus resulting in $\phi(0) > \phi_{diff}$. For the subsequent iterations, the central hole designed by A1 allows a large gain in efficiency, which reaches a factor 16 at the end.

The clear superiority of the solution found by A1 proves that, even for simple cases, performing topology optimization of particle propagators with a diffusion model is a bad idea. A procedure based on a modern transport code, capable to solve the Boltzmann equation with accuracy, has to be used whenever possible, upon pain of suboptimality. This recommendation will become even more evident in the following sections.

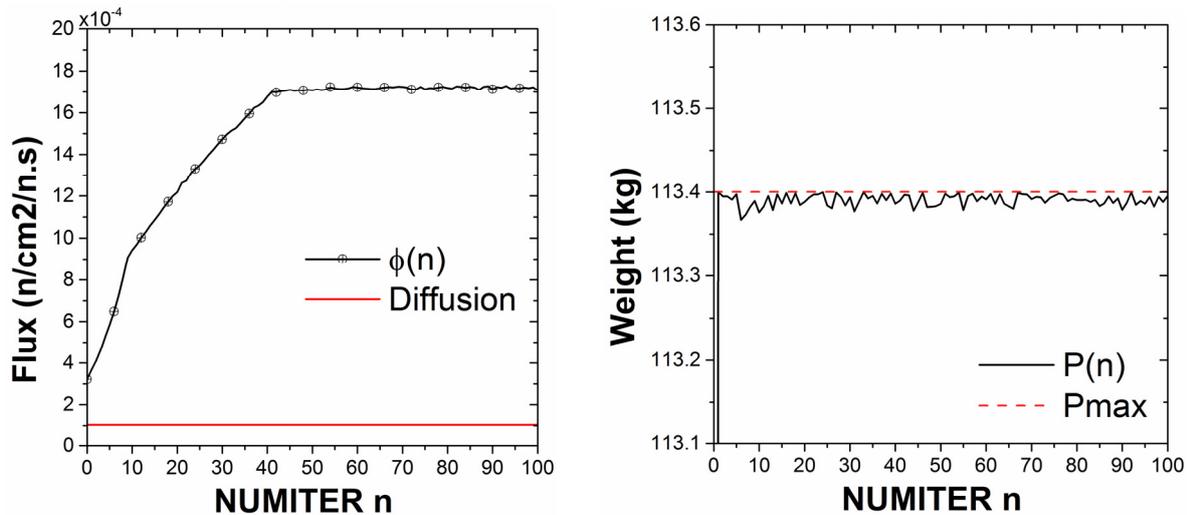

**Figure 3.** (Left) evolution of the neutron flux $\phi(n)$ in the detector cell with the iteration number $n$ (line + circles). The fluxes $\phi(n)$ are given in neutrons/cm$^2$ per source neutron. They are compared to the flux $\phi_{diff}$ calculated with MCNP for the shape obtained section 2.1 equations (2.1.6)-(2.1.8) (red line); (right) evolution of the weight of the structure, $P(n)$, in kg with $n$. The maximum permissible weight, $P_{max}$, is indicated by the red dashed line.

In fig. 2, we note that the matter excavated from the central hole is repositioned on the sides of the structure, along the axis SD. One could then wonder how these results would evolve if the material or the energies of the source neutrons were to be modified. In fig. 4, we hence present the structures calculated with algorithm A1 for: (i) on the left side, a material made of polyethylene, of density $\rho_{max} = 0.94$ g/cm$^3$, of weight $P_{max} = 9.4$ kg, fed by a thermal neutron source of energy $E_0 = kT = 2.53 \cdot 10^{-8}$ MeV; (ii) on the right side, a material made of natural lead, of density $\rho_{max} = 11.34$ g/cm$^3$, of weight $P_{max} = 113.4$ kg, fed by a 14 MeV neutron source typical of a D-T fusion neutron generator [33]. For both calculations, we took $NX = 20$, $NR = 10$, $NPS = 10^8$, $\rho_{min} = \rho_{max}/100$, $H = 20$ cm, and a linear quantization (2.2.1) of the density with $M = 100$. We observe that the optimal shape remains almost unchanged for the polyethylene block containing a thermal source. This result is consistent with the fact that the shape obtained fig. 1 in the framework of a monoenergetic diffusive transport is independent on the nature of the material, provided it is heavy and low-absorbing. On the other hand, in fig. 4 on the right, we note that the use of a 14 MeV source induces a slight asymmetry of the structure compared to the predictions of the model of section 2.1, which is not surprising since we are no longer in the context of a monoenergetic transport. In this case, we observe that an addition of matter is required in the area near the detector to send back towards D a part of the impinging neutrons.

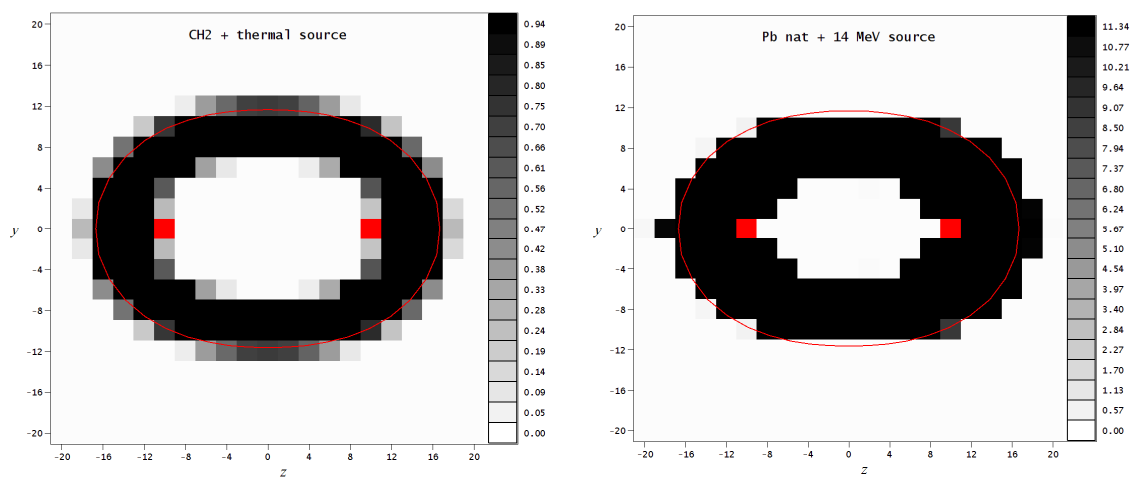

**Figure 4.** (Left) final structure obtained at iteration 150 of A1 for a polyethylene block fed by a thermal source; (right) final structure obtained at iteration 200 of A1 for a natural lead block fed by a 14 MeV source. The units of axes $y = \pm r$ and $z$ are cm. The gray scales of these figures give the values of the densities $\rho_i(n)$ in g/cm$^3$.

*Comment.* The iterative procedure of calculation of the densities $\rho_i(n)$ used at step 5 of A1 does not harness all information on the values of $C_i$, which seems to be a defect at first glance. Indeed, whatever the value of $C_i$ is, small or large, as soon as $C_i$ is greater than $\lambda$, we increase the density $\rho_i(n-1)$ of a constant value $\delta\rho$, independent on $C_i$. We could hence try to improve algorithm A1 by using a gradient-type method, instead of the quantization (2.2.1) or (2.2.2) of the density. However, we show in Supplemental Material G.3 that such a method gives poor results. Hence, in the rest of this study, we will use solely the quantization (2.2.1)-(2.2.2) of the density. This procedure has the advantage of simplicity, and will prove to be robust, as we will show in the following sections that it gives correct results for all the categories of problems addressed in this study. Moreover, from an engineering perspective, quantizing the density of the material is not an absurd idea. Machining a structure whose local density varies continuously from $\rho_{min}$ to $\rho_{max}$, by taking an infinity of intermediate values seems unrealistic. On the other hand, it should be possible to machine cells with intermediate densities $\rho_i$ in finite number, using for example, instead of the base solid material, honeycomb structures, powders, hollow and solid balls of variable sizes, or alloys and chemical compounds whose additional isotopes interact little with the neutrons.

2.3. To go further: optimization of a neutron shield.

In the previous section, we validated a topology optimization procedure, A1, capable of solving a problem of the type (2.1). We hence propose to use it here to address a classical problem in radiation protection. Suppose an intense source of 14 MeV neutrons is located in a building, lined with concrete walls. In an area D of the building are human operators, or fragile semiconductor detectors adversely affected by an excessive neutron fluence. It is therefore wished to reduce as much as possible the neutron fluence in D. To achieve this, the operators have a budget to buy a maximum volume $V_{max}$ of polyethylene (PE) [9]. Where to position this material and how to shape it to minimize the neutron fluence in D?

---

[9] Polyethylene $(CH_2)_n$ is an efficient neutron shield. Indeed, hydrogen is one of the isotopes that best slow down neutrons, because of its high scattering cross-section on one hand, and because its nuclear mass is close to that of the neutron on the other hand. Once thermalized, neutrons can be captured more easily.

In this section, we do not plan on performing a full engineering study, with a detailed modeling of a work area and its equipment for example. We rather want to see if algorithm A1 can succeed in managing complex neutron trajectories, over large distances, in materials such as concrete or PE, where the neutron energies can vary over 10 orders of magnitude, from 14 MeV down to a fraction of the thermal energy $kT = 2.53\ 10^{-8}$ MeV. We hence propose to make the following simplifications to speed up the calculations: (1) we impose an axial symmetry to the problem, and we retake the cylindrical coordinates used in section 2.2; (2) the neutron source is modeled by a point, placed on the axis SD, at the coordinates $r = 0$, $z = -H/2$; (3) the concrete walls are modeled by two cylinders, ~20 cm thick, 500 cm in radius, placed between the planes $z = -1.2H$ and $z = -H(1+1/NX)$ for the left wall, and between $z = H(1+1/NX)$ and $z = +1.2H$ for the right wall. The concrete used is a heavy one, of density 3.4 g/cm$^3$, whose composition is given in Supplemental Material H; (4) as in section 2.2, the space available is tiled using cylindrical rings $\Theta_i$, whose parametrization is given equation (2.2.3), with $H = 100$ cm, $NX = 20$ and $NR = 10$. The maximum radius of the available space, $R_{max} = H(1+1/2NR) \approx 100$ cm, is chosen smaller than the radius, 500 cm, of the concrete walls, to mimic a constraint on the maximum clutter of the work area; (5) the cell D, in which it is wished to minimize the neutron flux, is the cell $\Theta_{1+3NX/4}$, cf. (2.2.3), centered on the point $r = 0$, $z = H/2$. This cell is left empty, as well as the cell $\Theta_{1+NX/4}$ containing the source. For this study, we have taken a volume $V_{max}$ of PE equal to $1.25\ 10^6$ cm$^3$, with $\rho_{max} = 0.94$ g/cm$^3$ (standard volume density of PE). The calculations have been performed with $NPS = 10^8$, $\rho_{min} = \rho_{max}/1000$, and the linear quantization [10] (2.2.1) of the density with $M = 20$. Finally, in order to eliminate the statistically absurd values of the objective derivatives in the computations, we have filtered the values (1.11) of PERT$_i$ calculated by MCNP using the following procedure: if the statistical error estimated by MCNP on PERT$_i$ exceeds 95%, we replace PERT$_i$ by 0.

Fig. 5, we show the evolution of the density profiles $\rho_i(n)$ obtained by A1 between the iterations $n = 0$ and 400. A video, shield_concrete.mp4, given in Supplemental Material allows to better visualize the sequence of these profiles. The gray scale of the figures gives the density values in g/cm$^3$. The 3D density configurations are cut according to a plane containing the axis of symmetry SD, the squares drawn are thus cross-sections of the cylindrical rings used to tile the space. The white squares are cells containing PE at density $\rho_{min}$, the black ones contain PE at density $\rho_{max}$, and the red squares are the source (left) and detector (right) cells. The concrete walls are not shown in the figures, but they line the density maps directly on the left and on the right, cf. fig. 7. The computations were performed on a 16-CPU machine, and lasted ~45-60 min per iteration. They were carried out starting from an uniform initial configuration of densities, $\rho_i(0) = \rho(0)\ \forall i$, with $\rho(0)$ given in (2.2.5). In fig. 6, we also plot: (i) on the right side, the evolution of the weight $P(n)$ of the structures as a function of the iteration number $n$; (ii) on the left side, the evolution with $n$ of the neutron flux $\phi(n)$ in cell D, expressed in neutrons/cm$^2$ per source neutron. The statistical errors on $\phi(n)$ being lower than 1%, the associated error bars are not plotted. The fluxes $\phi(n)$ are compared with a flux $\phi_{int}$, calculated with MCNP for an intuitive solution of the problem addressed in this section. This intuitive solution is obtained by condensing all available PE in a cylinder of axis SD, of density $\rho_{max}$, of length $L = 2H(1+1/NX)$ (the cylinder is thereby in contact with both concrete walls) and of radius $R = (V_{max}/(\pi L))^{1/2}$.

---

[10] We also tested the logarithmic quantization (2.2.2) of the density, with poor results. As noted in section 2.2, this quantization is rigid, too much for the problem addressed here. Because of this rigidity, the calculated structure locks in a suboptimal configuration and no longer evolves after, whatever the iteration number is.

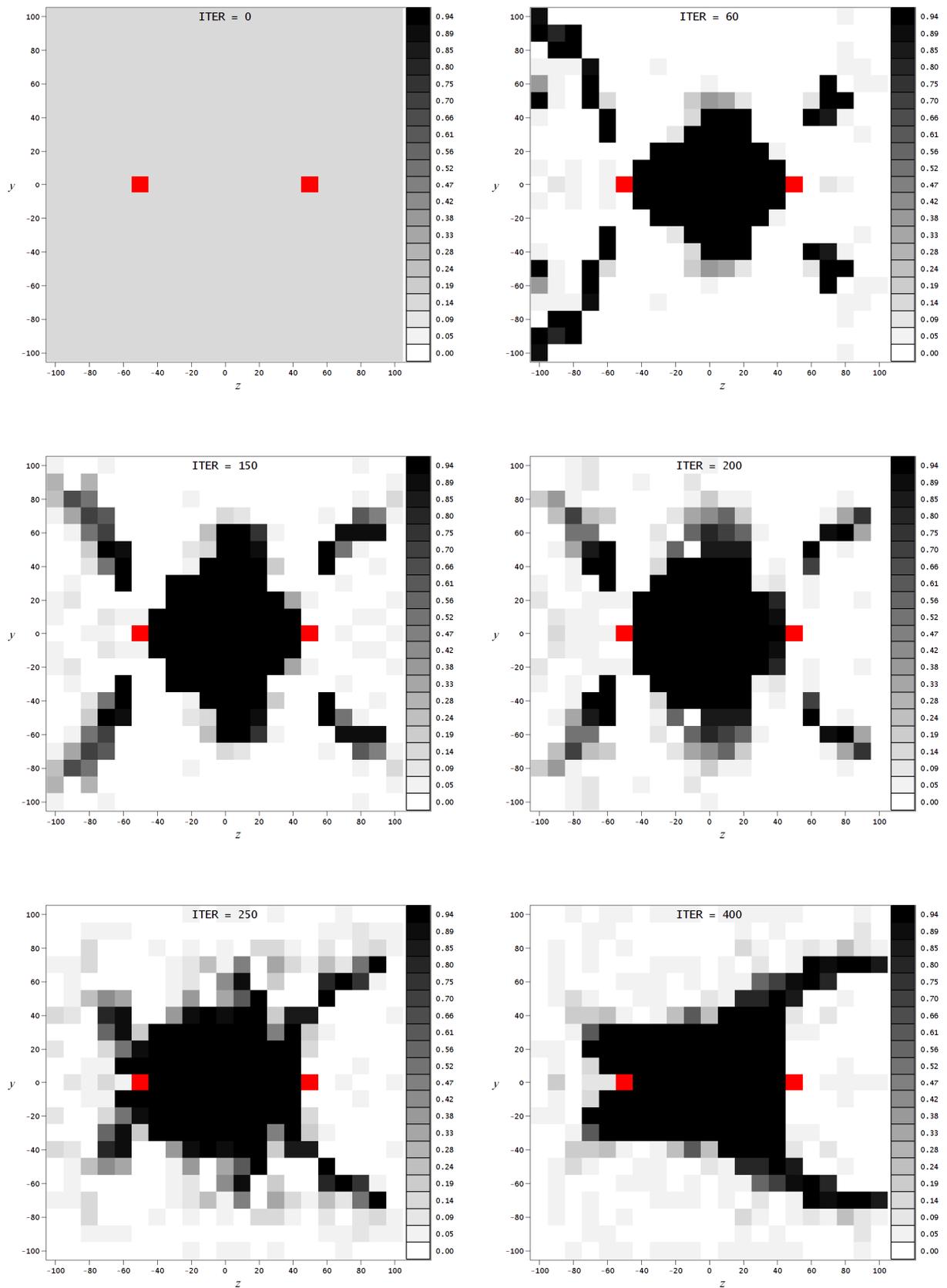

**Figure 5.** Density maps obtained at iterations 0 to 400 of algorithm A1. The units of the axes $y = \pm r$ and $z$ are cm. The gray scale gives the values of the PE densities $\rho_i(n)$ in g/cm$^3$.

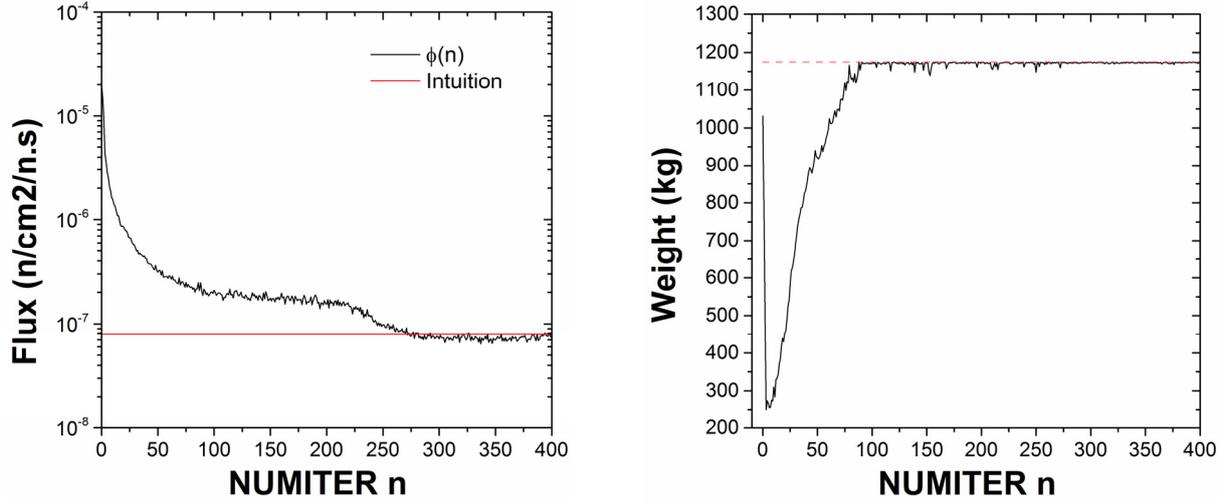

**Figure 6.** (Left) evolution of the neutron flux, $\phi(n)$, in the cell D with the iteration number $n$. The fluxes $\phi(n)$ are given in neutrons/cm² per source neutron, and are compared to the flux $\phi_{int}$ of the intuitive solution, indicated by the red line; (right) evolution of the weight of the structure, $P(n)$, in kg with $n$. The maximum permissible weight, $P_{max} = \rho_{max} V_{max}$, is indicated by the red dashed line.

The results obtained with algorithm A1 are convincing. First, we note in fig. 6 that the algorithm manages to reduce the neutron flux in the cell D by a factor 300, under the level $\phi_{int}$ obtained with a human intuitive solution. But the most intriguing result is the behavior of the algorithm itself between its iterations 0 and 100. We observe on the density maps shown in fig. 5, or better in the video shield_concrete.mp4, the construction and evolution of three "antennas" of matter, numbered $a_1$, $a_2$ and $a_3$, organized around the central block of PE that serves as the main shield. These antennas, which grow in symbiosis with the central shield, are a clever solution found by the algorithm to the optimization problem posed. By analyzing a posteriori this structure, in fig. 7, we note that these antennas suppress the neutron trajectories $T_1$, $T_2$ and $T_3$ that circumvent the central shield to reach the detector after having undertaken one or several series of collisions in the concrete walls. Algorithm A1 thus demonstrates here it can manage complex collision patterns, identifying and closing the most penalizing holes in the shield. It proves also adaptive, by eliminating the structures that have become obsolete as the shield efficiency is progressively improved. For example, the antenna $a_3$ is eliminated starting from iteration ~100, when the growth of the central shield closes by itself the trajectories $T_3$. Likewise, the antenna $a_1$ is absorbed by the central shield starting from iteration ~250, because its positioning at a large distance from the axis SD is too costly in volume (the volume of the cylindrical rings increase linearly with the distance to the axis SD). By bringing $a_1$ closer to the axis SD, the algorithm thus frees matter that can be repositioned in more useful places, e.g. for closing the antenna $a_2$ so that it touches the right-hand wall.

It is however important to temper these successes, no matter how interesting they are. Indeed, in this section, contrary to the problem addressed in sections 2.1-2.2, we do not have a reference solution. Therefore, we do not have a solid proof that algorithm A1 does converge towards the best structure possible. In Supplemental Material I, we indeed find two counterexamples that illustrate some of the limitations of algorithm A1. We show there that: (i) A1 has trouble with disassembling massive structures. This issue explains why we use throughout this study an uniform initial density configuration, as diluted as possible, with no matter already condensed in a suboptimal configuration; (ii) an overly coarse tiling of the available volume, whether it is chosen by default to reduce the computing time or due to constraints on the machining process of the object, can lead to a suboptimal structure. This limitation is a well-known issue in topology optimization.

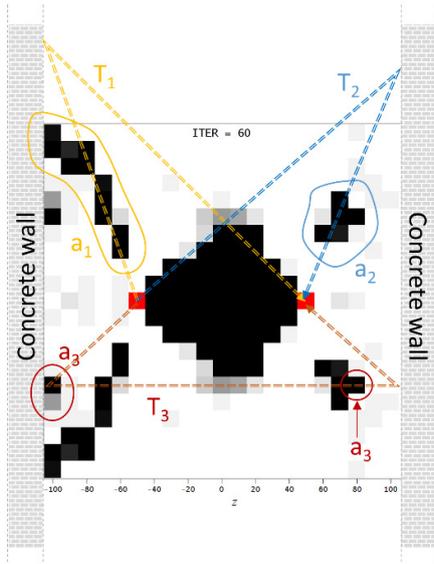

**Figure 7.** Antennas $a_1$, $a_2$, $a_3$, and trajectories $T_1$, $T_2$, $T_3$ for the density configuration obtained at iteration $n = 60$.

3. Problem n°2: optimization of an energy spectrum in a region of space

Suppose an agency operates a neutron source. Its users would like that the energy spectrum of the neutrons, $\phi(E)$, in an area D of the installation be equal to an objective spectrum, $\phi_{obj}(E)$. This objective spectrum could be a Gaussian centered on a resonance, or an epithermal spectrum, or the spectrum of a nuclear device of interest for example. To do this, the agency has obtained a budget to buy a quantity of materials. How to organize these materials so that $\phi(E)$ be as close as possible to $\phi_{obj}(E)$?

Answering this problem is interesting for two reasons: (i) it has obvious applications in nuclear science and engineering, e.g. for silicon doping, isotope production or for integral experiments, in which experimenters seek to measure nuclear reaction rates in realistic spectra [11] that are more complex or far from those obtainable with conventional neutron sources [12]. Solving this problem could thereby give access to customizable spectra that better agree the needs of users; (ii) in the smaller scope of this study, it is a good opportunity to study the behavior of A1 in the challenging case of a non-linear objective functional $O$.

In section 3.1, we will hence present the mathematical bases of a resolution procedure, then we will show how to adapt algorithm A1 to this problem. In section 3.2 and in Supplemental Material N, we will validate this resolution procedure on two particular problems for which we found analytical solutions. Finally, in section 3.3, we will use algorithm A1 to mimic the spectrum of a sodium-cooled fast reactor, using only a volume of lead cleverly machined and a layer of boron, a remarkable result opening interesting perspectives.

3.1. Description of the optimization procedure

Consider two energy spectra, $f(E)$ and $g(E)$. In order to compare them, we need a formula of the distance $d(f, g)$ that separates them, therefore we need a definition of their scalar product, $<f, g>$. In this study, we will use the following one:

---

[11] We can cite e.g. the neutron energy spectra in some nuclear reactors, which allow to diversify/constraint the information used to build the nuclear data libraries, or the neutron spectra generated in old stars or supernovas, whose replications are useful to refine our knowledge of nucleosynthesis by r-process or s-process.

[12] Among them are the usual 14 MeV D-T sources, Am-Be sources, $^{252}$Cf sources, etc.

$$\langle f, g \rangle = \int_{E=0}^{+\infty} f(E)g(E)\omega(E)dE \quad (3.1.1)$$

In (3.1.1), $\omega(E)$ is a weight function, real and positive. From this definition of the scalar product comes that of the norm of a spectrum $f$. If it exists, this norm, denoted $\|f\|$, is given by:

$$\|f\| = \sqrt{\langle f, f \rangle} \quad (3.1.2)$$

Now suppose that the neutrons have in the area D of the installation an energy spectrum $\phi(E, \underline{x})$. This spectrum is a function of a vector $\underline{x}$ of parameters, built from the densities $\underline{\rho}$ and/or the isotopic compositions $\underline{\chi}$ of the cells $\underline{\Theta}$ crossed by the neutrons. Using the definitions (3.1.1)-(3.1.2), solving the problem of this section thus amounts to finding the vector $\underline{x}_{opt}$ that minimizes the distance $d(\phi, \phi_{obj})$ between $\phi(E, \underline{x})$ and the objective spectrum $\phi_{obj}(E)$, given by:

$$d(\phi, \phi_{obj}) = \left\| \frac{\phi(E, \underline{x})}{\|\phi(E, \underline{x})\|} - \frac{\phi_{obj}(E)}{\|\phi_{obj}(E)\|} \right\| \quad (3.1.3)$$

Using the definitions (3.1.1)-(3.1.2), we note that:

$$d(\phi, \phi_{obj})^2 \propto 1 - \frac{\langle \phi, \phi_{obj} \rangle}{\|\phi\| \|\phi_{obj}\|} = 1 - \cos(\phi, \phi_{obj}) \quad (3.1.4)$$

Denoting $V_D$ the volume of the area D, the problem of this section can thus be rewritten:

$$\min_{\underline{x}} O\phi(\underline{x})$$
$$\text{s.t.} \quad B(\underline{x})\varphi(\underline{r}, E, \underline{\Omega}, \underline{x}) = Q(\underline{r}, E, \underline{\Omega})$$
$$P(\underline{x}) \leq P_{max}$$
$$\text{with} \quad O\phi(\underline{x}) = \frac{1}{2} - \frac{1}{2} \frac{\langle \phi(E, \underline{x}), \phi_{obj}(E) \rangle}{\|\phi(E, \underline{x})\| \|\phi_{obj}(E)\|} \quad (3.1.5)$$
$$\phi(E, \underline{x}) = \frac{1}{V_D} \int_{\underline{r} \in \mathfrak{R}^3} \int_{\underline{\Omega} \in 4\pi} \Theta[\underline{r} \in D] \varphi(\underline{r}, E, \underline{\Omega}, \underline{x}) d\underline{r} d\underline{\Omega}$$

We note that $O\phi \in [0, 1]\ \forall \phi$ and $\phi_{obj}$, its minimum 0 being reached if and only if $\phi(E, \underline{x}) \propto \phi_{obj}(E)$, which is the objective sought. Unlike the one used throughout section 2, we observe that the objective functional $O$ (3.1.5) is non-linear, due to the norms of the spectra $\phi$ and $\phi_{obj}$.
The solution $\underline{x}_{opt}$ of the problem (3.1.5), i.e. the optimal properties of the material to use so that $\phi(E, \underline{x})$ be as close as possible to $\phi_{obj}(E)$, will then obey the now usual system of equations:

$$L(\underline{x}, \lambda) = O\phi(\underline{x}) - \lambda(P(\underline{x}) - P_{max})$$
$$\frac{\partial L}{\partial \lambda} = 0, \quad \frac{\partial L}{\partial \underline{x}} = \underline{0} \Leftrightarrow C_i = \frac{\partial O\phi}{\partial x_i} \left( \frac{\partial P}{\partial x_i} \right)^{-1} = \lambda, \forall i \quad (3.1.6)$$

Equations (3.1.1)-(3.1.5) are valid for continuous energy spectra. However, a measured or a computed spectrum is often given in the form of a histogram, which can be written:

$$\phi(E) = \sum_{j=0}^{NBIN-1} \phi_j \Theta[E \in B_j], \quad B_j = [E_j, E_{j+1}] \quad (3.1.7)$$

In this case, denoting $\underline{\phi} = (\phi_0, \ldots, \phi_{NBIN-1})$ the values of the histogram of $\phi$, the terms that appear in the problem (3.1.5)-(3.1.6) can be rewritten:

$$\langle \underline{\phi}, \underline{\phi}_{obj} \rangle = \sum_{j=0}^{NBIN-1} \phi_j \phi_{obj\,j} \omega_j, \quad \|\underline{\phi}\| = \sqrt{\langle \underline{\phi}, \underline{\phi} \rangle}, \quad \omega_j = \int_{E=E_j}^{E_{j+1}} \omega(E) dE \quad (3.1.8)$$

With these notations, the derivatives of the objective appearing in (3.1.6) are given by:

$$\frac{\partial O\phi}{\partial x_i} = -\frac{1}{2\|\underline{\phi}_{obj}\| \|\underline{\phi}\|} \left( \sum_{j=0}^{NBIN-1} \frac{\partial \phi_j}{\partial x_i} \phi_{obj\,j} \omega_j - \frac{\langle \underline{\phi}, \underline{\phi}_{obj} \rangle}{\|\underline{\phi}\|^2} \sum_{j=0}^{NBIN-1} \frac{\partial \phi_j}{\partial x_i} \phi_j \omega_j \right) \quad (3.1.9)$$

Suppose now that the parameters $x_i$ are the densities $\rho_i$ of the cells $\Theta_i$ used to tile the space. To solve the optimization problem (3.1.5)-(3.1.6), one must compute (i) the bins $\phi_j$ of the spectrum $\underline{\phi}$, which can be done using the code MCNP with e.g. a tally F4 in the cell D and a card E, see [26] section 5.6.3, and (ii) the derivatives $\partial \phi_j / \partial \rho_i$ of the bins $\phi_j$ with respect to the densities $\rho_i$, which can be done by adding PERT cards in the MCNP input file, as described in section 1 and used in section 2. The command lines to add to the MCNP input file are the following:

F4 *integer designating the cell D*
E4 *list of the bin edges $E_j$ of the energy spectrum to be calculated*
PERT1:n  CELL=1  MAT=m  RHO=−$\rho_1$(1+$\varepsilon$)  METHOD=2
…
PERT*N*:n  CELL=*N*  MAT=m  RHO=−$\rho_N$(1+$\varepsilon$)  METHOD=2

After calculation, the MCNP output files contain the values of $\phi_j$ as well as lists of terms PERT*ij* = $(\partial \phi_j / \partial \rho_i) \varepsilon \rho_i$ with their statistical errors, which give access to the derivatives (3.1.9) sought. All that remains now is to write the optimization algorithm. Here, we must minimize $O\phi$, so we must position the available matter where an increase $\delta\rho$ of $\rho_i$ reduces $O\phi$, i.e. where $C_i < 0$, cf. (3.1.6). To solve a problem of the type (3.1.5), a simple solution is hence to use the algorithm A1 described in section 2.2, by simply replacing its terms $C_i$ by those given in (3.1.6)+(3.1.9).

3.2. Validation of the optimization procedure: mimicking an objective spectrum with a screen

Suppose that the neutron source operated by the aforementioned agency is a standard 14 MeV generator [33], positioned at a point $\underline{r}_s$. A thin cylindrical screen of thickness $dz$ and of axis SD is positioned between the source S and a detector D positioned at a point $\underline{r}_d$. A hole is machined at the center of the screen, along the axis SD, in order to allow the source neutrons emitted with directions $\propto \underline{r}_d - \underline{r}_s$ to pass through the screen without interaction. The screen is made of $N-1$ different isotopes, numbered from $k = 1$ to $N-1$, whose atomic fractions $\chi_k(\underline{r})$ can vary with the position $\underline{r}$ in the screen. The fraction $\chi_N$ is that of the vacuum, whose interaction cross-section $\Sigma_N$ is set to zero. The use of this vacuum fraction allows to vary locally the atomic density of the screen. How to optimize the distribution $\underline{\chi}(\underline{r}) = (\chi_1(\underline{r}), \ldots, \chi_N(\underline{r}))$ of matter in the screen so that the neutron spectrum $\phi(E)$ in D be as close as possible to an objective spectrum $\phi_{obj}(E)$? In section 3.2.1, we will find an analytical solution to this problem, which we will use in section 3.2.2 as a reference to test and validate the optimization procedure described in section 3.1.

3.2.1. Construction of a reference solution

For a screen sufficiently thin, the neutrons that interact in it will perform at most one collision. In this framework, by using an integral version of the Boltzmann equation, we demonstrate in Supplemental Material J that the flux $\phi_1(\underline{r}_d,E)$ of neutrons that have passed through the screen is at the point $\underline{r}_d$:

$$\phi_1(\underline{r}_d, E, \underline{\chi}) = \int_{\underline{r} \in V} \frac{\underline{\chi}(\underline{r}) \cdot \underline{\eta}(\underline{r}, E)}{\underline{\chi}(\underline{r}) \cdot \underline{1}} d\underline{r} \quad (3.2.1)$$

where the components of the vector $\underline{\eta}(\underline{r},E) = (\eta_1(\underline{r},E), \ldots, \eta_N(\underline{r},E))$ are given by:

$$\eta_{k \neq N}(\underline{r}, E) = \frac{Q_0 \Sigma_k(E_0 \to E, \mu(\underline{r}))}{8\pi^2 \|\underline{r} - \underline{r}_s\|^2 \|\underline{r} - \underline{r}_d\|^2}, \quad \eta_N(\underline{r}, E) = 0$$

$$\mu(\underline{r}) = \frac{\underline{r} - \underline{r}_s}{\|\underline{r} - \underline{r}_s\|} \cdot \frac{\underline{r}_d - \underline{r}}{\|\underline{r}_d - \underline{r}\|}$$

(3.2.2)

In this formula, $Q_0$ is the source intensity, $V$ the volume of the screen, and $\Sigma_k(E_0 \to E, \mu = \underline{\Omega} \cdot \underline{\Omega}')/(2\pi)$ the macroscopic cross-section of transition from the state $(E_0, \underline{\Omega}')$ to the state $(E, \underline{\Omega})$ after a collision with an isotope $k$. In formula (3.2.1), note that the uncollided 14 MeV flux is removed by construction, see Supplemental Material J.

We want to find the vector $\underline{\chi}_{opt}(\underline{r})$ that minimizes the distance between the flux $\phi_1(\underline{r}_d,E,\underline{\chi})$ at the output of the screen and an objective spectrum $\phi_{obj}(E)$. We must thus solve the optimization problem (3.1.5), in which the vector $\underline{x}$ is here the vector $\underline{\chi}(\underline{r})$. By using a well-chosen Lagrangian then by performing a calculus of variations similar to that used to obtain the result (2.1.5), we show in Supplemental Material K that the optimal vector $\underline{\chi}_{opt}(\underline{r})$ obeys a system of equations, given below:

$$\langle \eta_k(\underline{r}, E), \phi_1(\underline{r}_d, E, \underline{\chi}_{opt}) \rangle = \frac{\langle \eta_k(\underline{r}, E), \phi_{obj}(E) \rangle}{(1 - \lambda) \|\phi_{obj}\|}, \forall \underline{r} \in V, \forall k \in [1, N-1] \quad (3.2.3)$$

where $\lambda$ is a Lagrange multiplier and $\langle . \rangle$ designates the scalar product (3.1.1).
We now propose to use this result to design a test of the optimization procedure A1, applied to a type (3.1.5) problem. For this, we must find a physical configuration that is: (i) simple enough so we can manage to solve the equations (3.2.3); (ii) yet realistic enough so that the solution found can be finely compared with the predictions of algorithm A1, so with the output of a MCNP calculation. As the MCNP code incorporates state-of-the-art particle interaction models, we therefore have to solve the system (3.2.3) without making approximations. The aim of this section is to identify a physical configuration verifying these contradictory requirements.

In (3.2.2), the transition cross-sections $\Sigma_k$ can be rewritten as follows:

$$\Sigma_k(E_0 \to E, \mu(\underline{r})) = \Sigma_k(E_0) f_k(E_0 \to E, \mu(\underline{r})) \quad (3.2.4)$$

where function $f_k(E_0 \to E, \mu(\underline{r}) = \underline{\Omega}.\underline{\Omega}')/(2\pi)$ is the probability density of transition from a state $(E_0, \underline{\Omega}')$ to a state $(E, \underline{\Omega})$ after a scattering on an isotope $k$. This function has a simple expression when the neutron-nucleus scatterings are elastic and isotropic in the center of mass frame. In this case, $f_k$ is indeed given by, see [34] pp 436-437 [13]:

$$f_k(E_0 \to E, \mu(\underline{r})) = \frac{\Theta[\alpha_k \leq E/E_0 \leq 1]}{(1-\alpha_k)E_0} \delta\left(\mu(\underline{r}) - \left(\frac{A_k+1}{2}\sqrt{\frac{E}{E_0}} - \frac{A_k-1}{2}\sqrt{\frac{E_0}{E}}\right)\right)$$

$$\alpha_k = \left(\frac{A_k-1}{A_k+1}\right)^2, \quad A_k = \frac{M_k}{m_n} \quad (3.2.5)$$

where $M_k$ is the mass of isotope $k$ and $m_n$ the neutron mass.

However, for most nuclei, the formula (3.2.5) is valid only at energies $E$ lower than 10-100 keV. Indeed, at higher energy, inelastic scattering or even n-xn reaction channels open. Hence, for a conventional 14 MeV source, only scatterings on $^1$H, a nucleus devoid of nuclear structure, can be relatively accurately described by formula (3.2.5). Furthermore, only $^1$H has a coefficient $\alpha$ low enough to effectively slow down the impinging neutrons in a single collision. So, if one wants to simulate a general objective spectrum $\phi_{obj}$, which can have components at energies below the MeV, there is little choice: the screen must contain hydrogen. In order to use the result (3.2.3) as a precision test of the optimization procedure A1, we hence propose to make the simplest compromise, namely to take $N = 2$ in (3.2.3), with $\chi_1(\underline{r})$ being a hydrogen fraction and $\chi_2(\underline{r})$ being the void fraction.

Even in such a simple case, however, the calculation of $\chi_{1opt}$ remains lengthy and is therefore put in Supplemental Material L. By taking: (i) cylindrical coordinates $\underline{r} = (r,\theta,z)$, whose axis Oz is the axis SD, with $r$ being the distance from the point $\underline{r}$ to the axis SD; (ii) a thin screen of thickness $dz$ placed at $z = z_e$; (iii) a source positioned at the point of coordinates $r = z = 0$; (iv) a detector placed at the point of coordinates $r = 0, z = z_d$, we show that the optimal concentration $\chi_{1opt}(\underline{r})$ of hydrogen in the screen depends only on $r$ and is given by:

$$\chi_{1opt}(r) = C \frac{\Theta[\mu(r) \in [\sqrt{1-A^2}, 1]]}{2\mu(r)^2 + A^2 - 1} \frac{r^2 + z_e(z_d - z_e)}{\sqrt{r^2 + z_e^2}\sqrt{r^2 + (z_d - z_e)^2}} Z(\mu(r)) \quad (3.2.6)$$

with:

$$Z(x) = \phi_{obj}(E_0 g(x))g(x) + \phi_{obj}(E_0 g(-x))g(-x)$$

$$g(x) = \left(\frac{x + \sqrt{x^2 + A^2 - 1}}{A+1}\right)^2, \quad A = \frac{M}{m_n} \quad (3.2.7)$$

$$\mu(r) = \frac{z_e(z_d - z_e) - r^2}{\sqrt{r^2 + z_e^2}\sqrt{r^2 + (z_d - z_e)^2}}$$

---

[13] Contrary to what is indicated in ref. [34] pp 437, this formula is also valid for $A_k < 1$, not just for $A_k \geq 1$.

where $C$ is a constant and $M$ is the proton mass. For $N = 2$, we can finally note that $\chi_{1opt}(r)$ is directly proportional to the volume density $\rho_{opt}(r)$ of hydrogen to be used in the screen:

$$\rho_{opt}(r) = \beta \chi_{1opt}(r) \quad (3.2.8)$$

*Comment.* For the calculations performed in this section as well as in Supplemental Material N, we took $A = 0.99917$, the value given in the transport data library (JEFF-3.1) we used in MCNP. This value has a large number of significant figures, on one hand, and is very close to 1, on the other hand. One could therefore be tempted to delete some of its digits, or even more simply to take $A = 1$. Taking $A = 1$ would by the way greatly simplify the calculations of Supplemental Material L, as well as formulas (3.2.6)-(3.2.7). However, making these approximations would induce a large error in the results. Indeed, as shown in Supplemental Material M, the optimal density profile $\rho_{opt}(r)$ and the spectrum $\phi_1(E,\underline{r}_d)$ appear to be very sensitive to the value of $A$.

3.2.2. Resolution of the optimization problem with algorithm A1

We can now use the solution (3.2.6)-(3.2.8) as a reference for testing algorithm A1. For this task, we take a distance $H = 40$ cm between the source S and the detector D. The 14 MeV isotropic neutron source is positioned at the point of coordinates $r = 0$, $z = -H/2$. Detector D is modeled using a sphere of radius $R' = 0.5$ cm centered at the point of coordinates $r = 0$, $z = H/2$. We then model in the MCNP input a cylindrical screen of thickness $dz = 0.1$ cm comprised between the planes $z = -dz/2$ cm and $dz/2$, filled by a material made of 100% of $^1$H. This screen is subdivided into $NR$ cylindrical rings $\Theta_i(\underline{r})$, whose densities $\rho_i$ are to be optimized, which are parametrized as follows:

$$\Theta_i(\underline{r}) = \Theta[R_{i-1} \leq r < R_i] \times \Theta\left[-\frac{dz}{2} \leq z < \frac{dz}{2}\right]$$
$$R_i = R_{min} + \frac{(R_{max} - R_{min})}{NR}i, \quad i \in [1, NR] \quad (3.2.9)$$

For this study, we have taken $NR = 40$, $R_{min} = 16$ cm and $R_{max} = 20$ cm. These values are chosen to narrow the admissible area in which algorithm A1 can place hydrogen, based on the theoretical predictions (3.2.6)-(3.2.8). This reduction in the allowable zone, associated with the high $NR$ value taken here, allows to have a fine binning of the optimized density profile while keeping the computing time reasonable. For these calculations, we generate $NPS = 5.10^9$ source neutrons, whose directions are isotropically sampled in the solid angle delimited by the angles $\theta_{min} = \arctan(2R_{min}/H)$ and $\theta_{max} = \arctan(2R_{max}/H)$. This procedure allows to save computing time by forcing the neutrons to pass through the structure (3.2.9) while eliminating the 14 MeV peak of uncollided neutrons from the spectrum. The minimum and maximum volume densities of $^1$H in the rings $\Theta_i$ are taken equal to $\rho_{min} = 10^{-5}$ g/cm$^3$ and $\rho_{max} = 1$ g/cm$^3$.

To implement the resolution procedure described in section 3.1, we must now choose the energy binning of the histograms of spectra $\phi$ and $\phi_{obj}$, cf. (3.1.7). In this section, we take energy bins $B_j$ defined as follows:

$$E_j = E_{min}(E_{max}/E_{min})^{\frac{j-1}{NBIN-1}}$$
$$B_0 = [0, E_{min}], \quad B_{j \in [1, NBIN-1]} = [E_j, E_{j+1}] \quad (3.2.10)$$

with $NBIN = 41$, $E_{min} = 10^{-6}$ MeV and $E_{max} = 20$ MeV. For this test, we choose for objective spectrum $\phi_{obj}$ a thin Gaussian centered on the energy $E_{obj} = 100$ keV. Its histogram $\underline{\phi}_{obj}$ has for components:

$$\underline{\phi}_{obj_j} = \Theta[E_{obj} \in B_j], \forall j \quad (3.2.11)$$

We must now choose the weights $\omega_j$ in the expression (3.1.8) of $<\underline{\phi},\underline{\phi}_{obj}>$. For this test, we take $\omega_j = 1 \forall j$. We then take $P_{max} = +\infty$ in algorithm A1, as the maximum weight of the structure is in this case solely constrained by the volume of the usable space. Finally, we apply a filter on the PERT$ij$ values obtained in cells $\Theta_i$ and bins $B_j$: if the MCNP statistical error on PERT$ij$ exceeds 90%, we impose PERT$ij = 0$. With these parameters, the calculations lasted ~150 min per iteration on a 16-CPU machine.

The histograms $\rho_i(n)$ of the $^1$H densities obtained with algorithm A1 are shown in fig. 8 on the left for some iterations. These results are obtained using the logarithmic quantization (2.2.2) with $M = 100$, starting from an uniform initial density configuration $\rho_i(0) = 0.1$ g/cm$^3$ $\forall i$. The right side of fig. 8 gives the histograms of the spectra $\phi_1(E)$ obtained for the density profiles $\rho_i(n)$ plotted on the left side. The spectra are normalized to 1 and are compared to the histogram of $\phi_{obj}$ (gray area). For information, the total flux of neutrons in D is $1.4 \cdot 10^{-9}$ ($\pm 8.7\%$) n/cm$^2$ per source neutron at iteration $n = 133$. The joint evolution of the $^1$H density profiles $\rho_i(n)$ in the screen and of the spectra $\phi_1(E,n)$ obtained from iteration $n = 0$ to 135 is shown in a video, mimic_gauss_screen.mp4, given in Supplemental Material. In order to better appreciate the convergence of $\phi_1$ towards $\phi_{obj}$, we also give in fig. 9 on the left the evolution of the objective $O\phi$ (3.1.5) with $n$.

The results obtained in fig. 8-9 are excellent. We observe that algorithm A1 succeeds in configuring the screen so that the neutron energy spectrum $\phi_1$ in the area D is almost identical to the objective spectrum $\phi_{obj}$. Some differences inevitably remain between these two spectra: because of the non-zero sizes of the rings $\Theta_i$ and of the sphere used to model the point detector $\underline{r}_d$, the spectrum calculated with A1 is slightly noised, and slightly widened with respect to the spectrum $\phi_{obj}$. However, the final distance reached between these two spectra is very small, with $O\phi$ reaching 0.038 from $n = 79$. As for the density profile $\rho_i$, we note it converges well towards the reference optimal shape of the screen, which is a density peak at ~18 cm, as soon as the first iterations. It then evolves more slowly to eliminate the last superfluous areas of the screen. The final profile is reached at $n = 133$, with a maximum density equal to 0.4 g/cm$^3$, lower than $\rho_{max}$ probably to lower the probability that double collisions occur. Afterwards, we note a small vibration of the structure of period $n = 2$ between the final configuration at $n = 133+2k$, $k \in \mathbb{N}$, and a slightly suboptimal configuration at $n = 132+2k$, of almost identical shape but a little denser.

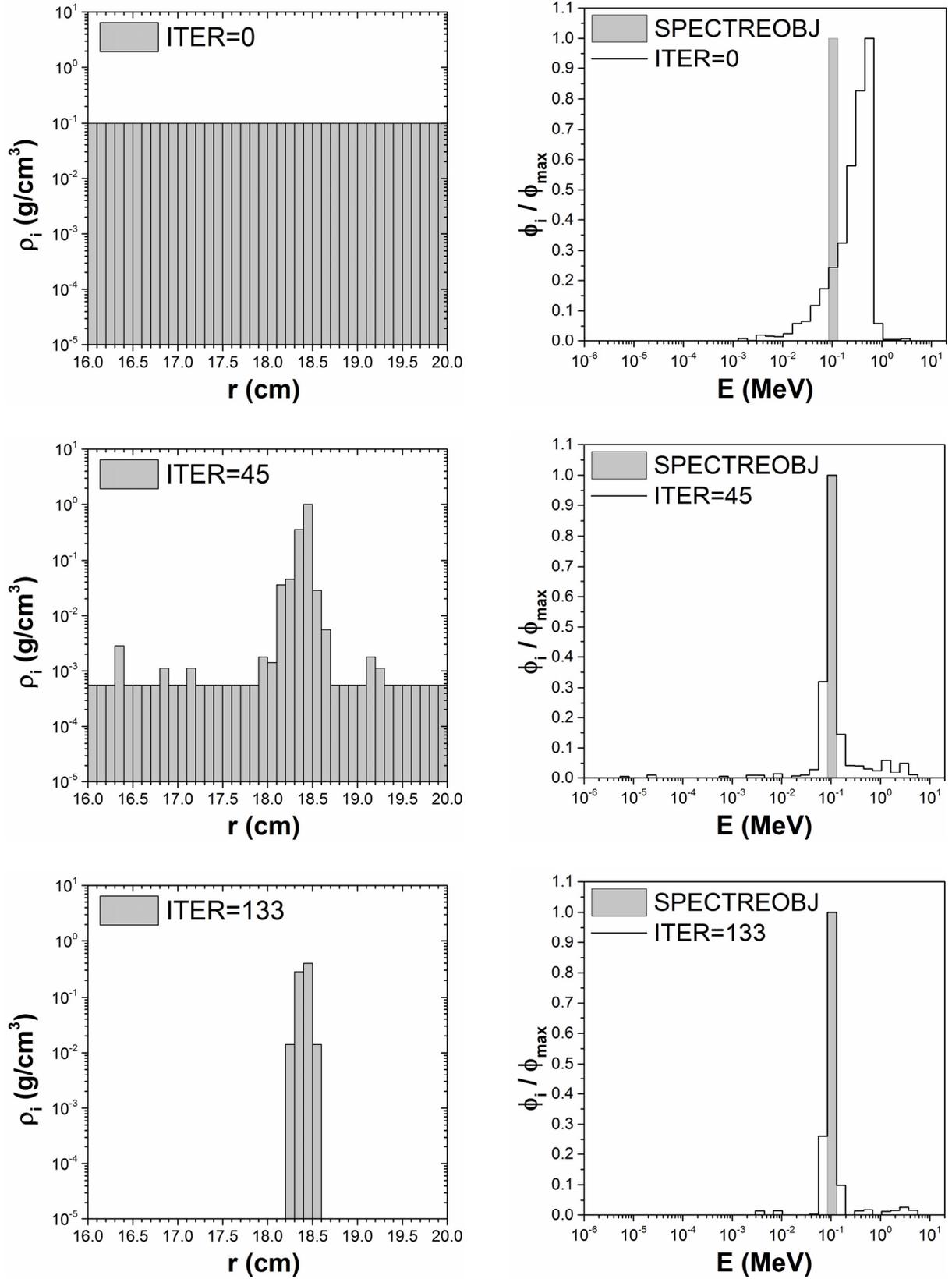

**Figure 8.** (Left) density histograms obtained at iterations 0, 45 and 133 of A1; (right) normalized histograms of the spectra $\phi_1(E)$ obtained in D using the density profiles plotted on the left-hand side, compared to the histogram of the objective spectrum (gray area).

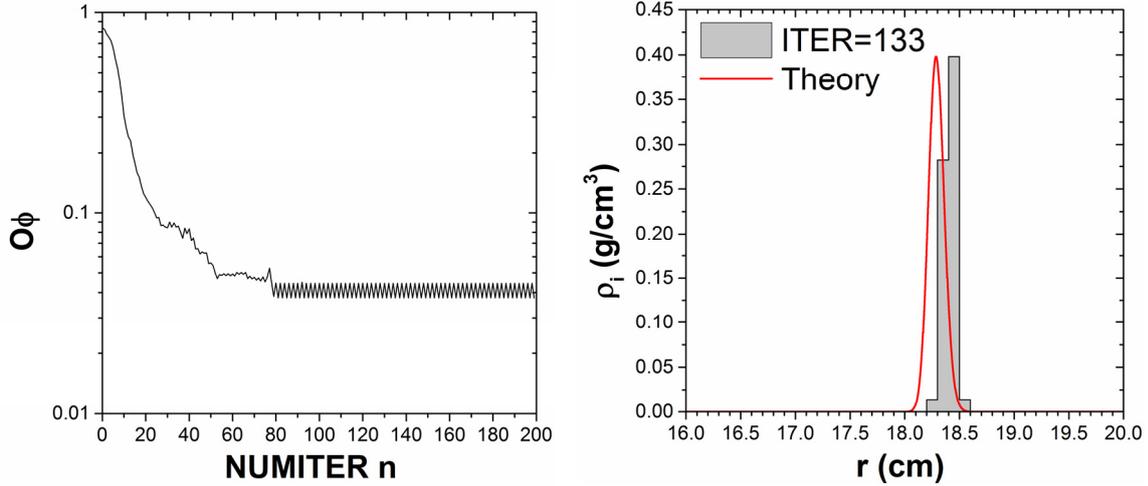

**Figure 9.** (Left) evolution of the objective $O\phi$ with the iteration number $n$; (right) density profile obtained with algorithm A1 at iteration $n = 133$, compared to the theoretical result (3.2.6)-(3.2.8).

In fig. 9 on the right-hand side, we can now compare the optimal density profile calculated with A1 to the theoretical density profile (3.2.8). The theoretical profile has been computed: (i) by adjusting the constant $\beta$ so that the maximum density of the theoretical profile be equal to that calculated with A1; (ii) by taking as objective spectrum a Gaussian, $\exp(-(E-E_{obj})^2/(2\sigma^2))$, centered on $E_{obj} = 100$ keV, whose width $\sigma$ is adjusted so that the total quantity of $^1$H in the theoretical screen be equal to that obtained with A1. Despite the non-zero width of the rings $\Theta_i$, we verify that the shapes of the A1 and theoretical density profiles are very close. However, interestingly, we note that these two profiles are shifted: the maximum of the A1 profile is reached at $r = 18.4$-$18.5$ cm, whereas it is reached at $r = 18.28$ cm for the theoretical profile. This slight shift in position, less than 1%, could be the result of a hypothesis made at the step (3.2.5) of the theoretical calculation. We assumed there that neutron-proton scatterings are isotropic in the center of mass frame. This is a good approximation, but it is not perfect. Indeed, the n-p differential scattering cross-section shows at high energy a slight anisotropy, ≲ 0.5% at 2-3 MeV and growing with energy, in the center of mass frame due to p-wave effects, see [35] pp 120-121. Another possible explanation for this shift in position could involve a slight discrepancy between the value of $A$ used in the evaluation, or in the processing by MCNP, of the scattering angle table and that written in the header of this file. An argument to support this hypothesis could be the fact that the energy spectrum obtained in fig. M.1 (Supplemental Material M) on the right using the density profile $A = 1$ drawn on the left is, coincidence or not, almost correctly centered on the objective energy $E_{obj} = 100$ keV. In fine, we note here that the topology optimization procedure A1, coupled to the computation of the objective derivatives by MCNP6.1, is numerically precise enough to be sensitive to fine details in the modeling of neutron-nuclei interactions.

*Comment.* In Supplemental Material N, we apply the resolution procedure described section 3.1 to solve another difficult problem. Suppose that the operated neutron source be a monoenergetic generator, installed at a location $\underline{r}_s$ of a work area. The source users have in stock a quantity of matter, and would like to machine it so that the neutrons that pass through it all have the same energy $E_{obj}$ at a point $\underline{r}_d$. Is such a monochromator of fast neutrons feasible? In Supplemental Material N, we answer positively to this question, showing that a possible solution is a thin layer of hydrogen deposited on a 3D surface, whose equation is given section N.1. This solution is then confirmed by algorithm A1. This satisfactory behavior of A1 on a 3D problem motivated the following study.

## 3.3. To go further: simulating a reactor spectrum with a block of lead

In section 3.2 and in Supplemental Material N, we demonstrated that algorithm A1 can solve a topology optimization problem of the type (3.1.5). However, the configurations tested so far in these sections were deliberately simple, in order to ease the comparison of the structures found with reference solutions. Moreover, although the solutions found (thin screen or hydrogenated surface monochromator) work on computer, they generate in output low usable neutron fluxes. Indeed, these structures are based on the principle that the particles that pass through them must make at most one collision inside them. To design a device able to mimic an arbitrary spectrum $\phi_{obj}$ with a good efficiency, i.e. with a reasonably high output flux, we should solve the problem (3.1.5) for a massive block of matter, capable of generating a large number of collisions, machined in a realistic material. But, can A1 manage a virtually infinite number of potential collision schemes? A positive element of answer to this question has already been obtained section 2.3, but this must be proved for a problem of the type (3.1.5).

Let us start by choosing a complex objective spectrum $\phi_{obj}$. By occupational bias, we propose a characteristic spectrum of a BN800 sodium-cooled fast nuclear reactor. This spectrum [36] is plotted in fig. 10 on the right side (gray area), using the binning (3.2.10) with $NBIN = 41$, $E_{min} = 10^{-9}$ MeV and $E_{max} = 20$ MeV. This spectrum varies over 6 orders of magnitude in intensity and 10 in energy. Still, in this section, we want to mime it for each of its bins in energy, without focusing solely on its main component around 0.1-1 MeV. To do this, we propose to use in the scalar product (3.1.8) the following weights $\omega_j$:

$$\omega_j = \frac{1}{\phi_{obj\,j}^2}, \quad \langle \underline{\phi}, \underline{\phi}_{obj} \rangle = \sum_{j=1}^{N} \frac{\phi_j}{\phi_{obj\,j}} \quad (3.3.1)$$

by replacing the values of the bins where $\phi_{obj} = 0$ by $\phi_{obj} = 10^{-100}$ n/cm²/s.n.

As before, we tile the volume that contains the structure to be optimized with a union of cells $\Theta_i$, whose parametrization is given in (2.2.3), with $H = 100$ cm, $NR = 10$, $NX = 20$. The cells $\Theta_i$ with $i \neq 1+NX/4$ and $i \neq 1+3NX/4$ contain natural lead. Their volume densities $\rho_i$ form the first components of the vector $\underline{x}$ of the problem (3.1.5) to be optimized with A1. To vary them, we choose the linear quantization (2.2.1), with $M = 40$, $\rho_{min} = 10^{-5}$ g/cm³ and $\rho_{max} = 11.34$ g/cm³ (lead natural density). The source cell S is the cell $\Theta_{1+NX/4}$. It contains at its center, at the coordinates $r = 0$, $z = -H/2$, an isotropic point source that emits $NPS = 5.10^8$ 14 MeV neutrons. Apart from this source, $\Theta_{1+NX/4}$ is empty. Finally, the last cell of the tiling, $\Theta_{1+3NX/4}$, is subdivided into two distinct cells, $\Theta_D$ et $\Theta_B$, whose parametrizations are:

$$\Theta_{1+3NX/4}(\underline{r}) = \Theta\left[0 \leq r < \frac{H}{2NR}\right] \times \Theta\left[\frac{H}{2} - \frac{H}{NX} \leq z < \frac{H}{2} + \frac{H}{NX}\right]$$

$$\Theta_D(\underline{r}) = \Theta\left[0 \leq r < \frac{H}{2NR} - \delta r\right] \times \Theta\left[\frac{H}{2} - \frac{H}{NX} + \delta z \leq z < \frac{H}{2} + \frac{H}{NX} - \delta z\right] \quad (3.3.2)$$

$$\Theta_B(\underline{r}) = \Theta_{1+3NX/4}(\underline{r}) - \Theta_D(\underline{r})$$

with $\delta r = \delta z = H/100$. Cell $\Theta_D$ is the detector cell, left empty. Cell $\Theta_B$ surrounds the detector and contains a volume density $\rho_{shield}$ of natural boron, which is the last component of the vector $\underline{x}$ to be optimized. To vary the density $\rho_{shield}$, we choose a logarithmic quantization (2.2.2), with $M = 40$, $\rho_{min} = 10^{-5}$ g/cm³ and $\rho_{max} = 2.355$ g/cm³ (boron natural density). Cell $\Theta_B$ allows to reduce the number of thermal neutrons that reach detector D, which facilitates the convergence of the algorithm [14]. In the absence of a weight constraint, finally, we retake $P_{max} = +\infty$ in A1. We also reuse the 90% filter on the PERT$ij$ values used section 3.2.2.

The density configurations $\rho_i(n)$ obtained with algorithm A1 are shown in fig. 10 on the left side for some iterations. These 3D structures are cut according to a plane containing the axis of symmetry SD, so the squares drawn are cross-sections of cells $\Theta_i$. The gray scale of the figures gives the lead density values in g/cm$^3$. These results are obtained: (i) for the lead cells, by starting from an uniform initial density configuration $\rho_i(0) = 1.134$ g/cm$^3$ $\forall i$; (ii) for the boron cell $\Theta_B$, by starting from $\rho_{shield}(0) = 0.2355$ g/cm$^3$. The calculations were performed on a 16-CPU machine, and lasted ~8-13 hours per iteration. The right part of fig. 10 gives the histograms of the neutron energy spectra in cell D obtained with the density configurations drawn on the left. These spectra are normalized so that their integrals be equal to 1, and are compared to the histogram of the spectrum $\phi_{obj}(E)$, normalized in the same way. A video, mimic_rnr.mp4, given in Supplemental Material shows the joint evolution of the density configurations and of the corresponding spectra obtained from $n = 0$ to 135. The total neutron flux obtained in D is $2 \cdot 10^{-4}$ ($\pm$ 0.05%) n/cm$^2$ per source neutron at $n = 140$: the gain in efficiency with respect to the structures obtained in sections 3.2.2 or N.2 is thus considerable. Finally, we give in fig. 11 the evolutions of the density $\rho_{shield}$ and of the objective $O\phi$ (3.1.5) with $n$.

The results shown in fig. 10-11 are convincing. We observe that the spectrum obtained in D approaches the objective spectrum over 6 orders of magnitude in energy and in intensity, with a distance $O\phi$ that reaches 0.22 at convergence. The solution found by A1 is also interesting in itself. We observe that the algorithm begins by excavating an area between the source and the detector and concentrates the matter thus liberated along the axis SD in order to cut the ~14 MeV component of the spectrum due to the source neutrons. It also accumulates a part of this matter at the periphery of the excavated area, in order to increase the number of collisions made by the neutrons, and thereby shift at lower energy the maximum of the spectrum. At the end of this first step, the distance $O\phi$ has already reached 0.42 at $n = 24$. Once this bulk of the work has been done, the algorithm then progressively shapes the remaining material so that the epithermal components of $\phi$ and $\phi_{obj}$ match as best as possible. During this phase, the distance $O\phi$ slowly decreases until it reaches its minimum, 0.22, starting from $n \approx 110$.

Still, even though it is promising, the solution shown in fig. 10 is perfectible. Indeed, differences remain between the spectrum at convergence and the objective spectrum, especially at low and high energies. Furthermore, the structure calculated by A1 oscillates with a period $n = 2$, as it can be seen in the video mimic_rnr.mp4. This oscillation is partly due to the fact that the algorithm seems to hesitate between two paths: improving the thermal component of the spectrum on one hand, or improving its fast component on the other hand, both ways offering comparable gains in terms of objective $O\phi$. These issues may be corrected: (i) by using weighs $\omega_j$ that better target the spectrum component(s) to be improved in priority; (ii) by decreasing the size of the cells, cf. Supplemental Material I.2, to eventually find a better solution; (iii) by increasing the number $M$ of levels in the density quantizations (2.2.1)-(2.2.2), to reduce the amplitude of the structure oscillations, visible e.g. fig. 11 on the left; (iv) and, finally, by using a composite material, made of several chemical species, cf. next section, to gain access to various neutron slowing-down speeds, and thus to improve the spectrum in its thermal and epithermal components. As these improvements require in practice just additional computing power, the optimization procedure presented in this section could help develop new types of spectrometers and sources better adapted to the needs of users.

---

[14] The improvement in convergence is mainly due to the fact that the objective spectrum is fast. However, we suppose that part of this improvement could be due to the fact that $\rho_{shield}$ is an additional variable that is not strongly dependent on the neutron transport. It can therefore give some slack to A1, by acting as a valve to the high pressure resting on it due to the strong correlation of the roles of the densities $\rho_i$ of the lead cells in the neutron transport.

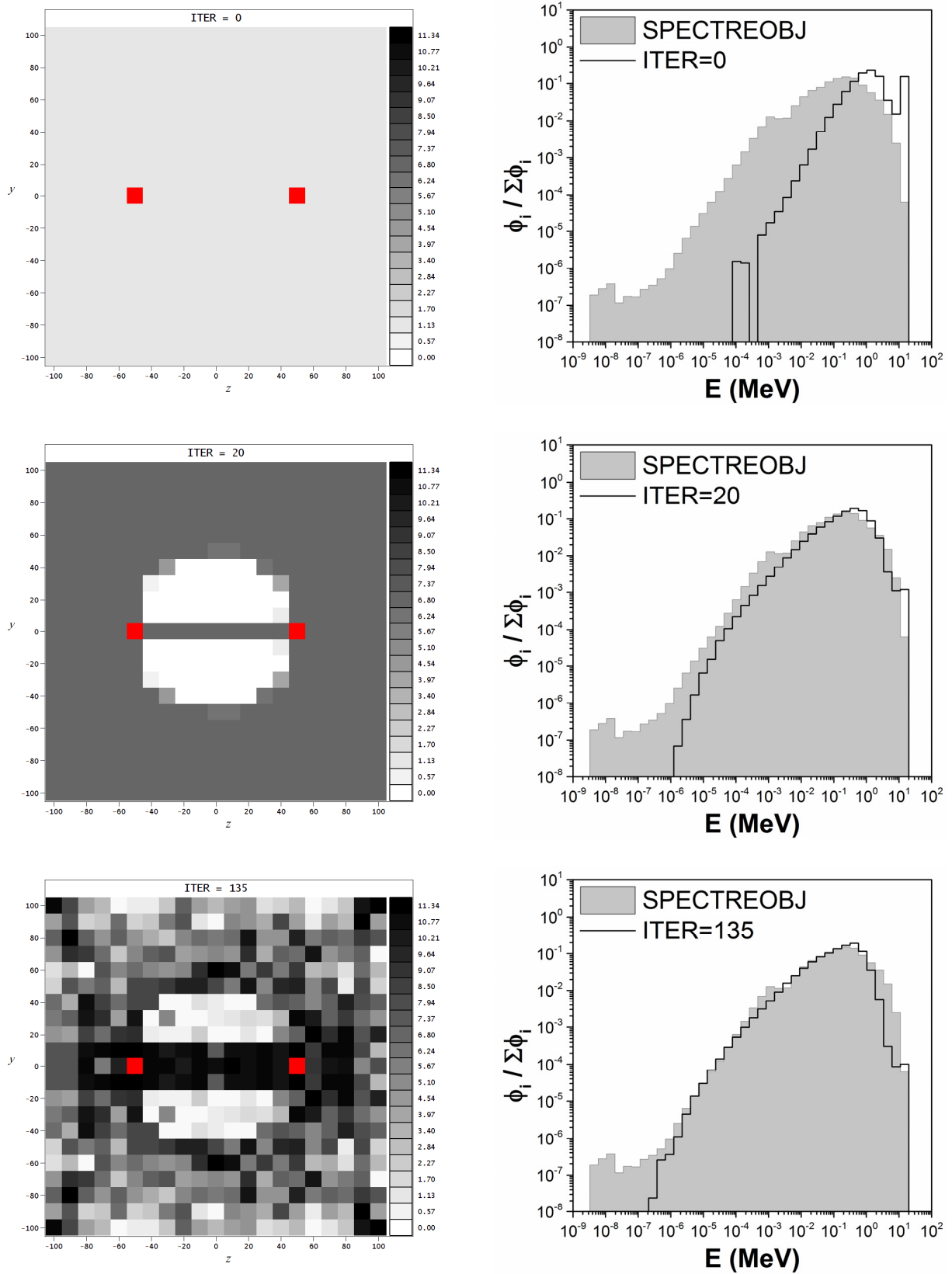

**Figure 10.** (Left) density maps obtained at iterations 0, 20 and 135 of algorithm A1. The units of the axes $y = \pm r$ and $z$ are cm. The gray scale gives the values of the lead densities $\rho_l(n)$ in g/cm$^3$ in the lead cells; (right) histograms of the neutron energy spectra $\phi(E)$ in cell D, obtained with the density profiles drawn on the left side, normalized and compared to the histogram of the objective spectrum $\phi_{obj}$ (gray area).

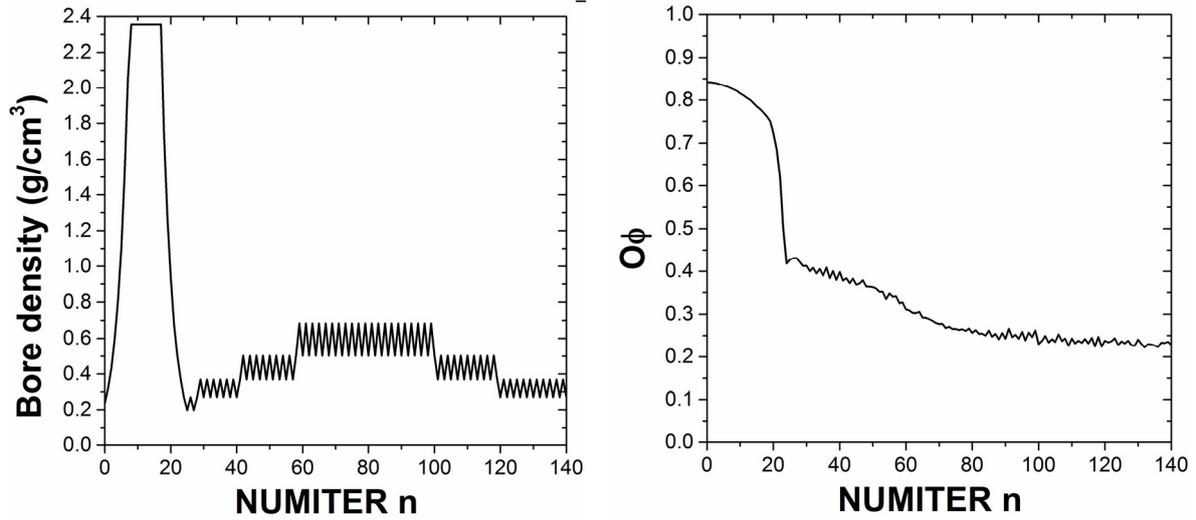

**Figure 11.** (Left) evolution of the density $\rho_{shield}$ with $n$, in g/cm³; (right) evolution of the objective $O\phi$ with $n$.

4. Problem n°3: optimization of the fuel distribution in a nuclear reactor

Suppose an operator has in stock a quantity of metallic uranium, enriched at 3.0at% (fuel made of 97at% of $^{238}$U + 3at% of $^{235}$U), to be used in a parallelepipedic nuclear reactor, whose height is $H$ and whose horizontal section is a square of side $A$. The fuel is to be inserted into $N$ vertical tubes of height $H$ and of square section of side $a$. The moderator is light water. How to position the fuel so that the reactor achieve criticality with the smallest amount of uranium possible? To model this problem, let us number the fuel tubes with an integer $i$, ranging from 1 to $N = (A/a)^2$. The tubes $\Theta_i$ can be parametrized as follows:

$$\Theta_i(\underline{r}) = \Theta[X_j \leq x < X_{j+1}] \times \Theta[Y_k \leq y < Y_{k+1}] \times \Theta[0 \leq z < H]$$
$$X_j = ja, \quad Y_k = ka, \quad M = A/a \quad (4.1)$$
$$i = Mj+k+1, \quad k \in [0, M-1], \quad j \in [0, M-1]$$

Each tube $\Theta_i$ contains a fraction $\chi_i$ of enriched uranium to be optimized, and a complementary fraction $1-\chi_i$ of water. The total mass of uranium in the reactor is thus:

$$M = \rho_U \sum_{i=1}^{N} \chi_i V_i = \rho_U \underline{\chi}.\underline{V} \quad (4.2)$$

where $\rho_U$ is the volume density of uranium enriched at 3at%, ~19.04 g/cm³, and $V_i = Ha^2$ is the volume of a tube $\Theta_i$. The fractions $\chi_i$ and the volumes $V_i$ form the components of the vectors $\underline{\chi}$ and $\underline{V}$. Let us now write a criterion for criticality: $k_{eff} = 1$. This coefficient $k_{eff}$ is the effective multiplication coefficient of the reactor, which is equal to the limit when $n \to +\infty$ of the number of fissions at generation $n+1$ over the number of fissions at generation $n$, see [25] section VIII.B. With these notations, the optimization problem to be solved can be written:

$$\min_{\underline{\chi}} \underline{\chi}.\underline{V} \text{ s.t. to } k_{eff} = 1 \quad (4.3)$$

Solving this problem is interesting for two reasons: (i) it models a major problem in nuclear engineering; (ii) it strongly differs from those studied so far in sections 2-3. First, the constraint $k_{eff}$ = 1 is more complex than a weight or volume constraint. Indeed, $k_{eff}$ depends in a complex way on the physicochemical properties and distribution of the fuel inside the reactor, but also on the shape of the reactor as well as the neutron interaction cross-sections and flux. Secondly, the material considered here is a composite, made of several chemical species, which influence each in a very specific way the transport of neutrons ($^{235}$U is a heavy element that can undergo fission, $^{238}$U can induce parasitic captures at low energy, $^{1}$H efficiently slows down neutrons). The resolution of this challenging problem can be anew performed with a variant of algorithm A1, called B0, described in Supplemental Material P. The fuel fractions $\chi_i(n)$ thus obtained are plotted in fig. 12 in the $xy$ plane for some iterations. They are indicated using a gray scale, each square in the figures representing a tube. A video, okeff.mp4, given in Supplemental Material shows the sequence of these results for $n$ ranging from 0 to 40. These calculations were performed taking $H$ = 100 cm, $A$ = 100 cm, $a$ = 2 cm, starting from an uniform initial fuel distribution $\chi_i(0)$ = 0.02 $\forall i$. As $A/a$ = 50, we needed $N$ = 2500 tubes to tile the section of the reactor. We used 5 $10^5$ neutrons per KCODE calculation, with 50 passive cycles and 100 active cycles. The calculations were performed on a 16-CPU machine, and lasted ~11 hours per iteration. Unlike the calculations in the previous sections, we did not use here the symmetries of the problem (4 planar symmetries) to reduce (by a factor 8) the number of tubes $i$ usable in the optimization procedure, thereby the computing time. We indeed wanted to observe the kind of structure vibrations that statistical fluctuations could generate for this class of problems. Finally, we show in fig. 13 the evolutions with the iteration number $n$ of the coefficient $k_{eff}(n)$ of the reactor (on the left) and of the mass $M(n)$ of uranium in the reactor (on the right), whose minimization is the objective of this section and whose formula is given in (4.2).

The convergence of the calculations is fast: only ~20 iterations are required to find the minimal critical mass, 145 kg of uranium enriched at 3at%, cf. fig. 13. The fuel distribution $\chi_{opt}$ found after convergence, shown in fig. 12 at $n$ = 40, has a hexagonal or octagonal rounded shape, almost cylindrical, banal given the symmetries of the problem. However, for the problem of this section, it is not the shape of the solution that matters, but rather the values of the fuel fractions obtained. We note that these fractions are low everywhere, at less than 4%. One could have thought that the optimal configuration would be a compact mass of pure uranium, similar to the heart of an atomic weapon, positioned at the center of the water pool. Instead, we observe that the solution obtained is a diluted configuration, which allows neutrons to make several collisions with water molecules (thus to thermalize) before colliding with a uranium nucleus (hence potentially inducing a fission). As the fission cross-section of $^{235}$U increases rapidly as the neutron energy decreases, this solution increases the fission rate, thus reducing the required critical mass. This design is actually used in pressurized water reactors, in which fuel rods are separated by water, which has the dual role of slowing down the neutrons and evacuating the heat generated by fissions. An additional discussion of the results shown in fig. 12-13 is given in Supplemental Material P.3.

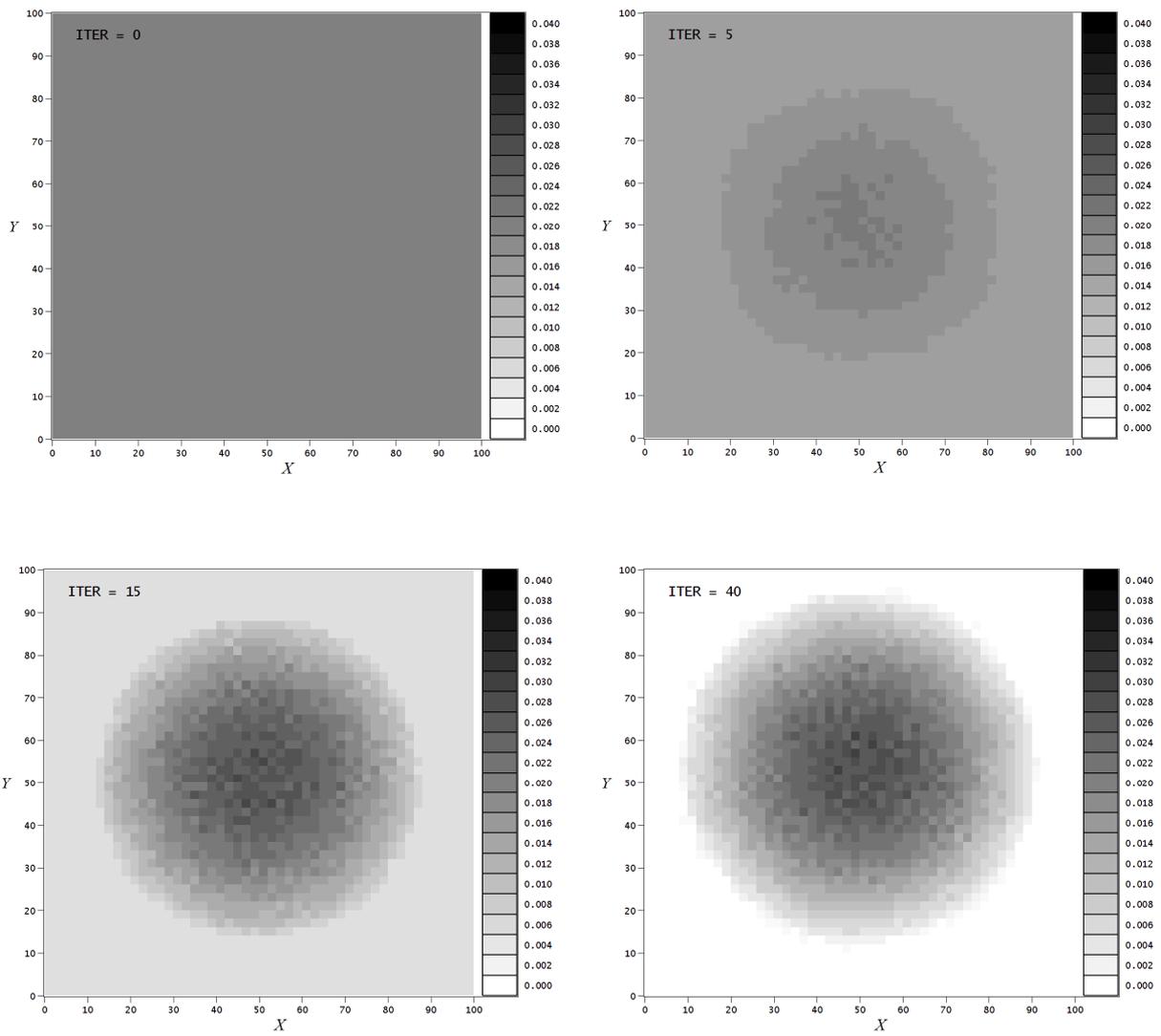

**Figure 12.** Uranium fractions $\chi_i(n)$ obtained at iterations 0 to 40 of B0. The units of the $x$ and $y$ axes are cm. The gray scales give the values of the fractions $\chi_i(n)$.

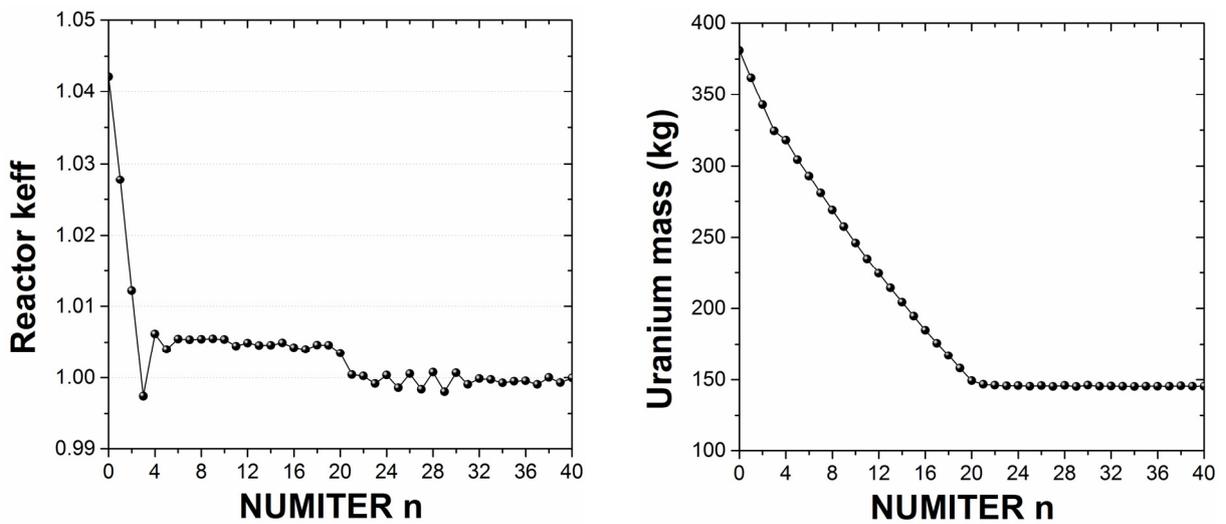

**Figure 13.** Evolutions with $n$ of the reactor's $k_{eff}$ (left) and of the uranium mass in kg in the reactor (right).

## 5. Problem n°4: optimal design of a gamma-ray collimator

By occupational bias of the author, the particles used so far in this study have been neutrons only. However, the tools and the solutions described in the previous sections are applicable to any particles whose transport can be modeled with the linear Boltzmann equation. Among them, one may cite e.g. X and gamma-rays, electrons or high energy ions in absence of strong EM fields, or more exotic particles such as positons or muons. By way of illustration, we hence propose in this section an example of application to the optimization of a gamma-ray collimator. The problem to be solved is the following. An operator manipulates a source of $^{60}Co$, contained in a cylinder of axis Oz, of radius 0.5 cm and of length 2 cm, drawn in red in fig. 14. This source S emits in 49.97% of times a gamma-ray of 1.1732 MeV, and in 50.03% of times a gamma-ray of 1.3325 MeV, both isotropically. It is used to irradiate a patient or an object located at a distance $H = 100$ cm from S. However, for radioprotection issues, it is desired to collimate these gamma-rays so that their fluence be: (i) maximum in a small disk $D_1$ of radius $R_{min} = 1$ cm, of axis Oz, located at $H = 100$ cm; (ii) as low as possible outside this disk, in a crown $D_2 - D_1$, where $D_2$ is a disk of radius $R_{max} = 100$ cm, of axis Oz, located at $H = 100$ cm. To do this, the operator can use a volume of lead, suspended to an articulated mechanism to increase its maneuverability. This mechanism cannot support a weight greater than $P_{max} = 100$ kg. Finally, the device must be contained in a cylinder $V_{dispo}$ of radius $R = 20$ cm and of length $L = 80$ cm to limit the clutter of the treatment area. With these constraints, how to machine the material at disposal to achieve the objectives (i) and (ii)?

As for previous studies, solving this problem is interesting for two reasons: (i) it models anew a practical problem, which has applications in radiological protection or nuclear medicine e.g.; (ii) from the narrower point of view of this study, it also offers a good opportunity to investigate an important topic: the impact that the choice of the objective functional $O$ may have on the shape and performances of the structure generated by the optimization procedure.

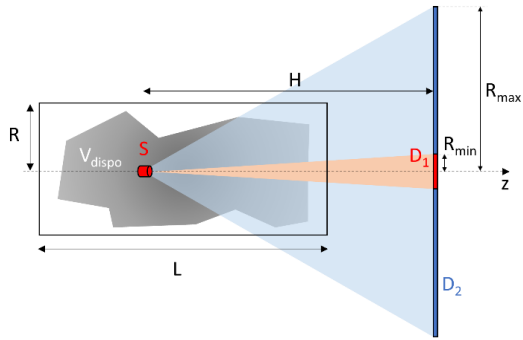

**Figure 14.** Modeling of the problem of section 5.

### 5.1. Resolution of the problem with a simplistic objective $O$

We start by tiling the available space $V_{dispo}$ using a union of cells $\Theta_i$, whose parametrization is:

$$\Theta_i(\underline{r}) = \Theta[R_{j-1} \leq r < R_j] \times \Theta[X_k \leq z < X_{k+1}]$$

$$X_k = \frac{L}{2}\left(\frac{2k-1}{NX} - 1\right), \quad R_j = \begin{cases} j\delta r/5 & \text{if } 0 \leq j \leq 5 \\ \delta r + \dfrac{R-\delta r}{NR-5}(j-5) & \text{if } j \geq 5 \end{cases} \quad (5.1.1)$$

$$i = (NX+1)j + k - NX, \quad k \in [0, NX], \quad j \in [1, NR]$$

with $\delta r = 0.5$ cm, $NX = 48$ and $NR = 24$. The coordinates used here are cylindrical coordinates, $\underline{r} = (r,\theta,z)$, where $r$ is the distance from the point $\underline{r}$ to the axis $Oz$.

In (5.1.1), the number of rings $\Theta_i$ contained in the area $r \leq \delta r$, radius of the source cylinder S, is increased in order to refine the shape of the collimator along the source-to-target line of sight, a zone one can reasonably assume to be important for the problem. The cylinder S is the union of the cells $i = (NX+1)j-3NX/4$ with $j$ ranging from 1 to 5. These cells are left empty and are not accessible to the optimization procedure. The $^{60}$Co source itself is modeled by a point placed at the center of S, at the coordinates $r = 0$, $z = -20$ cm. Finally, we define two additional cells $\Theta_{in}$ and $\Theta_{out}$, parametrized as follows:

$$\Theta_{in}(\underline{r}) = \Theta[0 \leq r < R_{min}] \times \Theta[z_{min} \leq z < z_{max}]$$
$$\Theta_{out}(\underline{r}) = \Theta[R_{min} \leq r < R_{max}] \times \Theta[z_{min} \leq z < z_{max}] \quad (5.1.2)$$

with $z_{min} = 79.5$ cm and $z_{max} = 80.5$ cm. The cell $\Theta_{in}$ is a cylinder of basis $D_1$, of axis $Oz$ and of length 1 cm, which will be used to model the target $D_1$. The cell $\Theta_{out}$ is a cylindrical ring, which will be used to model the external zone $D_2-D_1$. These two cells are also left empty and are not accessible to the optimization procedure.

The lead volume densities $\rho_i$ and the volumes $V_i$ of the other cells $\Theta_i$, eligible to the optimization procedure, form two vectors $\underline{\rho}$ and $\underline{V}$. With this parametrization, the problem to be solved in this section can thus be written:

$$\max_{\underline{\rho}} O\varphi(\underline{\rho})$$
$$\text{s.t.} \quad B(\underline{\rho})\varphi(\underline{r},E,\underline{\Omega},\underline{\rho}) = Q(\underline{r},E,\underline{\Omega}) \quad (5.1.3)$$
$$P(\underline{\rho}) = \underline{\rho}.\underline{V} \leq P_{max}$$

with, in first intuition:

$$O\varphi(\underline{\rho}) = \frac{\phi_0(\underline{\rho})}{\phi_1(\underline{\rho})^\beta}, \quad \phi_{0,1}(\underline{\rho}) = \frac{1}{\int_{\underline{r} \in \Re^3} \Theta_{in,out}(\underline{r})d\underline{r}} \int_{\underline{r} \in \Re^3} \int_{\underline{\Omega} \in 4\pi} \int_{E=0}^{+\infty} \Theta_{in,out}(\underline{r})\varphi(\underline{r},E,\underline{\Omega},\underline{\rho})d\underline{r}d\underline{\Omega}dE \quad (5.1.4)$$

In (5.1.3)-(5.1.4), $\phi_0$ is the total flux of gamma-rays in the cell $\Theta_{in}$ that models the target, and $\phi_1$ is the total flux of gamma-rays in the external zone, modeled by $\Theta_{out}$. In (5.1.4), we took $O\varphi = \phi_0/\phi_1^\beta$ in order to simultaneously maximize $\phi_0$ and minimize $\phi_1$. The coefficient $\beta$ allows to weight the relative importance of these objectives.

The problem (5.1.3)-(5.1.4) can be solved anew using algorithm A1, cf. Supplemental Material Q. The density maps $\rho_i(n)$ obtained are shown in fig. 15 for some iterations. As before, the 3D structures are generated by revolving these maps around the symmetry axis $y = 0$ of the problem. The gray scale of the figures gives the lead density values in g/cm$^3$. The red cell is the cylinder S that contains the $^{60}$Co source. The cells $\Theta_{in}$ and $\Theta_{out}$, located outside the available volume, are not represented. A video, gakn_simple_obj.mp4, given in Supplemental Material shows the evolution of the structure with the iteration number. The calculations were performed for $\beta = 1$, $NPS = 5.10^8$, $\rho_{min} = 10^{-5}$ g/cm$^3$ and $\rho_{max} = 11.34$ g/cm$^3$, using the linear quantization (2.2.1) with $M = 50$, starting from an uniform initial configuration, $\rho_i(0) = \rho(0) \forall i$, with $\rho(0)$ given in (2.2.5). The calculations were performed on a 16-CPU machine, and lasted ~2 hours per iteration. In fig. 16, we give on the left the evolution of the fluxes $\phi_0$ and $\phi_1$ with $n$, and on the right the evolution of the objective $O$ (5.1.4) with $n$.

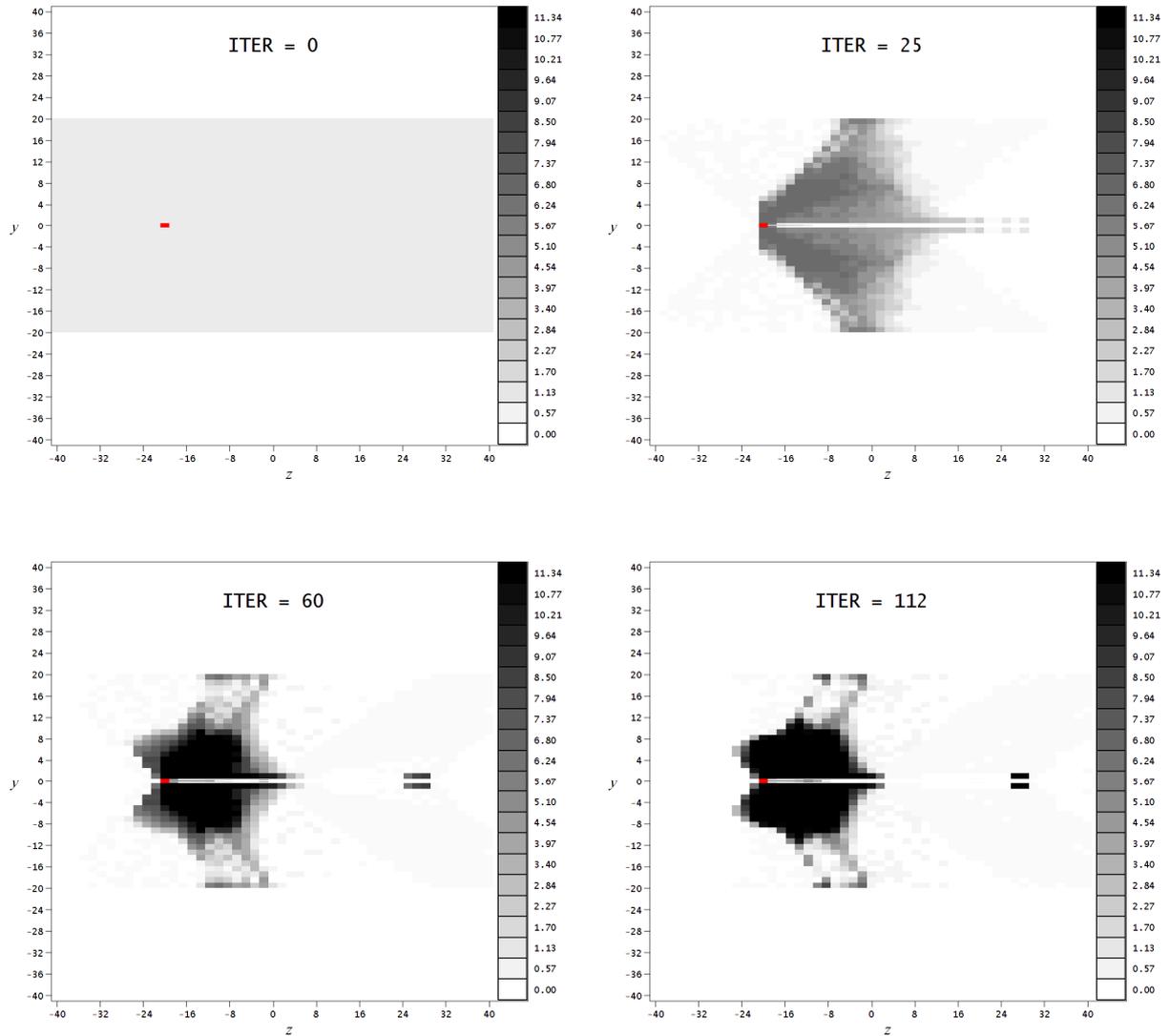

**Figure 15.** Density maps obtained at iterations 0 to 112 of A1 for the objective (5.1.4). The units of the axes $y = \pm r$ and $z$ are cm. The gray scale of the figures gives the density values $\rho_i(n)$ in g/cm$^3$ in the cells $\Theta_i$.

The solution found by the algorithm is interesting. As a human could have foreseen it, A1 opens a hole in the structure along the line of sight $y = 0$, to let the gamma-rays pass and increase $\phi_0$. Simultaneously, it adds matter in the area between the source and the external ring $\Theta_{out}$ to minimize $\phi_1$. To do this, it favors the cells that are close to the axis $y = 0$. Indeed, these cells have low radii therefore low volumes: their priority use allows to maximize the result obtained at constant weight. The algorithm thus creates a canon-shape structure, which turns out to be very effective: at iteration 112, the flux $\phi_1$ delivered outside the target is ~5400 times lower than the flux $\phi_0$ delivered in the target. The canon obtained is disjoint, with a part offset to the right of the structure. It is completed by a second, very thin, needle-shaped canon, positioned inside the structure, close to S. In the end, the optimal shape of the collimator obtained fig. 15 iteration 112 is complex: it would have been very hard to find with a parametric study, and it is inaccessible in its details to human intuition. The result is hence promising, even though it is not perfect. Among the problems identified, we note:
1) the irregular convergence of the algorithm when the flux $\phi_1$ becomes small. In this regime, the peripheral areas have been cleared, and the material have been correctly accumulated by A1 around the main structure. However, we observe the random appearance or disappearance of cells of density $\rho_{min}+\delta\rho$ in the normally empty zone, according to the statistical fluctuations on the values $C_i$ (Q.1). This noise on the structure is particularly visible in the video, especially around the canon. It randomly modifies the paths taken by the few gamma-rays that manage to reach the external area, which induces fluctuations on the flux $\phi_1$ thus on the objective $O\varphi$. As a result, the algorithm is periodically misled, taking a bad path every 20 or so iterations at the end, cf. fig. 16. This noise could have been prevented: (i) by increasing the number *NPS* of source gamma-rays sampled in the MCNP calculation, which would have reduced the statistical fluctuations on the values $C_i$, at the cost however of an even longer computing time; (ii) by increasing the number *M* of density levels in the quantization (2.2.1), in order to reduce $\delta\rho$, thus the effect of the random cells. The convergence would have been slower, but of better quality. We could have also tested the logarithmic quantization (2.2.2), to better clean the structure; (iii) by taking for objective min $\phi_1/\phi_0$ rather than max $\phi_0/\phi_1$ in the optimization problem (5.1.3). This choice would have prevented to have powers of $\phi_1$ at the denominator of $C_i$, therefore to divide by terms close to 0, with all the associated numerical stability problems.

2) Fig. 15, intuitively, the optimal collimator structure should not have many cells with non-zero densities to the left of the source S. Indeed, part of the gamma-rays emitted towards the left by S could be backscattered there, then increase the flux $\phi_1$ in the outer area $\Theta_{out}$, probably more than they would increase the flux $\phi_0$ in the target (this would require that they be scattered at 180° and repass by the canon of the structure, which is unlikely). Yet we note fig. 15 or in the video that A1 still puts a little bit of matter in some cells to the left of S. The positioning of this matter is, in part we think, an artifact of the optimization procedure A1. Since we cannot set $\rho_{min} = 0$ exactly under penalty of a MCNP error, the zone cleared to the left of S is not totally empty: it contains in fact lead at density $\rho_{min} = 10^{-5}$ g/cm$^3$. This residual lead sends back towards the external zone of the screen a small but non-zero fraction of the gamma-rays emitted by the source. So, to block these gamma-rays, the algorithm has to envelop the main structure in a halo of matter, which overflows slightly to the left of S, thus creating the cells behind this discussion. The construction or the removal of this halo throughout the calculations is partly the cause of the giant structure oscillation, of period ~20 iterations, observed fig. 16. This halo is admittedly an artifact, but it highlights an important point: the influence that the surroundings (walls, floor, roof, other objects in the work area, etc.) of the device to be optimized may have on its optimal shape, cf. section 2.3.

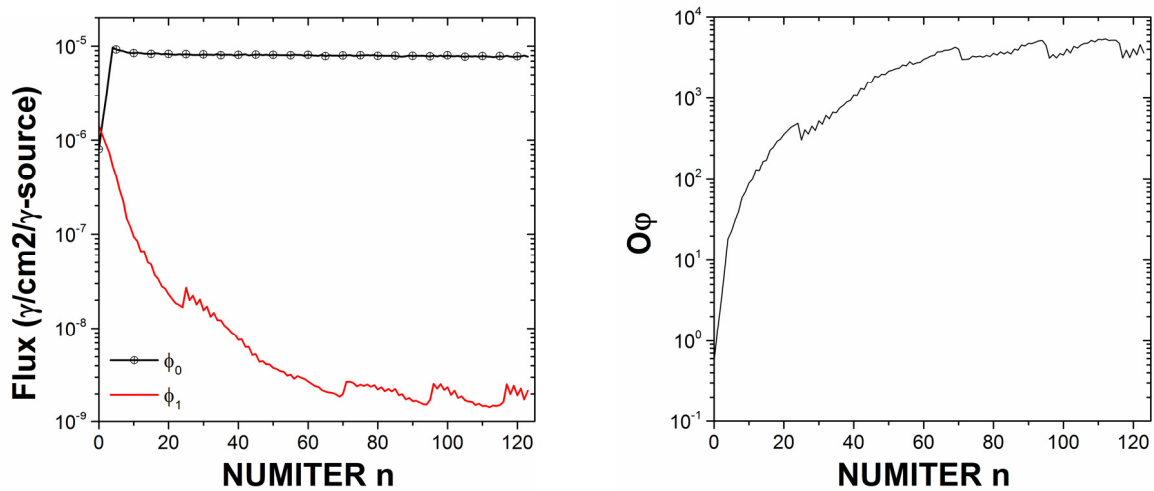

**Figure 16.** (Left) evolution of the gamma fluxes $\phi_0(n)$ (black line + circles) and $\phi_1(n)$ (red line) with the iteration number *n*. These fluxes are given in gamma/cm$^2$ per source gamma. Their statistical uncertainties being less than the percent, their error bars are not drawn; (right) evolution of the objective $O\varphi$ (5.1.4) with *n*.

## 5.2. Resolution of the problem with a smarter objective $O$

The results obtained fig. 15-16 have two other problems, more fundamental. First, they were obtained for a particular case, $\beta = 1$, of the functional $O$ (5.1.4). Since this value is no more justifiable than another, we cannot guarantee that the structure associated with it is indeed the best. One might be tempted to look at the influence of the choice of $\beta$ on the calculated shape. However, this would be a dead end, because the reasoning done section 5.1 has another, more important, problem that concerns the very formulation of the objective to be achieved. To wish that the fluence generated by the source be maximal in the target, $\Theta_{in}$, and minimal outside, in the cell $\Theta_{out}$, does not mean that we want to maximize $\phi_0$ in absolute. We rather want to find a structure for which ~100 % of the fluence is delivered in $\Theta_{in}$, and ~0 % in $\Theta_{out}$. It is hence desired to maximize the proportion of $\phi_0$ in the total $\phi_{tot}$ of the fluence deposited in the screen. If we define this total $\phi_{tot}$ as being the sum $\phi_0 + \phi_1$, we indeed retrieve the objective of section 5.1: to maximize $\phi_0/\phi_1$. However, this definition of $\phi_{tot}$ is rough, if only because of the difference in the volumes of the cells $\Theta_{in}$ and $\Theta_{out}$. We have to be more precise.

To do this, let us now divide the screen $\Theta_{in} + \Theta_{out}$ into $NBIN$ areas $B_j(\underline{r})$. For this study, we will use cylindrical rings parametrized as follows:

$$B_0(\underline{r}) = \Theta_{in}(\underline{r}) = \Theta[0 \leq r < R_{min}]\Theta[z_{min} \leq z < z_{max}]$$
$$B_{j \in [1, NBIN-1]}(\underline{r}) = \Theta[r_j \leq r < r_{j+1}]\Theta[z_{min} \leq z < z_{max}] \quad (5.2.1)$$

In this section, we chose a logarithmic binning, similar to the binning (3.2.10):

$$r_j = R_{min}(R_{max}/R_{min})^{\frac{j-1}{NBIN-1}} \quad (5.2.2)$$

Let us then denote $\phi_j$ the flux in the area $B_j$ of the screen, and gather the $\phi_j$ values in a vector $\underline{\phi} = (\phi_0, \phi_1, \ldots, \phi_{NBIN-1})$. In this form, the optimization problem of this section now resembles that of section 3. Using the definitions (3.1.8) of the scalar product and of the norm, we note that the distribution of the flux in each bin $B_j$ can be given by $\phi_j \omega_j^{1/2}/\|\underline{\phi}\|$, with:

$$\|\underline{\phi}\| = \sqrt{\sum_{j=0}^{NBIN-1} \phi_j^2 \omega_j} \quad (5.2.3)$$

where the coefficients $\omega_j$ are weights. With this notation, maximizing the proportion of the flux in the target is equivalent to minimizing the distance, $d(\underline{\phi}, \underline{\phi}_{obj})$, between $\underline{\phi}$ and an objective flux $\underline{\phi}_{obj}$, which is here $\underline{\phi}_{obj} = (1, 0, \ldots, 0)$. The problem to be solved is thus identical to that of section 3, the only difference being that the bins $B_j$ are here bins in position, not bins in energy. Using the results of section 3.1, the optimization problem of this section can hence be rewritten:

$$\min_{\underline{\rho}} O\varphi(\underline{\rho})$$
$$\text{s.t.} \quad B(\underline{\rho})\varphi(\underline{r}, E, \underline{\Omega}, \underline{\rho}) = Q(\underline{r}, E, \underline{\Omega}) \quad (5.2.4)$$
$$P(\underline{\rho}) = \underline{\rho}.\underline{V} \leq P_{max}$$

with:

$$O\varphi(\underline{\rho}) = 1 - \frac{\langle \underline{\phi}(\underline{\rho}), \underline{\phi}_{obj}\rangle}{\|\underline{\phi}(\underline{\rho})\|\|\underline{\phi}_{obj}\|} = 1 - \frac{\phi_0(\underline{\rho})\sqrt{\omega_0}}{\|\underline{\phi}(\underline{\rho})\|} \quad (5.2.5)$$

In (5.2.5), by taking the simplest binning $NBIN = 2$, and by choosing for norm of $\underline{\phi}$ the $L^1$ norm, i.e. $\|\underline{\phi}\|_1 = \Sigma \phi_i \omega_j^{1/2} = \phi_0 \omega_0^{1/2} + \phi_1 \omega_1^{1/2}$, instead of the $L^2$ norm (5.2.3), we observe that we retrieve the objective (5.1.4), since minimizing $1 - \phi_0 \omega_0^{1/2}/\|\underline{\phi}\|_1$ amounts to maximizing $\phi_0/\phi_1$. The results obtained fig. 15-16 are thus not false in the strict sense of the term, they are just crude.
The practical resolution of the problem (5.2.4) with algorithm A1 was described in section 3.1. To obtain the derivatives $\partial \phi_{j=0...NBIN-1}/\partial \rho_{i=1...N}$ of (3.1.9) required to calculate the coefficients $C_i$ used in A1, we can add the following lines to the MCNP input file, cf. section 1:

F4:p *list of the NBIN numbers that define the cells $B_j$ in the MCNP input*
PERT1:p  CELL=1  MAT=$m$  RHO=−1.05×$\rho_1$  METHOD=2
…
PERT$N$:p  CELL=$N$  MAT=$m$  RHO=−1.05×$\rho_N$  METHOD=2

The density configurations $\rho_i(n)$ obtained by A1 with the objective (5.2.5) are given in fig. 17 on the left side for some iterations. On the right side of these figures, we give the fractions $\phi_j \omega_j^{1/2}/\|\underline{\phi}\|$ of the gamma-ray flux delivered in each ring $B_j$ of the screen, calculated with MCNP for the density configurations plotted on the left side. The edges of the bins of these histograms indicate the position of the edges of the rings $B_j$, the distance $r$ being the distance to the axis $y = 0$. The joint evolution of the volume densities $\rho_i$ and of the distributions $\omega_j^{1/2}\phi_j/\|\underline{\phi}\|$ of the gamma-ray fluxes in the screen throughout the iterations of algorithm A1 are shown in a video, gakn_improved_obj.mp4, given in Supplemental Material. These calculations turned on a 16-CPU machine, taking ~2h per iteration. They were performed taking $NBIN = 20$, $\omega_j = 1$ $\forall j$, as well as the parameters $NPS$, $\rho_{min}$ and $\rho_{max}$ used section 5.1. Like in fig. 15-16, we used the linear quantization (2.2.1) with $M = 50$, starting anew from an uniform initial configuration $\rho_i(0) = \rho(0)$ $\forall i$, with $\rho(0)$ given in (2.2.5). Finally, we filtered the PERT$ij$ values by setting PERT$ij$ = 0 when the MCNP statistical error on PERT$ij$ exceeds 90%.
In fig. 18, we give on the left the evolution of the objective $O\varphi$ (5.2.5) with the iteration number $n$. On the right, we compare the distributions $\underline{\phi}_{opt}/\|\underline{\phi}_{opt}\|$ of the fluxes delivered in the screen obtained after convergence of A1 for the objective (5.1.4) (structure drawn fig. 15 iteration 112) and for the objective (5.2.5) (structure drawn fig. 17 iteration 138). Finally, to allow a better visualization of the solutions found by A1 sections 5.1 and 5.2, we draw in fig. 19 the 3D shapes of the optimal collimators obtained with the objective (5.1.4) at iteration 112, and with the objective (5.2.5) at iteration 138. To improve the readability of these structures, we only show the cells whose density is greater than $\rho_{max}/10$.

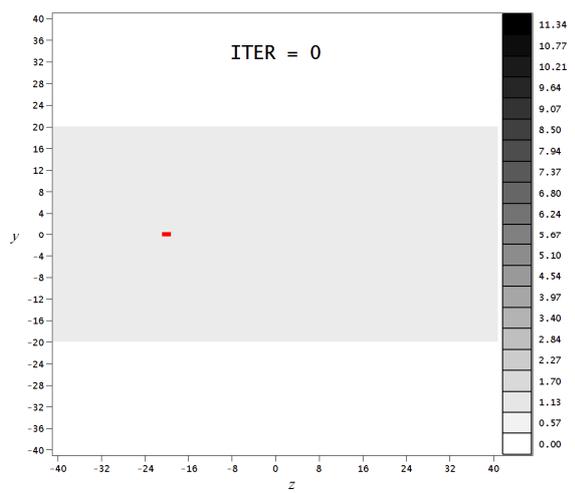
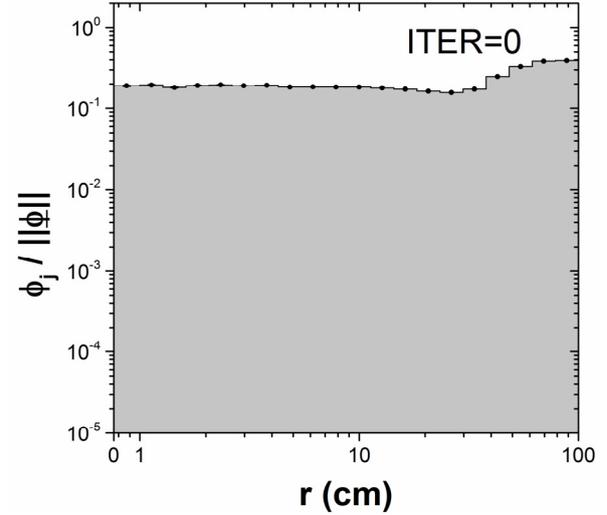
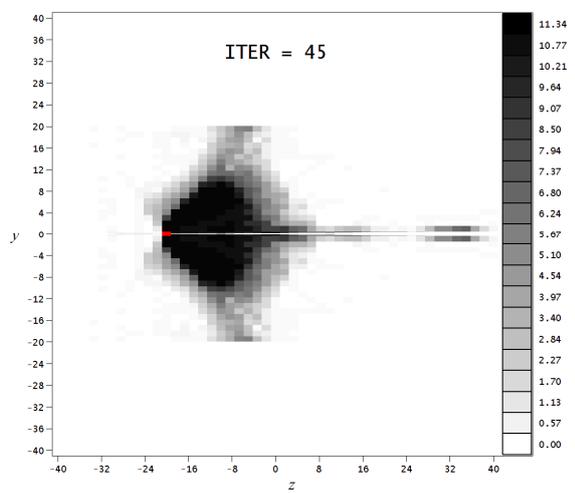
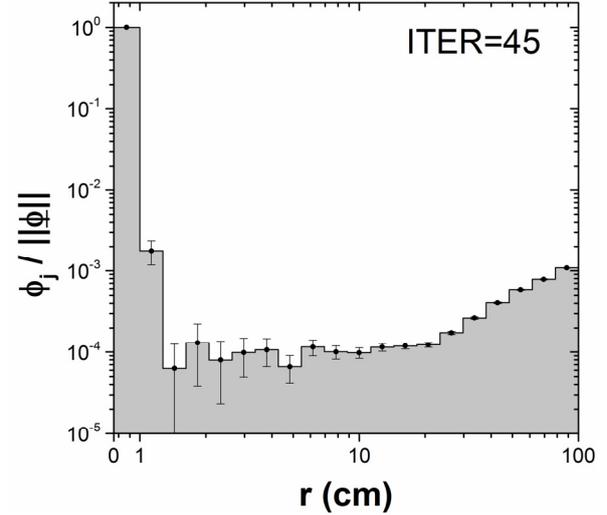
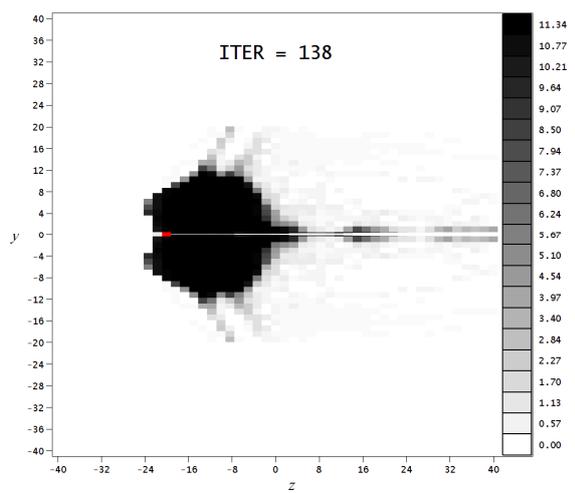
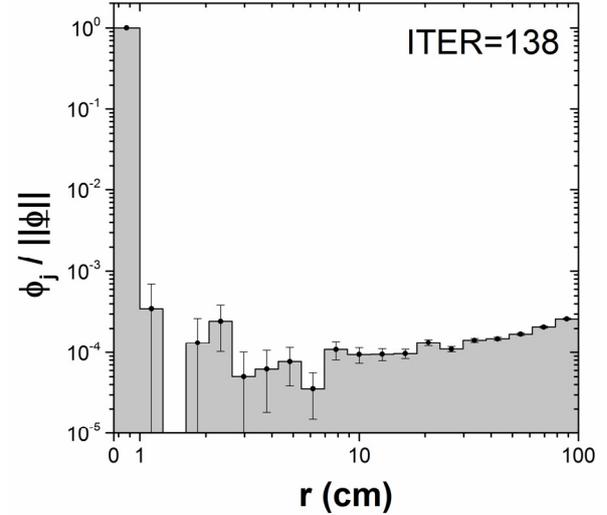

**Figure 17.** (Left) density maps obtained at iterations 0, 45 and 138 of A1 with the objective (5.2.5). The units of the axes $y = \pm r$ and $z$ are cm. The gray scale gives the density values $\rho_i(n)$ in g/cm$^3$; (right) distributions $\phi_j/\|\phi\|$ of the gamma-ray fluxes $\phi_j$ delivered in each ring $B_j$ of the screen, calculated for the density configurations plotted on the left. The axis $r$ is in log scale between 1 and 100 cm.

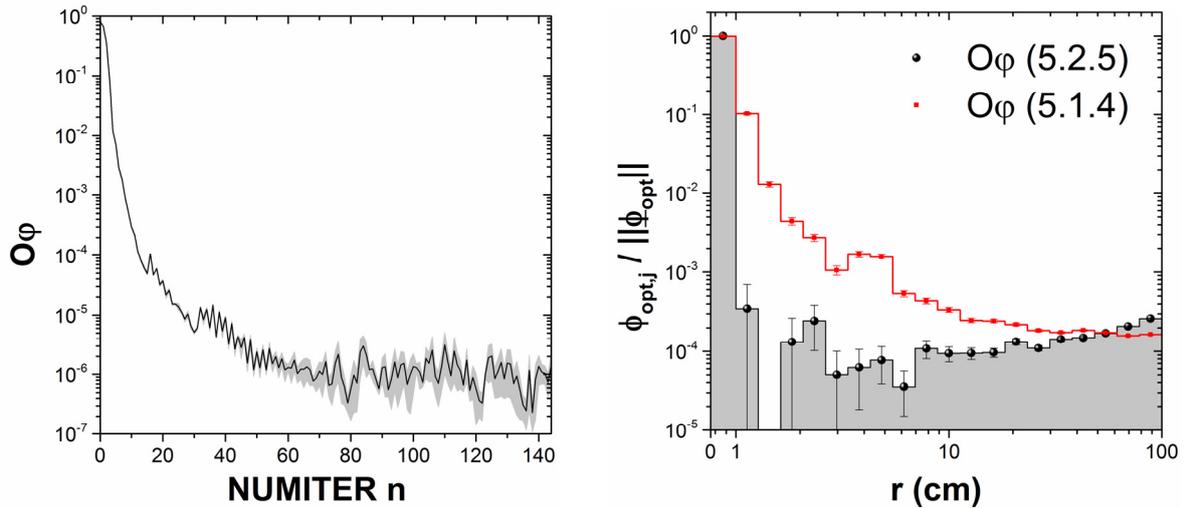

**Figure 18.** (Left) evolution of objective $O\varphi$ (5.2.5) with $n$. The gray area is the statistical error; (right) comparison of the optimal distributions, $\phi_{opt}/\|\phi_{opt}\|$, of the flux in the screen obtained for the objectives (5.1.4) (red squares) and (5.2.5) (black circles). The error bars are obtained by propagating the MCNP statistical errors on the fluxes $\phi_j$.

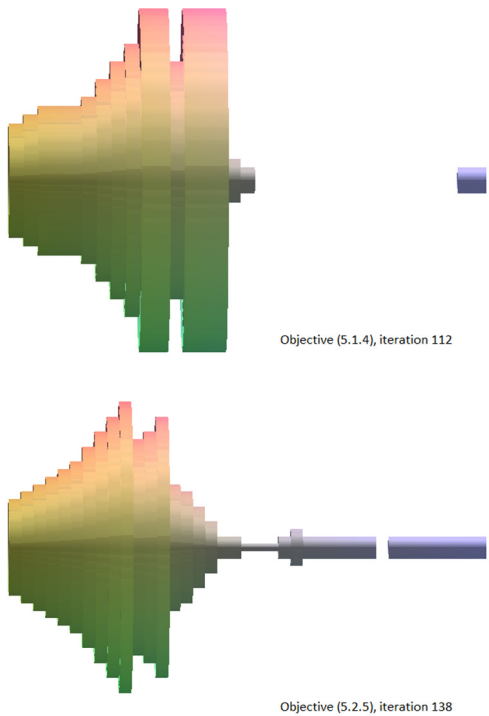

**Figure 19.** Collimator designs obtained by A1 with the objectives (5.1.4) (top) and (5.2.5) (bottom).

If the solution found in section 5.1 was promising, the shape of the collimator obtained with the objective (5.2.5) proves to be better. Fig. 17 or 19, A1 finds anew a canon-like shape. However, the canon is now longer and thinner (especially near S) than the one obtained with (5.1.4), and it has no longer the central needle visible in fig. 15. The canon still emerges from a central body that envelops the source and blocks the gamma-rays emitted towards the external area. This result may be further improved by reducing e.g. the thickness of the cylindrical rings $\Theta_i$ (5.1.1) used to tile the available space. This may lead to a thinner barrel, with a narrower opening near S. One could also choose weights $\omega_j \neq 1$ in (5.2.5), for example $\omega_j = (v_j/v_{tot})^2$, with $v_j$ being the volume of a ring $B_j$ of the screen and $v_{tot}$ the total volume of the screen. In this way, the differences in size of the rings $B_j$ would be better taken into account. However, the shape found fig. 17 or 19, obtained for a modest amount of lead, 100 kg i.e. 8.81 dm$^3$, already delivers a gamma-ray beam of extreme cleanness: the proportion $\omega_0^{1/2}\phi_0/\|\phi\|$ of gamma-rays that hit the screen outside the target is lower than $3.10^{-5}$% at iteration 138 in $L^2$ norm, cf. fig. 18 (to within the statistical error bars). This superior collimation comes with a price: the flux $\phi_0$ obtained with the structure drawn fig. 17 iteration 138, $2.9\ 10^{-6}$ gamma/cm$^2$ per source gamma, is lower than that obtained with the structure drawn fig. 15 iteration 112, $7.8\ 10^{-6}$ gamma/cm$^2$/s.g. It will therefore be necessary to compensate for this ~3-fold reduction of the flux in the target by increasing the duration of irradiation by the same amount. Even so, the gain in cleanness of the beam remains in favor of the structure obtained with objective (5.2.5). Indeed, we note fig. 18 that the fluxes $\phi_{j>0}/\|\phi\|$ obtained with (5.1.4) are several orders of magnitude higher than those obtained with (5.2.5) in the vicinity of the target. They become comparable only in the outer regions of the screen, when $r > 50$ cm. However, the fluxes $\phi_j/\|\phi\|$ deposited in these outer rings are negligible compared to that in the target, $\phi_0/\|\phi\|$, for both objectives. The collimation efficiency of the structure found by algorithm A1 using the objective (5.2.5) is thus much greater than that obtained with the simple objective (5.1.4).

This result illustrates the impact that the formulation of the objective can have on the shape and performances of the object to be optimized. In this study, it is not the first place where a human choice appears: the parametrization of the tiling or the choice of the materials to use, e.g., are manual. However, these are default choices, made to save the scarce computing resources at our disposal. In a different context of abundance of computing resources, it should be possible to automate them, if only by brute force, by reducing the size of the cells down to a sufficiently small level so that their shape have no more importance, or by testing all materials listed in the transport data libraries using the procedures described in section 4. At the opposite, for complex problems, the choice and formulation of the objective to reach appear fundamental. As often in these cases, since there is no perfect solution, the sharper the question the better the answer.

6. Conclusions and perspectives

6.1. Conclusion

In this study, we proved the feasibility of a topology optimization procedure in the framework of the linear Boltzmann equation. This procedure builds on functions of modern Monte-Carlo particle transport codes, such as MCNP, and the data libraries they use. After having presented the mathematical framework of this method in section 1, we tested it on a series of problems, chosen for their practical applications but also for the challenge posed by their resolution. At the end of these tests, presented in sections 2-5, we noticed the efficiency and versatility of this tool, which proved capable of finding solutions that outperform human design capabilities while consuming computing resources already modest in 2018. With the increase in computing power in the coming years, this tool could be developed to assist or automate the design of particle optics or nuclear equipment. As an example, we used it in section 2 to design neutron concentrators and shields, whose obtained shapes agree with theoretical calculations or intuition. Section 3, we showed how this tool can organize a block of ordinary matter to recreate in a region of space a spectrum of interest in nuclear engineering. This promising result could help develop new types of sources and spectrometers. In section 4, we used this tool to optimize in a simple way the distribution of fuel in a schematic nuclear reactor. Finally, in section 5, we optimized the design of a gamma-ray collimator. This last example rose an interesting point, the influence that the formulation of the objective to reach can have on the performances of the calculated structure.

6.2. Limits, areas for improvements, perspectives

Although the optimization procedure developed in this study yields promising results, it is certainly perfectible. At the end of the tests carried out in sections 2 to 5, we identified some of its limits, and suggested several areas for improvement. Among the main issues encountered, we can quote the:

*Noise on the calculated structures.* When the number of particle trajectories simulated is small, the statistical fluctuations on the derivatives of the objective can induce some vibrations in the calculated structures, visible e.g. fig. 5. This noise is inherent in the Monte-Carlo method used to solve the Boltzmann equation. It can obviously be dampened by increasing the number *NPS* of source particles emitted. However, a 2-fold reduction in statistical fluctuations roughly requires a 4-fold increase in *NPS*, hence in computing time. The latter will therefore rapidly diverge when the desired spatial accuracy on the design of the structure gets high. In this case, methods of structure regularization may be employed [37-39]. However, these techniques should be used with caution, as they may render the calculated structures suboptimal because of the strong inter-dependence of the cells in the particle transport.

*Long calculation times.* The long computing times required for complex structures or objectives could be reduced by using cuts (in energy, in time, in collision number, etc.) and/or convergence acceleration methods, e.g. the weight window generator or the Russian roulette, see [25] section VII and [26] section 5.7, in the Monte-Carlo calculations. However, these ruses could bias the values of derivatives obtained, thus leading to suboptimal structures. If the computing resources available are limited and do not allow to do without, it will be necessary to test these techniques case by case, without bestowing them blind trust, a basic recommendation for any MCNP study.

*Difficulty to evolve a compact structure.* We observed this problem in Supplemental Material I.1. A possible solution, not tested in this study, could be to vary the importance of the particles, see [26] section 5.7.1, with the densities of the cells they cross. The idea would be to multiply the particles that come from a high mass cell when they pass in a low mass cell.

*Effect of a too crude tiling.* When the size and/or the macroscopic interaction cross-sections of the cells used to tile the space are too large, the resulting structure may be suboptimal, as seen in Supplemental Material I.2. Ideally, it should be ensured that the size of cells used is small compared to the mean free path of the particles, which is not obvious when the energy, the cross-sections or the atomic densities of the materials crossed by the particles vary greatly throughout their transport and the iterations of the algorithm. On the other hand, if the size of cells is excessively reduced, the problems of the large computing time or of the noise on the structures resurface.

However, after the tests carried out in sections 2-5, we note that the main limiting factor of the optimization procedure proposed in this study, and the common denominator of all problems mentioned above, is undoubtedly − and unsurprisingly − the computing power at disposal. For objects with an axial symmetry, made of a few thousands cells and a small number of materials, we have shown that a 16-CPU machine from the early 2010s is sufficient. But for symmetry-free problems involving a very large number of cells and materials, with complex objectives or constraints, these computations would require heavy industry or state resources, e.g. computer clusters and supercomputers, until the cost of computer components drop sufficiently for these calculations to democratize.

Finally, there are important questions that have not been addressed in this study. We can cite for example the problem posed by the calculation of the derivatives of the objective when the modeling of the system to be optimized requires the coupling of the transport physics with those of the materials used, e.g. hydraulics or heat transfer physics in a nuclear reactor. If the codes used for these supplementary calculations can compute sensitivities in the same way as MCNP does, however, this task should be feasible. Another interesting problem is the calculation of the impact that uncertainties on the transport data may have on the calculated optimal structure. In this respect, it will not suffice to quantify the resulting uncertainty on the derivatives of the objective. It will be necessary to propagate these uncertainties throughout the iterations of the algorithm, and see if a slight modification of a transport datum, e.g. a parameter of a resonance or the value of a cross-section, can cause a strong modification of the final shape, i.e. induce a topological phase transition. This problem looks thrilling, but very difficult. Indeed, emergence phenomena in iterative systems, even simple (which is not the case here), remain difficult if not impossible to predict analytically, see for example the classical counterexample of Langton's ant [40-42]. At first, it will therefore be necessary to solve this problem by brute force, using a Full Monte-Carlo method, which will require a considerable computing power.

# Topology optimization in the framework of the linear Boltzmann equation – supplemental material

Sébastien Chabod
Univ. Grenoble Alpes, CNRS, Grenoble INP, LPSC-IN2P3, 38000 Grenoble, France
Postal address: LPSC, 53 avenue des Martyrs, 38026 Grenoble, France
Email address: sebastien.chabod@lpsc.in2p3.fr
ORCID ID: https://orcid.org/0000-0003-2154-2012

**Abstract**. In this Supplemental Material file are gathered several discussions and demonstrations performed for the study, required for its comprehension but too lengthy to be included in the main body of the manuscript.

## Section A. Explicit formulation of the angular flux as a function of the density

By transforming the Boltzmann equation (0.1)-(0.2) into an integral equation, one obtains in the case where $\Sigma_{fixed} = 0$ for example, see [27] pp 95-96:

$$\varphi(\underline{r}, E, \underline{\Omega}) = \varphi_0(\underline{r}, E, \underline{\Omega})$$

$$+ \int_{s=0}^{+\infty}\int_{E'=0}^{+\infty}\int_{\underline{\Omega}' \in 4\pi} \left[ \begin{array}{l} n(\underline{r} - s\underline{\Omega})\sigma_s(E')f(E' \to E, \underline{\Omega}' \to \underline{\Omega})\varphi(\underline{r} - s\underline{\Omega}, E', \underline{\Omega}') \\ \times \exp\left(-\sigma_t(E)\int_{s'=0}^{s} n(\underline{r} - s'\underline{\Omega})ds'\right) \end{array} \right] dE'\, d\underline{\Omega}'\, ds \quad (A.1)$$

$$\varphi_0(\underline{r}, E, \underline{\Omega}) = \int_{s=0}^{+\infty} Q(\underline{r} - s\underline{\Omega}, E, \underline{\Omega})\exp\left(-\sigma_t(E)\int_{s'=0}^{s} n(\underline{r} - s'\underline{\Omega})ds'\right)ds$$

In (A.1), $\varphi_0(\underline{r},E,\underline{\Omega})$ is the flux of particles having performed 0 scattering. With this formulation, one can note that the flux $\varphi_{n+1}$ of particles having performed $n+1$ scatterings can be expressed as a function of the flux $\varphi_n$ of particles having performed $n$ scatterings, in the following way:

$$\varphi_{n+1}(\underline{r}, E, \underline{\Omega}) =$$

$$\int_{s=0}^{+\infty}\int_{E'=0}^{+\infty}\int_{\underline{\Omega}' \in 4\pi} \left[ \begin{array}{l} n(\underline{r} - s\underline{\Omega})\sigma_s(E')f(E' \to E, \underline{\Omega}' \to \underline{\Omega})\varphi_n(\underline{r} - s\underline{\Omega}, E', \underline{\Omega}') \\ \times \exp\left(-\sigma_t(E)\int_{s'=0}^{s} n(\underline{r} - s'\underline{\Omega})ds'\right) \end{array} \right] dE'\, d\underline{\Omega}'\, ds \quad (A.2)$$

The angular flux $\varphi(\underline{r},E,\underline{\Omega})$ of particles is then given by:

$$\varphi(\underline{r}, E, \underline{\Omega}) = \sum_{n=0}^{+\infty} \varphi_n(\underline{r}, E, \underline{\Omega}) \quad (A.3)$$

The particle flux is thus a sum of multiple integrals involving an increasing number of products $n(\underline{r})n(\underline{r}')\ldots$ of the density to be optimized, multiplied by an increasing number of exponentials of $n(\underline{r})$, to be calculated over an infinite number of possible trajectories. The particle flux thus appears as a non-linear, complicated function of the density $n(\underline{r})$.

# Section B. Discarded approaches for computing the derivatives of $O\varphi$

In sections B.1 and B.2, we present two methods that are conventionally used for computing derivatives of an objective: (i) the direct sensitivity analysis method, (ii) and the adjoint problem method. For both, we discuss why they cannot be used in practice in our study. We then describe in section B.3 an approach we have developed, ultimately abandoned too in favor of the solution described in section 1, which is already implemented in the reference transport code MCNP.

## B.1. Computation of the derivatives of $O\varphi$ using a direct sensitivity analysis: impossible today

Suppose we have a computer code that can solve the linear Boltzmann equation. We can use it to compute the objective functional $O\varphi(\underline{\rho})$, for an initial guess of the densities $\underline{\rho}$ (that will be most likely suboptimal). By slightly increasing the component $\rho_i$ of $\underline{\rho}$ by a quantity $\delta\rho_i$ in the associated cell $\Theta_i$, then reusing the code to compute the new objective functional $O\varphi(\ldots, \rho_i+\delta\rho_i, \ldots)$, we can estimate the derivative $\partial O\varphi/\partial\rho_i$ at the point $\underline{\rho}$ by doing:

$$\frac{\partial O\varphi}{\partial \rho_i}(\underline{\rho}) \approx \frac{O\varphi(\ldots, \rho_i+\delta\rho_i, \ldots) - O\varphi(\ldots, \rho_i, \ldots)}{\delta\rho_i} \quad (B.1)$$

To solve the optimization problem (1.8), the next step would be the construction of an algorithm that would make these derivatives converge towards the isovalue $\lambda$ that verifies the weight constraint, cf. (1.10). This procedure of resolution by direct calculation seems simple on paper. However, for realistic problems, e.g. with a large number of cells $\Theta_i$, it is unusable in practice with the computing resources available in 2018, for two main reasons:
1) the computing time: solving the Boltzmann equation for a heterogeneous configuration of densities takes time. Yet, assuming that the available space is divided into a reasonable number of volumes, e.g. 20×20×20 = 8000 to achieve a spatial accuracy of 5%, it will be necessary to solve 8001 times the transport problem to calculate all the derivatives (B.1). This number must then be multiplied by the number of iterations required to converge towards the isovalue $\lambda$. This is a procedure that is hardly doable in a human-compatible time, even for simple problems;
2) the numerical precision: in order for the derivative (B.1) not to be biased by second-order terms, $\delta\rho_i$ must be a small variation of density, which in turn imposes that $O\varphi(\rho_i+\delta\rho_i)$ is likely close to $O\varphi(\rho_i)$. But the computations of $O\varphi(\rho_i+\delta\rho_i)$ and $O\varphi(\rho_i)$ cannot be performed with an unlimited numerical precision. These values will be affected by statistical and numerical errors, potentially non-negligible compared to the difference $O\varphi(\rho_i+\delta\rho_i)-O\varphi(\rho_i)$. Thus, the estimated derivatives (B.1) will be noised. This problem has no simple solution: if the accuracy of the computations is increased to lower this noise, the issue of the computing time may take over.

## B.2. Computation of the derivatives of $O\varphi$ using the adjoint problem: interesting but difficult

The practical difficulty of computing the derivatives of $O\varphi$ with the direct method (B.1) is not exclusive to our specific problem of the design of an optimal particle propagator. To circumvent it, there is a reference solution, omnipresent in the publications about topology optimization, which makes use of the adjoint problem, see e.g. [5], [6] pp 16-18, [8] pp 26-27.
First, let us define the scalar product $<u,v>$ of two functions $u(\underline{r},E,\underline{\Omega})$ and $v(\underline{r},E,\underline{\Omega})$ as follows:

$$\langle u,v \rangle_{\underline{r}E\underline{\Omega}} = \int_{\underline{\Omega}\in 4\pi} \int_{\underline{r}\in\mathfrak{R}^3} \int_{E=0}^{+\infty} u(\underline{r},E,\underline{\Omega})^* v(\underline{r},E,\underline{\Omega}) d\underline{r}dEd\underline{\Omega} \quad (B.2)$$

and suppose that the objective functional $O$ is given by the formula (1.1). Now, let us take the Lagrangian (1.9), and insert into it the equation of the problem, here the Boltzmann equation, as a constraint. With the notation (B.2), this Lagrangian becomes:

$$L(\underline{\rho},\lambda) = \langle fgh, \varphi \rangle - \lambda(\underline{\rho}.\underline{V} - P_{max}) - \langle \omega, B\varphi - Q \rangle \quad (B.3)$$

In (B.3), function $\omega(\underline{r},E,\underline{\Omega})$ is a weight function, which is nondescript yet and which absorbs the Lagrange multiplier associated to the constraint $B\varphi = Q$. Now derivate $L$ with respect to $\rho_i$:

$$\frac{\partial L}{\partial \rho_i} = \left\langle fgh, \frac{\partial \varphi}{\partial \rho_i} \right\rangle - \lambda V_i - \underbrace{\left\langle \frac{\partial \omega}{\partial \rho_i}, B\varphi - Q \right\rangle}_{=0} - \left\langle \omega, \frac{\partial B}{\partial \rho_i}\varphi \right\rangle - \left\langle \omega, B\frac{\partial \varphi}{\partial \rho_i} \right\rangle = 0 \quad (B.4)$$

Let us then introduce the adjoint Boltzmann operator, $B^+$, defined by:

$$\langle u, Bv \rangle = \langle B^+u, v \rangle \quad (B.5)$$

For functions $u$ and $v$ that vanish when $\|\underline{r}\| \to +\infty$, using integrations by parts for the differential term $\underline{\Omega}.\underline{\nabla}$ of B, cf. (0.2), and making the changes of variables $E \leftrightarrow E'$ and $\underline{\Omega} \leftrightarrow \underline{\Omega}'$ for the integral term of B, one can obtain the expression of $B^+$, which is written:

$$B^+u(\underline{r},E,\underline{\Omega}) = -\underline{\Omega}.\underline{\nabla}u(\underline{r},E,\underline{\Omega}) + (\sigma_t(E)n(\underline{r}) + \Sigma_{fixed,t}(\underline{r},E))u(\underline{r},E,\underline{\Omega})$$

$$- n(\underline{r})\sigma_s(E) \int_{E'=0}^{+\infty} \int_{\underline{\Omega}'\in 4\pi} f(E \to E', \underline{\Omega} \to \underline{\Omega}')u(\underline{r},E',\underline{\Omega}')dE'\,d\Omega' \quad (B.6)$$

$$- \Sigma_{fixed,s}(\underline{r},E) \int_{E'=0}^{+\infty} \int_{\underline{\Omega}'\in 4\pi} f_{fixed}(\underline{r},E \to E', \underline{\Omega} \to \underline{\Omega}')u(\underline{r},E',\underline{\Omega}')dE'\,d\Omega'$$

By using this adjoint operator $B^+$ in (B.4), one finds:

$$\frac{\partial L}{\partial \rho_i} = \left\langle fgh - B^+\omega, \frac{\partial \varphi}{\partial \rho_i} \right\rangle - \left\langle \omega, \frac{\partial B}{\partial \rho_i}\varphi \right\rangle - \lambda V_i = 0 \quad (B.7)$$

where the calculation of $\partial B/\partial \rho_i$ is given in section B.3. Now let us choose the weight function $\omega$ in such a way that:

$$B^+\omega = fgh \quad (B.8)$$

The derivatives of $L$ are then rewritten:

$$\frac{\partial L}{\partial \rho_i} = -\left\langle \omega, \frac{\partial B}{\partial \rho_i}\varphi \right\rangle - \lambda V_i = 0 \quad \Rightarrow \quad -\frac{1}{V_i}\left\langle \omega, \frac{\partial B}{\partial \rho_i}\varphi \right\rangle = \lambda \quad (B.9)$$

With the direct computation procedure, (B.1), of the derivatives, for a tiling of the usable space in 20×20×20 cells, we have seen it was necessary to carry out 8001 resolutions of the Boltzmann equation per iteration of the convergence algorithm towards the isovalue $\lambda$. Here, by using the adjoint problem, only two computations per iteration are required, that of $\varphi$ and that of $\omega$. These computations are of course more complex than those of section B.1: if the computer code used to solve the Boltzmann equation is for example the MCNP code [15], this method amounts to replacing 8001 computations with 1 tally card (that of $O\varphi$) by 2 computations with 8000 tally cards each (those of $\omega$ and of $(\partial B/\partial \rho_i)\varphi$ in each cell $\Theta_i$). As these computations could still be faster than the previous 8001 ones, the use of the adjoint problem may be an interesting solution. However, for practical issues, this solution is also not usable for our study. Indeed:
1) if the code used to solve the Boltzmann equation is a Monte-Carlo code, e.g. MCNP, it cannot be used as is to solve the adjoint problem: it must first be adapted to operate in reverse transport [16]. Another issue is that the interaction cross-sections that appear in $B^+$ are not the same as those that appear in the direct problem, $B\varphi = Q$. To create them, transport data libraries would have to be re-processed, which would require a sizable amount of work. Finally, even if $\omega$ and $(\partial B/\partial \rho_i)\varphi$ could be computed with a Monte-Carlo code, this task would have to be done with a fine discretization in energy and solid angle to estimate the scalar product (B.9) with a good accuracy, which would require the sampling of a large number of source particles, so a large computing time. Calculation tricks, such as the IFP method [17], could possibly unlock this situation, but we are not aware of such a solution published for our problem.
2) If the code used is a deterministic code, e.g. PARTISN [18], it may be used as is to solve the adjoint problem. However, deterministic codes work with potentially unacceptable constraints for this study: (i) a precise calculation of the integrals $<\omega, (\partial B/\partial \rho_i)\varphi>$ will require anew a fine discretization in energy and solid angle of $\varphi$ and of $\omega$, so a large computing time as well as the handling of large matrices; (ii) in order to work, deterministic codes make use of simplifications and truncations, on the geometry, on the number of groups in energy, on the number of terms in the multipolar development of the flux, etc., which may lead to suboptimal solutions; (iii) as a matter of fact, the optimal structures obtained in the following sections are heterogeneous and have voids, which deterministic codes have difficulty managing.
For these reasons, we will not use in this study the adjoint problem method for computing derivatives. We will use it only in section E, to obtain the equation of the surface of a device that maximizes the neutron flux in a region of space.

B.3. Computation of the derivatives of $O\varphi$ using MCNP cell flag cards: not ready

Suppose that the objective functional $O$ is linear, as in section 2 for example. Then the derivative of the objective $O\varphi$ is simply:

$$\frac{\partial O\varphi}{\partial \rho_i} = O\frac{\partial \varphi}{\partial \rho_i} \quad \text{(B.10)}$$

In the case $O$ is non-linear, suppose otherwise that the derivatives of $O\varphi$ can be expressed as a combination of derivatives $\partial \phi/\partial \rho_i$ of some total particle fluxes, as in sections 3 and 5. In these two cases, one can observe that the computation of the derivatives of the objective only requires to compute the derivatives $\partial \varphi/\partial \rho_i$ of the flux.
By differentiating the Boltzmann equation, $B\varphi = Q$, with respect to the volume density $\rho_i$ in cell $\Theta_i$, we note that:

$$\frac{\partial B}{\partial \rho_i}\varphi + B\frac{\partial \varphi}{\partial \rho_i} = 0 \quad \text{(B.11)}$$

Using the expression (0.2) of B and the parametrization (1.6) of $\rho(\underline{r})$, we then note that:

$$\frac{\partial B}{\partial \rho_i}\varphi = \beta\left[\sigma_t(E)\varphi(\underline{r},E,\underline{\Omega}) - \int_{E'=0}^{+\infty}\int_{\underline{\Omega}'\in 4\pi}\sigma_s(E')f(\underline{r},E'\to E,\underline{\Omega}'\to\underline{\Omega})\varphi(\underline{r},E',\underline{\Omega}')dE'\,d\underline{\Omega}'\right]\Theta_i(\underline{r}) \quad (B.12)$$

where $\beta$ is a proportionality factor between the atomic density $n(\underline{r})$ and the volume density $\rho(\underline{r})$ of the material. For reminder, function $\Theta_i(\underline{r})$ returns 1 if $\underline{r} \in \Theta_i$ and 0 otherwise. Observing that: (i) $\Sigma_{fixed}$ should be zero in a cell $\Theta_i$ usable in the optimization procedure; (ii) $\Theta_i(\underline{r})^2 = \Theta_i(\underline{r})$, the result (B.12) can be simplified noticing that:

$$[Q(\underline{r},E,\underline{\Omega}) - \underline{\Omega}\nabla\varphi(\underline{r},E,\underline{\Omega})]\Theta_i(\underline{r}) = \rho_i\frac{\partial B}{\partial \rho_i}\varphi = [Q(\underline{r},E,\underline{\Omega}) - div(\underline{j})]\Theta_i(\underline{r}) \quad (B.13)$$

where $\underline{j}(\underline{r},E,\underline{\Omega}) = \underline{\Omega}\varphi(\underline{r},E,\underline{\Omega})$ is the particle current.
In the same way as $\Sigma_{fixed} = 0$ in the cells $\Theta_i$, we have most often $Q = 0$ in $\Theta_i$. Indeed, the volume that contains the source is most often not usable in the optimization procedure. For example, if one inserts a neutron generator or a gamma-ray source in a volume of matter whose density profile $\rho$ is to be optimized, the material of the source is fixed, independent on $\rho$. In this case, by imposing that $\rho_i \neq 0$ [1], we have in the material:

$$\frac{\partial B}{\partial \rho_i}\varphi = -\frac{div(\underline{j})}{\rho_i}\Theta_i(\underline{r}) \quad (B.14)$$

For a volume $\Theta_i$ sufficiently small, using the divergence theorem, we have:

$$div(\underline{j})\Theta_i(\underline{r}) \approx \frac{\Theta_i(\underline{r})}{V_i}\int_{\underline{r}\in\Theta_i}div(\underline{j})d\underline{r} \approx \frac{\Theta_i(\underline{r})}{V_i}\oint_{\underline{r}\in S_i}\underline{j}.\underline{n}dS \quad (B.15)$$

where $S_i$ and $V_i$ are the surface and volume of cell $\Theta_i$, and $\underline{n}$ is an outgoing unitary normal vector of $S_i$. Using result (B.11), we thus observe that the derivatives $\partial\varphi/\partial\rho_i$ of the flux sought obey Boltzmann equations, whose sources are effective sources $Q_i$ localized in $\Theta_i$ and given by:

$$B\frac{\partial\varphi}{\partial\rho_i} = Q_i, \quad Q_i \approx \frac{\Theta_i(\underline{r})}{V_i\rho_i}\oint_{\underline{r}\in S_i}\underline{j}.\underline{n}dS \quad (B.16)$$

The derivatives $\partial\varphi/\partial\rho_i$ are therefore computable with the solver of the Boltzmann equation at our disposal. We then note that the effective sources $Q_i$ are related to the balance of particles entering and leaving the cells $\Theta_i$. So, if the solver used is of the Monte-Carlo type, by using the registry of the collisions undertaken by the particles and flagging cards, it may be possible to compute the derivatives $\partial\varphi/\partial\rho_i$ in one direct calculation.
However, we have not developed further on this method, as the solution described in section 1, based on MCNP PERT cards, is already available and convenient.

---

[1] This is feasible by imposing a non-zero minimum bound on the density $\rho(\underline{r})$, a constraint that we have actually adopted in this study.

## Section C. Optimal surface $S_{opt}$ for a monoenergetic diffusive transport

In this section, we will solve analytically the optimization problem (2.1.4).
The Lagrangian of this problem can be written:

$$L(S,\lambda) = \phi(\underline{r}_s \to \underline{r}_d, S) - \lambda \left( \int_{\underline{r} \in \Re^3} \Theta[S(\underline{r}) \geq 0] d\underline{r} - V_{max} \right) \quad (C.1)$$

The optimal surface $S_{opt}(\underline{r})$, solution of (2.1.4), thus obeys to the following system of equations:

$$\left. \frac{\partial L}{\partial S} \right|_{S=S_{opt}} = 0, \quad \left. \frac{\partial L}{\partial \lambda} \right|_{S=S_{opt}} = 0 \quad (C.2)$$

C.1. Formulation of the neutron flux as an explicit function of the surface

To solve (2.1.4), we must first reformulate the neutron flux as an explicit function of the surface $S$. For this task, we can use a Green's function, $G(\underline{r};\underline{r}')$, of the Laplacian, cf. [28] pp 74, which is by definition a solution [2] of the equation:

$$-\Delta_r G(\underline{r};\underline{r}') = \delta(\underline{r}-\underline{r}') \quad (C.1.1)$$

Using Green's second identity, see [28] pp 73, the differential problem (2.1.1) and its boundary condition (2.1.3) can be transformed into a single, integral equation, which is written, cf. [28] pp 75:

$$\phi(\underline{r}_s \to \underline{r}, S) = \frac{1}{D} \int_{\underline{r}' \in V} G(\underline{r}';\underline{r}) Q(\underline{r}') d\underline{r}' + \oint_{\underline{r}' \in S} G(\underline{r}';\underline{r}) \underline{n}' \cdot \nabla_{r'} \phi(\underline{r}_s \to \underline{r}', S) dS' \quad (C.1.2)$$

with $V$ and $S$ being the volume and the surface of the material to be optimized, $d\underline{r}'$ and $dS$ being elements of this volume and this surface, and $\underline{n}'(\underline{r}')$ being an outgoing unit normal vector of the surface $S$ at the point $\underline{r}'$. This equation holds regardless of the Green's function $G(\underline{r};\underline{r}')$ used. It is thus most often simplified by taking the homogeneous Green's function, denoted $G_H(\underline{r};\underline{r}')$, which vanishes at the surface $S$, i.e. such that $G_H(\underline{r};\underline{r}') = 0 \; \forall \underline{r} \in S$. This choice allows to eliminate the surface integral in (C.1.2), which leads directly to a solution of the problem (2.1.1)-(2.1.3). For simple volumes, e.g. a cube or a sphere, it is then possible to formulate $G_H(\underline{r};\underline{r}')$ by using the spectral representation of the Green's functions, which makes use of the eigenfunctions and eigenvalues of the Laplacian (see [28] pp 67-69 or [29] pp 32-33). For the problem considered in this section, however, the shape $S$ of the material to be optimized is still unknown. The usual resolution with the homogeneous Green's function will therefore lead to a dead end here, as this function will implicitly depend on $S$ on the one hand, and as it will not possess any useful analytical expression on the other hand.
To circumvent this problem, one could look for a Green's function that is both independent on $S$ and analytically expressible. A function satisfying these conditions is the Green's function of the infinite system. This specific Green's function, which can be denoted $G_\infty(\underline{r};\underline{r}')$, cf. [28] pp 78-79, is the unique solution of the system:

---
[2] The equation (C.1.1) admitting an infinity of solutions, globally one per imaginable boundary conditions, there is an infinity of eligible Green's functions.

$$-\Delta_r G_\infty(\underline{r};\underline{r}') = \delta(\underline{r}-\underline{r}')$$
$$G_\infty(\underline{r};\underline{r}') \xrightarrow[\|\underline{r}\| \to +\infty]{} 0 \quad \text{(C.1.3)}$$

This system can be solved using a 3D Fourier transform, see [28] pp 78-79, to obtain:

$$G_\infty(\underline{r};\underline{r}') = \frac{1}{4\pi\|\underline{r}-\underline{r}'\|} \quad \text{(C.1.4)}$$

By using this Green's function, the equation (C.1.2) becomes:

$$\phi(\underline{r}_s \to \underline{r}, S) = \int_{\underline{r}' \in V} \frac{Q(\underline{r}')d\underline{r}'}{4\pi D\|\underline{r}-\underline{r}'\|} + \oint_{\underline{r}' \in S} \frac{\underline{n}'\cdot\nabla_{r'}\phi(\underline{r}_s \to \underline{r}', S)}{4\pi\|\underline{r}-\underline{r}'\|} dS' \quad \text{(C.1.5)}$$

By reinjecting in (C.1.5) the expression (2.1.2) of the source $Q(\underline{r})$, and remembering that the volume must contain the source under penalty of nullity of the neutron fluence, cf. comment 1 section 2.1, this equation is then rewritten:

$$\phi(\underline{r}_s \to \underline{r}, S) = \phi^{(0)}(\underline{r}_s \to \underline{r}) + \oint_{\underline{r}' \in S} \frac{\underline{n}'\cdot\nabla_{r'}\phi(\underline{r}_s \to \underline{r}', S)}{4\pi\|\underline{r}-\underline{r}'\|} dS' \quad \text{(C.1.6)}$$

where the field $\phi^{(0)}$ is independent on the surface $S(\underline{r})$ sought and is given by:

$$\phi^{(0)}(\underline{r}_s \to \underline{r}) = \frac{Q_0}{4\pi D\|\underline{r}-\underline{r}_s\|} \quad \text{(C.1.7)}$$

We can make an extra effort to take out the surface equation sought, $S(\underline{r})$, from the equation (C.1.6). By applying the divergence theorem, which is written:

$$\oint_{\underline{r}' \in S} \underline{n}'\cdot\underline{F}(\underline{r}')dS' = \int_{\underline{r}' \in V} div_{r'}(\underline{F}(\underline{r}'))d\underline{r}' \quad \text{(C.1.8)}$$

with $\underline{F}(\underline{r}')$ being a vector field, to the integral term in (C.1.6), we obtain:

$$\phi(\underline{r}_s \to \underline{r}, S) = \phi^{(0)}(\underline{r}_s \to \underline{r}) + \int_{\underline{r}' \in V} div_{r'}\left(\frac{\nabla_{r'}\phi(\underline{r}_s \to \underline{r}', S)}{4\pi\|\underline{r}-\underline{r}'\|}\right) d\underline{r}' \quad \text{(C.1.9)}$$

We can finally remove $S(\underline{r})$ from the bounds of the integral using a function $\Theta[]$, to obtain below an equation of the neutron flux drastically more manageable than the initial equations (2.1.1)-(2.1.3). Indeed, this equation now depends explicitly on the function $S(\underline{r})$ sought, which will allow us to perform a calculus of variations on it.

$$\phi(\underline{r}_s \to \underline{r}, S) = \phi^{(0)}(\underline{r}_s \to \underline{r}) + \int_{\underline{r}' \in \Re^3} \Theta[S(\underline{r}') \geq 0] div_{r'}\left(\frac{\nabla_{r'}\phi(\underline{r}_s \to \underline{r}', S)}{4\pi\|\underline{r}-\underline{r}'\|}\right) d\underline{r}' \quad \text{(C.1.10)}$$

C.2. Resolution of the optimization problem using a calculus of variation

To obtain the optimal shape $S_{opt}$ of the material that maximizes the neutron flux at the point $\underline{r}_d$, we must now solve the optimization problem (C.1)-(C.2). To calculate the derivatives (C.2) of the Lagrangian (C.1), we can use a perturbation method. Let us take the optimal surface $S_{opt}(\underline{r})$ and perturb it slightly. We obtain a new surface, $S(\underline{r}) = S_{opt}(\underline{r}) + dS(\underline{r})$, that is slightly suboptimal. This perturbation $dS$ of the surface induces in turn a perturbation $d\phi$ of the neutron flux,

$$\phi(\underline{r}_s \to \underline{r}, S_{opt} + dS) = \phi(\underline{r}_s \to \underline{r}, S_{opt}) + d\phi(\underline{r}) \quad (C.1.11)$$

as well as a perturbation $dL$ of the Lagrangian, which is given by:

$$L(S_{opt} + dS, \lambda) = L(S_{opt}, \lambda) + dS \frac{\partial L}{\partial S} = L(S_{opt}, \lambda) + dL$$

$$= \phi(\underline{r}_s \to \underline{r}_d, S_{opt}) + d\phi(\underline{r}_d) - \lambda \left( \int_{\underline{r} \in \Re^3} \Theta[S_{opt}(\underline{r}) + dS(\underline{r}) \geq 0] d\underline{r} - V_{max} \right) \quad (C.1.12)$$

where $\phi(\underline{r}_s \to \underline{r}_d, S_{opt})$ and $L(S_{opt}, \lambda)$ are given by:

$$\phi(\underline{r}_s \to \underline{r}, S_{opt}) = \phi^{(0)}(\underline{r}_s \to \underline{r}) + \int_{\underline{r}' \in \Re^3} \Theta[S_{opt}(\underline{r}') \geq 0] div_{\underline{r}'} \left( \frac{\nabla_{\underline{r}'} \phi(\underline{r}_s \to \underline{r}', S_{opt})}{4\pi \|\underline{r} - \underline{r}'\|} \right) d\underline{r}'$$

$$L(S_{opt}, \lambda) = \phi(\underline{r}_s \to \underline{r}_d, S_{opt}) - \lambda \left( \int_{\underline{r} \in \Re^3} \Theta[S_{opt}(\underline{r}) \geq 0] d\underline{r} - V_{max} \right) \quad (C.1.13)$$

Let us first calculate $dL$. Observing that the derivative of $\Theta[x \geq 0]$ is the Dirac function, $\delta(x)$, we obtain using a first-order expansion:

$$\Theta[S_{opt}(\underline{r}) + dS(\underline{r}) \geq 0] = \Theta[S_{opt}(\underline{r}) \geq 0] + dS(\underline{r})\delta(S_{opt}(\underline{r})) \quad (C.1.14)$$

Reinjecting this result in (C.1.12), we find:

$$dL = dS \frac{\partial L}{\partial S} = d\phi(\underline{r}_d) - \lambda \int_{\underline{r} \in \Re^3} dS(\underline{r})\delta(S_{opt}(\underline{r})) d\underline{r} \quad (C.1.15)$$

Let us now calculate the term $d\phi(\underline{r}_d)$ in (C.1.15). Using the result (C.1.10), we have:

$$\phi(\underline{r}_s \to \underline{r}, S_{opt} + dS) = \phi(\underline{r}_s \to \underline{r}, S_{opt}) + d\phi(\underline{r})$$

$$= \phi^{(0)}(\underline{r}_s \to \underline{r}) + \int_{\underline{r}' \in \Re^3} \begin{pmatrix} (\Theta[S_{opt}(\underline{r}') \geq 0] + dS(\underline{r}')\delta(S_{opt}(\underline{r}'))) \\ \times \left( div_{\underline{r}'} \left( \frac{\nabla_{\underline{r}'} \phi(\underline{r}_s \to \underline{r}', S_{opt})}{4\pi \|\underline{r} - \underline{r}'\|} \right) + div_{\underline{r}'} \left( \frac{\nabla_{\underline{r}'} d\phi(\underline{r}')}{4\pi \|\underline{r} - \underline{r}'\|} \right) \right) \end{pmatrix} d\underline{r}' \quad (C.1.16)$$

Neglecting the second-order term in $dSd\phi$ in (C.1.16), we obtain an equation of the perturbation $d\phi(\underline{r})$ sought:

$$d\phi(\underline{r}) = d\phi_0(\underline{r}) + \int_{\underline{r}'\in\mathfrak{R}^3} \Theta[S_{opt}(\underline{r}') \geq 0] div_{\underline{r}'}\left(\frac{\nabla_{\underline{r}'} d\phi(\underline{r}')}{4\pi\|\underline{r}-\underline{r}'\|}\right) d\underline{r}'$$

$$d\phi_0(\underline{r}) = \int_{\underline{r}'\in\mathfrak{R}^3} dS(\underline{r}')\delta(S_{opt}(\underline{r}')) div_{\underline{r}'}\left(\frac{\nabla_{\underline{r}'}\phi(\underline{r}_s\to\underline{r}',S_{opt})}{4\pi\|\underline{r}-\underline{r}'\|}\right) d\underline{r}'$$

(C.1.17)

We verify that the equation of the perturbation $d\phi(r)$ is identical to that, (C.1.13), of the neutron flux $\phi(\underline{r}_s\to\underline{r},S_{opt})$. Hence, by considering the fields $\phi^{(0)}(\underline{r}_s\to\underline{r})$ and $d\phi_0(\underline{r})$ as source terms in this mutual equation, we can solve it directly by using one of its Green's function, $g(\underline{r};\underline{r}'')$, which is by definition a solution of:

$$g(\underline{r};\underline{r}'') = \delta(\underline{r}-\underline{r}'') + \int_{\underline{r}'\in\mathfrak{R}^3} \Theta[S_{opt}(\underline{r}') \geq 0] div_{\underline{r}'}\left(\frac{\nabla_{\underline{r}'} g(\underline{r}';\underline{r}'')}{4\pi\|\underline{r}-\underline{r}'\|}\right) d\underline{r}' \quad (C.1.18)$$

With this Green's function and the result (C.1.7), the flux $\phi(\underline{r}_s\to\underline{r},S_{opt})$ and its perturbation $d\phi$ are indeed given by:

$$\phi(\underline{r}_s\to\underline{r},S_{opt}) = \int_{\underline{r}''\in\mathfrak{R}^3} \phi^{(0)}(\underline{r}_s\to\underline{r}'')g(\underline{r};\underline{r}'')d\underline{r}'' = \frac{Q_0}{4\pi D}\int_{\underline{r}''\in\mathfrak{R}^3} \frac{g(\underline{r};\underline{r}'')}{\|\underline{r}_s-\underline{r}''\|}d\underline{r}''$$

$$d\phi(\underline{r}) = \int_{\underline{r}''\in\mathfrak{R}^3} d\phi_0(\underline{r}'')g(\underline{r};\underline{r}'')d\underline{r}'' \quad (C.1.19)$$

$$= \int_{\underline{r}'\in\mathfrak{R}^3}\int_{\underline{r}''\in\mathfrak{R}^3} g(\underline{r};\underline{r}'')dS(\underline{r}')\delta(S_{opt}(\underline{r}')) div_{\underline{r}'}\left(\frac{\nabla_{\underline{r}'}\phi(\underline{r}_s\to\underline{r}',S_{opt})}{4\pi\|\underline{r}'-\underline{r}''\|}\right) d\underline{r}''d\underline{r}'$$

We have thereby, using (C.1.15) and making the change of variable $\underline{r}\leftrightarrow\underline{r}'$ in its integral term:

$$dL = d\phi(\underline{r}_d) - \lambda \int_{\underline{r}'\in\mathfrak{R}^3} dS(\underline{r}')\delta(S_{opt}(\underline{r}'))d\underline{r}'$$

$$= \int_{\underline{r}'\in\mathfrak{R}^3} dS(\underline{r}')\delta(S_{opt}(\underline{r}'))\left[\int_{\underline{r}''\in\mathfrak{R}^3} g(\underline{r}_d;\underline{r}'') div_{\underline{r}'}\left(\frac{\nabla_{\underline{r}'}\phi(\underline{r}_s\to\underline{r}',S_{opt})}{4\pi\|\underline{r}'-\underline{r}''\|}\right) d\underline{r}'' - \lambda\right] d\underline{r}'$$

(C.1.20)

But, cf. (C.2) and (C.1.15), in order for the surface $S_{opt}$ to be optimal, it is required that $dL = 0$ $\forall dS$. We must therefore have:

$$\delta(S_{opt}(\underline{r}'))\left[\int_{\underline{r}''\in\mathfrak{R}^3} g(\underline{r}_d;\underline{r}'') div_{\underline{r}'}\left(\frac{\nabla_{\underline{r}'}\phi(\underline{r}_s\to\underline{r}',S_{opt})}{4\pi\|\underline{r}'-\underline{r}''\|}\right) d\underline{r}'' - \lambda\right] = 0, \forall \underline{r}'$$

$$\Rightarrow \int_{\underline{r}''\in\mathfrak{R}^3} g(\underline{r}_d;\underline{r}'') div_{\underline{r}'}\left(\frac{\nabla_{\underline{r}'}\phi(\underline{r}_s\to\underline{r}',S_{opt})}{4\pi\|\underline{r}'-\underline{r}''\|}\right) d\underline{r}'' = \lambda, \forall \underline{r}'/S_{opt}(\underline{r}') = 0, \text{i.e. } \forall \underline{r}'\in S_{opt}$$

(C.1.21)

We obtain here a first equation of the optimal surface $S_{opt}$, which can be further simplified by noticing that the *div* operator in (C.1.21) only acts on the coordinates $\underline{r}'$. We thereby have:

$$div_{\underline{r}'}\left(\underline{\nabla}_{\underline{r}'}\phi(\underline{r}_s \to \underline{r}', S_{opt})\int_{\underline{r}''\in\Re^3} \frac{g(\underline{r}_d;\underline{r}'')}{4\pi\|\underline{r}'-\underline{r}''\|}d\underline{r}''\right) = \lambda, \forall \underline{r}' \in S_{opt} \quad (C.1.22)$$

But, cf. (C.1.19), we observe that:

$$\int_{\underline{r}''\in\Re^3} \frac{g(\underline{r}_d;\underline{r}'')}{\|\underline{r}'-\underline{r}''\|}d\underline{r}'' = \frac{4\pi D}{Q_0}\phi(\underline{r}' \to \underline{r}_d, S_{opt}) \quad (C.1.23)$$

The surface equation hence becomes:

$$div_{\underline{r}'}\left(\phi(\underline{r}' \to \underline{r}_d, S_{opt})\underline{\nabla}_{\underline{r}'}\phi(\underline{r}_s \to \underline{r}', S_{opt})\right) = \lambda Q_0/D, \forall \underline{r}' \in S_{opt} \quad (C.1.24)$$

Noting that $div(u\underline{\nabla}v) = u\Delta v + \underline{\nabla}u.\underline{\nabla}v$, then using the starting equations (2.1.1)-(2.1.2), we can further simplify this result, to obtain:

$$\underline{\nabla}_{\underline{r}'}\phi(\underline{r}' \to \underline{r}_d, S_{opt})\underline{\nabla}_{\underline{r}'}\phi(\underline{r}_s \to \underline{r}', S_{opt}) - \frac{Q_0}{D}\phi(\underline{r}' \to \underline{r}_d, S_{opt})\delta(\underline{r}'-\underline{r}_s) = \frac{\lambda Q_0}{D}, \forall \underline{r}' \in S_{opt}$$
(C.1.25)

For a diffusive transport, the optimal surface cannot contain the point $\underline{r}_s$. Indeed, if the source is slightly moved outside the object, the neutron flux at $\underline{r}_d$ immediately falls to 0, cf. comment 1 of section 2.1. At the opposite, if the source is slightly moved inside the volume, half of the emitted neutron will be lost due to the boundary condition, which is suboptimal. We must thus have $\delta(\underline{r}'-\underline{r}_s) = 0$ if $\underline{r}' \in S_{opt}$, and it remains:

$$\underline{\nabla}\phi(\underline{r}_s \to \underline{r}, S_{opt}).\underline{\nabla}\phi(\underline{r} \to \underline{r}_d, S_{opt}) = \lambda Q_0/D, \forall \underline{r} \in S_{opt} \quad (C.1.26)$$

We can simplify a bit more this result using Green's second identity, which is written:

$$\int_{\underline{r}\in V}(u\Delta v - v\Delta u)d\underline{r} = \int_{\underline{r}\in S}\underline{n}.(u\underline{\nabla}v - v\underline{\nabla}u)dS \quad (C.1.27)$$

where $u$ et $v$ are unspecified functions. By taking $u(\underline{r}) = \phi(\underline{r}_1 \to \underline{r})$ and $v(\underline{r}) = \phi(\underline{r}_2 \to \underline{r})$ in (C.1.27), then using the starting equation (2.1.1) in the left-hand term, and the boundary condition (2.1.3), $\phi(\underline{r}_1 \to \underline{r}) = \phi(\underline{r}_2 \to \underline{r}) = 0 \ \forall \underline{r} \in S$, in the right-hand term, one can show that $\phi(\underline{r}_1 \to \underline{r}_2) = \phi(\underline{r}_2 \to \underline{r}_1)$. The equation of the surface $S_{opt}$ can thus be rewritten:

$$\underline{\nabla}\phi(\underline{r}_s \to \underline{r}, S_{opt}).\underline{\nabla}\phi(\underline{r}_d \to \underline{r}, S_{opt}) = \lambda Q_0/D, \forall \underline{r} \in S_{opt} \quad (C.1.28)$$

In the framework of a monoenergetic diffusive transport, we thereby show that the surface of the propagator that maximizes the neutron flux at a point $\underline{r}_d$ is given by an isovalue of the scalar product of the gradients of $\phi(\underline{r}_s \to \underline{r})$ (flux generated at $\underline{r}$ by a point source positioned at $\underline{r}_s$) and $\phi(\underline{r}_d \to \underline{r})$ (flux generated at $\underline{r}$ by a point source placed at $\underline{r}_d$). The value of the Lagrange multiplier $\lambda$, hence of the isovalue to be used, is set by the volume condition, $V = V_{max}$.

Finally, for a diffusive transport, one can note that the current of neutrons, $\underline{J}(\underline{r}' \to \underline{r})$, generated at the point $\underline{r}$ by a point source placed at $\underline{r}'$ is:

$$\underline{J}(\underline{r}' \to \underline{r}, S_{opt}) = -D\underline{\nabla}\phi(\underline{r}' \to \underline{r}, S_{opt}) \quad (C.1.29)$$

We thus obtain an alternative formulation of the optimal surface $S_{opt}$, which is:

$$\underline{J}(\underline{r}_s \to \underline{r}, S_{opt}) \underline{J}(\underline{r}_d \to \underline{r}, S_{opt}) = \lambda Q_0 D, \forall \underline{r} \in S_{opt} \quad (C.1.30)$$

C.3. Resolution of the surface equation for a monoenergetic diffusive transport

Although the equations (C.1.28) or (C.1.30) of the optimal surface are short, using them in practice does not seem simple. Indeed, how to calculate the flux $\phi(\underline{r}_s \to \underline{r})$ or the current $\underline{J}(\underline{r}_s \to \underline{r})$ in an object whose shape is still unknown? One could propose an iterative procedure, in which a finite-volume type software could for example be used to carry out these calculations for an arbitrary initial geometry $S_0$. One would then have to find a way to iterate the shape of the object so that its surface $S_n$ obtained at the iteration $n$ of the procedure converge towards $S_{opt}$ when $n \to +\infty$. This seems difficult, and uselessly computationally intensive.

As often, a simpler solution can be found by performing a physical analysis of the problem. By better observing the initial equations (2.1.1)-(2.1.3), cf. [28] pp 81, one can note that the neutron flux obeys the same equations as those of an electrostatic potential, $V(\underline{r})$, generated by a point electric charge $Q_0$ placed at the point $\underline{r}_s$ inside a grounded conducting cavity of surface $S$. In this case, Poisson's equation is indeed written:

$$-\Delta V = \rho/\varepsilon_0 = Q_0 \delta(\underline{r} - \underline{r}_s)/\varepsilon_0$$
$$V(\underline{r}) = 0, \forall \underline{r} \in S \quad (C.1.31)$$

Here, the permittivity $\varepsilon_0$ is $D$, and we have $V(\underline{r}) = \phi(\underline{r}_s \to \underline{r})$. Now, applying Gauss's theorem on the surface of the cavity, one can obtain the following classical result in electromagnetism [30]:

$$\underline{E}(\underline{r}) = -\underline{\nabla}V = \frac{\sigma(\underline{r})}{\varepsilon_0}\underline{n}, \forall \underline{r} \in S \quad (C.1.32)$$

where $\underline{E}$ is the electric field, $\underline{n}$ is a unit normal vector at the surface of the cavity, and $\sigma$ is the surface charge density induced on $S$ by the point charge $Q_0$ positioned at $\underline{r}_s$. By analogy, we can therefore write:

$$-\underline{\nabla}\phi(\underline{r}_{s,d} \to \underline{r}, S_{opt}) = \frac{\sigma_{s,d}(\underline{r})}{D}\underline{n}, \forall \underline{r} \in S \quad (C.1.33)$$

Applying this result to our problem, the surface equation (C.1.28) is then rewritten:

$$\sigma_s(\underline{r})\sigma_d(\underline{r}) = C, \forall \underline{r} \in S_{opt} \quad (C.1.34)$$

where $C$ is a constant. The optimal surface $S_{opt}$ sought in this section is thus also the surface of a conducting cavity for which the product of the electric charge surface densities $\sigma_s(\underline{r})$ and $\sigma_d(\underline{r})$, respectively induced on the surface of the cavity by point charges placed at $\underline{r}_s$ and $\underline{r}_d$, is constant at every point of the surface. It only remains now to find the equations of $\sigma_s$ and $\sigma_d$.

By injecting the expression (C.1.33) into the equations (C.1.6)-(C.1.7), we note that:

$$\phi(\underline{r}_{s,d} \to \underline{r}, S_{opt}) = \frac{Q_0}{4\pi D \|\underline{r} - \underline{r}_{s,d}\|} - \oint_{\underline{r}' \in S_{opt}} \frac{\sigma_{s,d}(\underline{r}')}{4\pi D \|\underline{r} - \underline{r}'\|} dS' \quad (C.1.35)$$

Then, taking $\underline{r} \in S_{opt}$ in (C.1.35) and using the boundary condition (2.1.3), we obtain two simple integral equations on $\sigma_s(\underline{r})$ and $\sigma_d(\underline{r})$, which are written:

$$\frac{Q_0}{\|\underline{r} - \underline{r}_{s,d}\|} = \oint_{\underline{r}' \in S_{opt}} \frac{\sigma_{s,d}(\underline{r}')}{\|\underline{r} - \underline{r}'\|} dS', \forall \underline{r} \in S_{opt} \quad (C.1.36)$$

For a monoenergetic diffusive transport, the complex optimization problem (2.1) hence reduces to the much simpler following problem:

Find $S_{opt}$ such as

$$\frac{Q_0}{\|\underline{r} - \underline{r}_{s,d}\|} = \oint_{\underline{r}' \in S_{opt}} \frac{\sigma_{s,d}(\underline{r}')}{\|\underline{r} - \underline{r}'\|} dS', \forall \underline{r} \in S_{opt}$$
$$\sigma_s(\underline{r})\sigma_d(\underline{r}) = C, \forall \underline{r} \in S_{opt} \quad (C.1.37)$$
$$V_{max} = \int_{\underline{r} \in \Re^3} \Theta[S_{opt}(\underline{r}) \geq 0] d\underline{r}$$

Now let us think about the shape that the optimal surface $S_{opt}$ should have. Our problem has two symmetries: (1) an axial symmetry around the axis SD passing by the points $\underline{r}_s$ and $\underline{r}_d$. To parametrize the problem, it would therefore be wise to use cylindrical or spherical coordinates; (2) a planar symmetry. Indeed, as the equation (C.1.28) of $S_{opt}$ and the integral equations of the densities $\sigma_s$ and $\sigma_d$ are symmetric in $\underline{r}_s$ and $\underline{r}_d$, the optimal surface should be symmetrical with respect to the plane perpendicular to the axis SD that contains the point $M$ at the middle of the segment SD. In addition to these symmetries, we noted in the comment 1 of section 2.1 that the optimal propagating medium should not be discontinuous nor contain a hole. Following these conclusions, let us use spherical coordinates, take for origin $\underline{0} = (0,0,0)$ of these coordinates the point $M$, take for axis $z$ of the spherical coordinates the axis SD, denote $r = \|\underline{r}\|$ the distance between $\underline{0}$ and the point $\underline{r} = (r,\theta,\varphi)$, and denote $\theta$ the angle between the axis $z$ and the line $(\underline{0},\underline{r})$. Observing that the optimal surface should not be wrinkled [3], and denoting $H$ the distance between the points $\underline{r}_s$ and $\underline{r}_d$, we can parametrize the surface $S_{opt}$, the densities $\sigma_{s,d}$, the elements of volume $d\underline{r}$ and of surface $dS'$, the points $\underline{r}_s = (r_s,\theta_s,\varphi_s)$ and $\underline{r}_d = (r_d,\theta_d,\varphi_d)$, and the distances $\|\underline{r}-\underline{r}_s\|$ and $\|\underline{r}-\underline{r}'\|$, as follows:

---

[3] At constant volume, a wrinkle induces an increase of the surface therefore an increase in neutron leakage, which is suboptimal.

$$S_{opt}(\underline{r}) = f(\theta) - r = 0$$
$$\sigma_{s,d}(\underline{r}) = \sigma_{s,d}(\theta)$$
$$\underline{r}_s = (H/2, \pi, 0), \quad \underline{r}_d = (H/2, 0, 0), \quad \underline{r} = (f(\theta), \theta, \varphi)$$

$$\oint_{\underline{r}' \in S_{opt}} \frac{\sigma_{s,d}(\underline{r}')}{\|\underline{r} - \underline{r}'\|} dS' = \int_{\theta'=0}^{\pi} \int_{\varphi'=0}^{2\pi} \frac{\sigma_{s,d}(\theta')}{\|\underline{r} - \underline{r}'\|} f(\theta') \sin\theta' \sqrt{f(\theta')^2 + \left(\frac{df}{d\theta'}\right)^2} d\theta' d\varphi' \quad \text{(C.1.38)}$$

$$\int_{\underline{r} \in \Re^3} \Theta[S_{opt}(\underline{r}) \geq 0] d\underline{r} = \int_{\theta=0}^{\pi} \int_{\varphi=0}^{2\pi} \int_{r=0}^{f(\theta)} r^2 \sin\theta \, dr \, d\theta \, d\varphi$$

$$\|\underline{r} - \underline{r}_s\| = \sqrt{f(\theta)^2 + f(\theta) H \cos\theta + (H/2)^2}$$
$$\|\underline{r} - \underline{r}'\| = \sqrt{f(\theta)^2 + f(\theta')^2 - 2f(\theta)f(\theta')(\cos\theta\cos\theta' + \sin\theta\sin\theta'\cos(\varphi - \varphi'))}$$

The planar symmetry also requires that:

$$\sigma_d(\theta) = \sigma_s(\pi - \theta), \forall \theta \quad \text{(C.1.39)}$$

In the particular case where $\underline{r}_s = \underline{r}_d$, due to the central symmetry, the optimal shape is a sphere of radius $R$ given by:

$$R = \left(\frac{3V_{max}}{4\pi}\right)^{1/3} \quad \text{(C.1.40)}$$

Using this radius $R$, we can make the equations (C.1.37)-(C.1.38) dimensionless by taking:

$$f(\theta) = RF(\theta), \quad \sigma_s(\theta) = \frac{Q_0 u(\theta)}{R^2} \quad \text{(C.1.41)}$$

Finally, by using: (i) the planar symmetry to reduce the range of variations of $\theta$ to $[0, \pi/2]$; (ii) the axial symmetry to eliminate the variable $\varphi$, taking $\varphi = 0$; (iii) the results (C.1.38) to (C.1.41) to simplify the equations, the problem (C.1.37) can be rewritten in the following simple way:

Find $F(\theta)$ such as

$$\frac{1}{\sqrt{F(\theta)^2 + F(\theta)\beta\cos\theta + (\beta/2)^2}} =$$
$$\int_{\theta'=0}^{\pi} \int_{\varphi'=0}^{2\pi} \frac{u(\theta') F(\theta') \sin\theta' \sqrt{F(\theta')^2 + (dF/d\theta')^2} \, d\theta' d\varphi'}{\sqrt{F(\theta)^2 + F(\theta')^2 - 2F(\theta)F(\theta')(\cos\theta\cos\theta' + \sin\theta\sin\theta'\cos\varphi')}}, \forall \theta \in \left[0, \frac{\pi}{2}\right] \quad \text{(C.1.42)}$$
$$u(\theta) u(\pi - \theta) = \alpha, \forall \theta \in [0, \pi/2]$$
$$\int_{\theta=0}^{\pi} F(\theta)^3 \sin\theta \, d\theta = 2$$

where $\alpha$ is a constant and where $\beta$ is given by:

$$\beta = \frac{H}{R} = H\left(\frac{4\pi}{3V_{max}}\right)^{1/3} \quad (C.1.43)$$

The surface $S_{opt}$ sought is then given by the equation $r/R = F(\theta)$, cf. (C.1.38) and (C.1.41).

## Section D. Numerical resolution of the system (2.1.7)

To solve the system (2.1.7), we must find a way to evaluate the double integral that appears in the right-hand side of its first equation. To do this, we eliminate its variable $\varphi'$, noticing that:

$$\int_{\varphi'=0}^{2\pi} \frac{d\varphi'}{\sqrt{F(\theta)^2 + F(\theta')^2 - 2F(\theta)F(\theta')(\cos\theta\cos\theta' + \sin\theta\sin\theta'\cos\varphi')}}$$

$$= \frac{4K\left(2\sqrt{\dfrac{F(\theta)F(\theta')\sin\theta\sin\theta'}{F(\theta)^2 + F(\theta')^2 - 2F(\theta)F(\theta')\cos(\theta+\theta')}}\right)}{\sqrt{F(\theta)^2 + F(\theta')^2 - 2F(\theta)F(\theta')\cos(\theta+\theta')}} \quad (D.1)$$

where function $K$ is the complete elliptic integral of the first kind, defined by:

$$K(k) = \int_{t=0}^{1} \frac{dt}{\sqrt{1-t^2}\sqrt{1-k^2 t^2}} = \int_{\theta=0}^{\pi/2} \frac{d\theta}{\sqrt{1-k^2 \sin(\theta)^2}} \quad (D.2)$$

Then, instead of solving the first and the second equations of (2.1.7) for all angles $\theta$, we will solve them for a finite number of angles $\theta_i$, chosen as follows:

$$\theta_i = i \times d\theta, \quad d\theta = \frac{\pi}{N}, \quad i \in [0, N] \quad (D.3)$$

So we pose:

$$F(\theta_i) = F_i, \quad u(\theta_i) = u_i \quad (D.4)$$

With these simplifications, the problem (2.1.7) can be rewritten:

Find $\underline{F} = (F_0, ..., F_{N/2})$ such as

$$\frac{1}{\sqrt{F_i^2 + F_i \beta \cos\theta_i + (\beta/2)^2}} = \sum_{j=0}^{N-1} \int_{\theta'=\theta_j}^{\theta_{j+1}} g_i(\theta') d\theta', \forall i \in \left[0, \frac{N}{2}\right]$$

$$u_i u_{N-i} = \alpha, \forall i \in [0, N/2] \quad (D.5)$$

$$\sum_{j=0}^{N-1} \int_{\theta'=\theta_j}^{\theta_{j+1}} F(\theta')^3 \sin\theta' \, d\theta' = 2$$

with:

$$g_i(\theta') = \frac{4u(\theta')F(\theta')\sin\theta'}{\sqrt{F_i^2 + F(\theta')^2 - 2F_iF(\theta')\cos(\theta_i + \theta')}}$$
$$\times \sqrt{F(\theta')^2 + \left(\frac{dF}{d\theta'}\right)^2} K\left(2\sqrt{\frac{F_iF(\theta')\sin\theta_i \sin\theta'}{F_i^2 + F(\theta')^2 - 2F_iF(\theta')\cos(\theta_i + \theta')}}\right) \quad \text{(D.6)}$$

The integrals appearing in the problem (D.5) can be approximated using the trapezoidal rule. We begin by discretizing the angles $\theta'$ in each interval $[\theta_j, \theta_{j+1}]$ as follows:

$$\theta_{jk} = \theta_j + \delta\theta \times (k + 1/2), \quad \delta\theta = \frac{d\theta}{M}, \quad k \in [0, M-1] \quad \text{(D.7)}$$

In (D.7), the use of $k+1/2$ instead of $k$ serves to avoid the case $\theta_{jk} = \theta_j$, which induces some zero values at the denominator of $g_i$. Afterwards, by approximating the functions $F(\theta')$ and $u(\theta')$ in each interval $[\theta_j, \theta_{j+1}]$ by linear functions,

$$F(\theta') = F_j + \frac{F_{j+1} - F_j}{d\theta}(\theta' - \theta_j), \quad u(\theta') = u_j + \frac{u_{j+1} - u_j}{d\theta}(\theta' - \theta_j), \quad \theta' \in [\theta_j, \theta_{j+1}] \quad \text{(D.8)}$$

and using the symmetry $F(\theta) = F(\pi - \theta)$, the problem (D.5) becomes:

Find $F$ such as

$$\frac{1}{\sqrt{F_i^2 + F_i\beta\cos\theta_i + (\beta/2)^2}} = \delta\theta \sum_{j=0}^{N-1}\sum_{k=0}^{M-1} g_{ijk}, \quad \forall i \in \left[0, \frac{N}{2}\right]$$
$$u_i u_{N-i} = \alpha, \quad \forall i \in [0, N/2] \quad \text{(D.9)}$$
$$\delta\theta \sum_{j=0}^{N-1}\sum_{k=0}^{M-1} F_{jk}^3 \sin\theta_{jk} = 2$$

with:

$$g_{ijk} = \frac{4u_{jk} F_{jk} \sin\theta_{jk} \sqrt{F_{jk}^2 + dF_j^2}}{\sqrt{F_i^2 + F_{jk}^2 - 2F_i F_{jk}\cos(\theta_i + \theta_{jk})}} K\left(2\sqrt{\frac{F_i F_{jk}\sin\theta_i \sin\theta_{jk}}{F_i^2 + F_{jk}^2 - 2F_i F_{jk}\cos(\theta_i + \theta_{jk})}}\right)$$
$$u_{jk} = u_j + \frac{u_{j+1} - u_j}{d\theta}(\theta_{jk} - \theta_j)$$
$$G_j = \begin{cases} F_j & \text{if } j < \frac{N}{2} \\ F_{N-j} & \text{else} \end{cases} \quad \text{(D.10)}$$
$$F_{jk} = G_j + dF_j(\theta_{jk} - \theta_j), \quad dF_j = \frac{G_{j+1} - G_j}{d\theta}$$

Now, consider $N+1$ numbers $x_i$, whose variance $\sigma^2$ is:

$$\sigma^2 = \frac{1}{N+1}\sum_{i=0}^{N} x_i^2 - \left(\frac{1}{N+1}\sum_{i=0}^{N} x_i\right)^2 \geq 0 \quad (D.11)$$

One can note that the equality $\sigma^2 = 0$ is reached iff the numbers $x_i$ are all equal. Solving the equation $u_i u_{N-i} = \alpha \ \forall i$ of (D.9) is thus equivalent to solving:

$$\left(\frac{\sigma}{M}\right)^2 = \frac{1}{(N+1)M^2}\sum_{i=0}^{N}(u_i u_{N-i})^2 - 1 = 0, \quad M = \frac{1}{N+1}\sum_{i=0}^{N} u_i u_{N-i} \quad (D.12)$$

In (D.12), we renormalized by $M^2$ to make this equation dimensionless.
To obtain the distances $F_i$, i.e. the shape of the propagator that maximizes the neutron flux at the point $\underline{r}_d$, we hence have to solve the following optimization problem:

$$\min_{\underline{F}} (\sigma/M)^2$$

s.t. 
$$\frac{1}{\sqrt{F_i^2 + F_i \beta \cos\theta_i + (\beta/2)^2}} = \delta\theta \sum_{j=0}^{N-1}\sum_{k=0}^{M-1} g_{ijk}, \ \forall i \in \left[0, \frac{N}{2}\right] \quad (D.13)$$

$$\delta\theta \sum_{j=0}^{N-1}\sum_{k=0}^{M-1} F_{jk}^3 \sin\theta_{jk} = 2$$

For this task, we used the NLPSolve function contained in the Optimization Package of the Maple code [31], whose arguments are:

NLPSolve(*OBJ*, *CONSTR*, *VAR*, initialpoint={*CI*})

For our calculation, we took for objective *OBJ* to be minimized, for constraints *CONSTR* and for variables *VAR* of the optimization procedure:

$$OBJ = (\sigma/M)^2$$

$$CONSTR = \begin{cases} \delta\theta\sum_{j=0}^{N-1}\sum_{k=0}^{M-1} F_{jk}^3 \sin\theta_{jk} = 2, \\ u_i u_{N-i} = \alpha \text{ for } i \in [0, N/2], \\ \dfrac{1}{\sqrt{G_i^2 + G_i\beta\cos\theta_i + (\beta/2)^2}} = \delta\theta\sum_{j=0}^{N-1}\sum_{k=0}^{M-1} g_{ijk} \text{ for } i \in [0, N] \end{cases} \quad (D.14)$$

$$VAR = \{F_{i=0...N/2}, u_{i=0...N}, \alpha\}$$

In the *CONSTR* vector, we introduced some redundancy to better constrain the calculations. Then we chose the ranges and the list *CI* of initial values of the variables *VAR* in the NLPSolve command. The calculations were carried out with $N = 20$ and $M = 10$. The results obtained are shown in fig. 1.

## Section E. Optimal surface $S_{opt}$ for a monoenergetic Boltzmann transport

By using the Lagrangian derivation method presented in section B.2, it is possible to derive the equation (2.1.5) of the surface $S_{opt}$ directly from the Boltzmann equation. Indeed, for neutrons that propagate in a heavy low-absorbing material, undertaking collisions that are mostly elastic and isotropic in the laboratory frame while keeping an energy constant throughout their transport, the Boltzmann equation in the problem (2.1) can be rewritten [4], parametrizing the atomic density $n(\underline{r})$ using the last equation of (2.1), $n(\underline{r}) = n_0 \Theta[S(\underline{r}) \geq 0]$:

$$B\varphi(\underline{r},\underline{\Omega}) = Q(\underline{r},\underline{\Omega}) = \underline{\Omega}.\underline{\nabla}\varphi(\underline{r},\underline{\Omega}) + \Sigma_s \Theta[S(\underline{r}) \geq 0]\left(\varphi(\underline{r},\underline{\Omega}) - \frac{1}{4\pi}\int_{\underline{\Omega}' \in 4\pi}\varphi(\underline{r},\underline{\Omega}')d\underline{\Omega}'\right) \quad (E.1)$$

with $\Sigma_s = n_0 \sigma_s$. For an isotropic neutron source positioned at the point $\underline{r}_s$, we have:

$$Q(\underline{r},\underline{\Omega}) = \frac{Q_0}{4\pi}\delta(\underline{r} - \underline{r}_s) \quad (E.2)$$

Let us define the scalar product, $<u,v>$, of two functions $u(\underline{r},\underline{\Omega})$ and $v(\underline{r},\underline{\Omega})$ as follows:

$$\langle u,v \rangle_{\underline{r}\underline{\Omega}} = \int_{\underline{\Omega} \in 4\pi}\int_{\underline{r} \in \Re^3}u(\underline{r},\underline{\Omega})^* v(\underline{r},\underline{\Omega})d\underline{r}d\underline{\Omega} \quad (E.3)$$

With this notation, the total neutron flux $\phi(\underline{r}_s \to \underline{r}_d, S)$ generated at $\underline{r}_d$ by a source placed at $\underline{r}_s$ can be written:

$$\phi(\underline{r}_s \to \underline{r}_d) = \int_{\underline{\Omega} \in 4\pi}\varphi(\underline{r}_s \to \underline{r}_d, \underline{\Omega})d\underline{\Omega} = \langle \delta(\underline{r} - \underline{r}_d), \varphi \rangle \quad (E.4)$$

As done in section B.2, let us add to the Lagrangian (C.1) of the problem (2.1) the Boltzmann equation as a constraint. We obtain:

$$L(S,\lambda) = \langle \delta(\underline{r} - \underline{r}_d), \varphi \rangle - \lambda\left(\int_{\underline{r} \in \Re^3}\Theta[S(\underline{r}) \geq 0]d\underline{r} - V_{max}\right) - \langle \omega, B\varphi - Q \rangle \quad (E.5)$$

where $\omega(\underline{r},\underline{\Omega})$ is an unspecified weight function.

As done in section C.2, let us now calculate the perturbation $dL$ of the Lagrangian induced by a perturbation $dS$ of the optimal surface $S_{opt}$. Following the calculation steps of section B.2 and using the result (C.1.14), we find:

$$dL = \langle \delta(\underline{r}-\underline{r}_d) - B^+\omega, d\varphi \rangle - \langle \omega, dB\varphi \rangle - \lambda\int_{\underline{r} \in \Re^3}dS(\underline{r})\delta(S_{opt}(\underline{r}))d\underline{r} \quad (E.6)$$

---

[4] See [27] pp 99. The Boltzmann equation (E.1) is obtained by eliminating the variable $E$ in (0.2) as the transport is monoenergetic, by taking $\Sigma_s(\underline{\Omega}' \to \underline{\Omega}) = \Sigma_s/4\pi$ as the material is heavy so the elastic collisions can be considered isotropic in the laboratory frame, and by taking $\Sigma_t = \Sigma_s$ as the material is assumed low-absorbing.

where the adjoint operator B$^+$ and the perturbation $d\mathrm{B}$ induced by the perturbation $dS$ are given by:

$$\mathrm{B}^+\omega(\underline{r},\underline{\Omega}) = -\underline{\Omega}.\underline{\nabla}\omega(\underline{r},\underline{\Omega}) + \Sigma_s \Theta[S_{opt}(\underline{r}) \geq 0]\left(\omega(\underline{r},\underline{\Omega}) - \frac{1}{4\pi}\int_{\underline{\Omega}'\in 4\pi}\omega(\underline{r},\underline{\Omega}')d\underline{\Omega}'\right)$$

$$d\mathrm{B}\varphi(\underline{r}_s \to \underline{r},\underline{\Omega}) = \Sigma_s dS(\underline{r})\delta(S_{opt}(\underline{r}))\left(\varphi(\underline{r}_s \to \underline{r},\underline{\Omega}) - \frac{1}{4\pi}\phi(\underline{r}_s \to \underline{r})\right)$$
(E.7)

By choosing the weight function $\omega(\underline{r},\underline{\Omega})$ such that:

$$\mathrm{B}^+\omega(\underline{r},\underline{\Omega}) = \delta(\underline{r}-\underline{r}_d) \quad (\text{E.8})$$

we obtain:

$$dL = -\langle \omega, d\mathrm{B}\varphi \rangle - \lambda \int_{\underline{r}\in\Re^3} dS(\underline{r})\delta(S_{opt}(\underline{r}))d\underline{r}$$

$$= -\int_{\underline{r}\in\Re^3} dS(\underline{r})\delta(S_{opt}(\underline{r}))\left[\Sigma_s \int_{\underline{\Omega}\in 4\pi} \omega(\underline{r},\underline{\Omega})^*\left(\varphi(\underline{r}_s \to \underline{r},\underline{\Omega}) - \frac{1}{4\pi}\phi(\underline{r}_s \to \underline{r})\right)d\underline{\Omega} + \lambda\right]d\underline{r}$$
(E.9)

But the optimal surface $S_{opt}$ sought is given by, cf. (C.2):

$$dL = \frac{\partial L}{\partial S}dS = 0, \forall dS(\underline{r}) \quad (\text{E.10})$$

The equation of the optimal surface $S_{opt}$ is therefore:

$$-\frac{\lambda}{\Sigma_s} = \int_{\underline{\Omega}\in 4\pi} \omega^*(\underline{r},\underline{\Omega})\left(\varphi(\underline{r}_s \to \underline{r},\underline{\Omega}) - \frac{1}{4\pi}\phi(\underline{r}_s \to \underline{r})\right)d\underline{\Omega}, \forall \underline{r}/S_{opt}(\underline{r}) = 0 \text{ i.e. } \forall \underline{r} \in S_{opt}$$
(E.11)

Now let us expand the flux $\varphi(\underline{r},\underline{\Omega})$ on the basis of spherical harmonics [32]:

$$\varphi(\underline{r}_s \to \underline{r},\underline{\Omega}) = \sum_{l=0}^{+\infty}\sum_{m=-l}^{l}\varphi_{lm}(\underline{r}_s \to \underline{r})Y_{lm}(\underline{\Omega}) \quad (\text{E.12})$$

The harmonics $Y_{lm}(\underline{\Omega})$ forming an orthonormal basis of functions, one has [32]:

$$\int_{\underline{\Omega}\in 4\pi} Y^*_{l'm'}(\underline{\Omega})Y_{lm}(\underline{\Omega})d\underline{\Omega} = \delta_{mm'}\delta_{ll'}, \quad \delta_{nn'} = \begin{cases} 1 \text{ if } n = n' \\ 0 \text{ else} \end{cases} \quad (\text{E.13})$$

Since the propagating medium is made of heavy nuclei, we can use the P1 approximation, see [27] pp 99-104, which consists in stopping the expansion (E.12) at the order $l = 1$. In this case, the flux can be written more simply:

$$\varphi(\underline{r}_s \to \underline{r}, \underline{\Omega}) \approx \sum_{l=0}^{1} \sum_{m=-l}^{l} \varphi_{lm}(\underline{r}) Y_{lm}(\underline{\Omega}) \approx F(\underline{r}) + \underline{\Omega} \cdot \underline{G}(\underline{r}) \quad \text{(E.14)}$$

We can give a physical meaning to the functions $F(\underline{r})$ and $\underline{G}(\underline{r})$ by integrating the expression (E.14), then the neutron current $\underline{j}(\underline{r},\underline{\Omega}) = \underline{\Omega}\varphi(\underline{r},\underline{\Omega})$, over $\underline{\Omega}$. We obtain:

$$\underline{\Omega} = (\sin\theta\cos\varphi, \sin\theta\sin\varphi, \cos\theta), \quad d\underline{\Omega} = \sin\theta d\theta d\varphi, \quad \int_{\underline{\Omega}\in 4\pi} = \int_{\theta=0}^{\pi}\int_{\varphi=0}^{2\pi}$$

$$\Rightarrow \phi(\underline{r}_s \to \underline{r}) = \int_{\underline{\Omega}\in 4\pi} \varphi(\underline{r}_s \to \underline{r}, \underline{\Omega}) d\underline{\Omega} = 4\pi F(\underline{r}) \quad \text{(E.15)}$$

$$\Rightarrow \underline{J}(\underline{r}_s \to \underline{r}) = \int_{\underline{\Omega}\in 4\pi} \underline{j}(\underline{r}_s \to \underline{r}, \underline{\Omega}) d\underline{\Omega} = \frac{4\pi}{3} \underline{G}(\underline{r})$$

So we observe that:

$$\varphi(\underline{r}_s \to \underline{r}, \underline{\Omega}) \approx \frac{1}{4\pi}\left[\phi(\underline{r}_s \to \underline{r}) + 3\underline{\Omega}\cdot\underline{J}(\underline{r}_s \to \underline{r})\right] \quad \text{(E.16)}$$

where $\underline{J}(\underline{r}_s \to \underline{r})$ is the neutron current at the point $\underline{r}$ generated by the source placed at $\underline{r}_s$.
By making the change of variables, $\underline{\Omega}$ in $-\underline{\Omega}$, in (E.7)-(E.8), we then note that:

$$B\omega(\underline{r}, -\underline{\Omega}) = \delta(\underline{r} - \underline{r}_d) \quad \text{(E.17)}$$

By comparing this equation with the equation (E.1)-(E.2) of the flux $\varphi(\underline{r}_s \to \underline{r}, \underline{\Omega})$, we note that the function $\omega(\underline{r}, -\underline{\Omega})$ is the neutron flux generated at the point $\underline{r}$ by an isotropic point source of intensity $4\pi/Q_0$ placed at $\underline{r}_d$. We thus have:

$$\omega(\underline{r},\underline{\Omega}) = \frac{4\pi}{Q_0}\varphi(\underline{r}_d \to \underline{r}, -\underline{\Omega}) \approx \frac{1}{Q_0}\left[\phi(\underline{r}_d \to \underline{r}) - 3\underline{\Omega}\cdot\underline{J}(\underline{r}_d \to \underline{r})\right] \quad \text{(E.18)}$$

Finally, by reinjecting the results (E.16) and (E.18) in the equation (E.11) of $S_{opt}$, we obtain:

$$-\frac{\lambda}{\Sigma_s} \approx \frac{3}{4\pi}\underline{J}(\underline{r}_s \to \underline{r}) \cdot \int_{\underline{\Omega}\in 4\pi} \underline{\Omega}\omega(\underline{r},\underline{\Omega})d\underline{\Omega}, \forall \underline{r} \in S_{opt}$$

$$\Rightarrow \frac{\lambda Q_0}{3\Sigma_s} \approx \underline{J}(\underline{r}_s \to \underline{r}) \cdot \underline{J}(\underline{r}_d \to \underline{r}), \forall \underline{r} \in S_{opt} \quad \text{(E.19)}$$

We retrieve the equation (2.1.5) of the surface $S_{opt}$, previously obtained in the more restrictive framework of a diffusive transport, as well as the correct value of the diffusion coefficient, $D = 1/(3\Sigma_s)$, for a heavy low-absorbing material.
We can even do a little better. By taking the result (E.18) then expanding in it the angular flux $\varphi(\underline{r}_d \to \underline{r}, \underline{\Omega})$ on the basis of spherical harmonics, cf. (E.12), without stopping at the order $l = 1$, we find, observing that $Y_{lm}(-\underline{\Omega}) = (-1)^l Y_{lm}(\underline{\Omega})$:

$$\omega(\underline{r},\underline{\Omega}) = \frac{4\pi}{Q_0}\varphi(\underline{r}_d \to \underline{r},-\underline{\Omega}) = \frac{4\pi}{Q_0}\sum_{l'=0}^{+\infty}\sum_{m'=-l'}^{l'}(-1)^{l'}\varphi_{l'm'}(\underline{r}_d \to \underline{r})Y_{l'm'}(\underline{\Omega}) \quad (E.20)$$

By injecting this expansion as well as the expansion (E.12) of $\varphi(\underline{r},\underline{\Omega})$ in the equation (E.11) of $S_{opt}$, we then find:

$$-\frac{\lambda Q_0}{\Sigma_s 4\pi} = \sum_{l=0}^{+\infty}\sum_{m=-l}^{l}\sum_{l'=0}^{+\infty}\sum_{m'=-l'}^{l'} (-1)^{l'}\varphi_{lm}(\underline{r}_s \to \underline{r})\varphi_{l'm'}(\underline{r}_d \to \underline{r}) \times \left( \int_{\underline{\Omega}\in 4\pi} Y^*_{l'm'}(\underline{\Omega})Y_{lm}(\underline{\Omega})d\underline{\Omega} - \frac{1}{4\pi}\int_{\underline{\Omega}\in 4\pi} Y^*_{l'm'}(\underline{\Omega})d\underline{\Omega} \int_{\underline{\Omega}\in 4\pi} Y_{lm}(\underline{\Omega})d\underline{\Omega}\right), \quad (E.21)$$

$\forall \underline{r} \in S_{opt}$

Since $Y_{00}(\underline{\Omega}) = 1/(4\pi)^{1/2}$, using the orthogonality (E.13) of the spherical harmonics, we have:

$$\int_{\underline{\Omega}\in 4\pi} Y_{lm}(\underline{\Omega})d\underline{\Omega} = \sqrt{4\pi}\delta_{l0}\delta_{m0}, \quad \int_{\underline{\Omega}\in 4\pi} Y^*_{l'm'}(\underline{\Omega})Y_{lm}(\underline{\Omega})d\underline{\Omega} = \delta_{ll'}\delta_{mm'} \quad (E.22)$$

Injecting these results in (E.21), we obtain a more precise equation of the optimal surface $S_{opt}$, valid in the framework of the monoenergetic Boltzmann equation, for a heavy, low-absorbing, material, in which the neutrons-nuclei collisions are assumed to be mostly elastic and isotropic:

$$-\frac{\lambda Q_0}{4\pi \Sigma_s} = \sum_{l=1}^{+\infty}\sum_{m=-l}^{l}(-1)^l \varphi_{lm}(\underline{r}_s \to \underline{r})\varphi_{lm}(\underline{r}_d \to \underline{r}), \forall \underline{r} \in S_{opt} \quad (E.23)$$

## Section F. General equation of the optimal surface $S_{opt}$

Using a method similar to that developed in section C, we can formulate a general equation of the optimal surface $S_{opt}$, valid in the framework of the Boltzmann equation, with no restriction on the physicochemical properties of the propagating medium nor on the neutron energy, for an objective functional $O$ that is linear. To do this, take a Green's function, $G(\underline{r},E,\underline{\Omega};\underline{r}',E',\underline{\Omega}';S)$, of the Boltzmann equation, $B\varphi = Q$. This function is by definition solution of:

$$B_{\underline{r}E\underline{\Omega}}(S)G(\underline{r},E,\underline{\Omega};\underline{r}',E',\underline{\Omega}';S) = \delta(\underline{r}-\underline{r}')\delta(E-E')\delta(\underline{\Omega}-\underline{\Omega}') \quad (F.1)$$

We thereby directly have:

$$\varphi(\underline{r},E,\underline{\Omega},S) = \int_{\underline{r}'\in\Re^3}\int_{\underline{\Omega}'\in 4\pi}\int_{E'=0}^{+\infty} Q(\underline{r}',E',\underline{\Omega}')G(\underline{r},E,\underline{\Omega};\underline{r}',E',\underline{\Omega}';S)d\underline{r}'dE'd\underline{\Omega}' \quad (F.2)$$

The optimal surface $S_{opt}$ obeys the equations (C.2), with a Lagrangian $L$ given by:

$$L(S,\lambda) = O\varphi - \lambda\left(\int_{\underline{r}\in\Re^3}\Theta[S(\underline{r})\geq 0]d\underline{r} - V_{max}\right) \quad (F.3)$$

To calculate the derivatives (C.2) of *L*, we can use the same calculation steps as those described in section C.2. Let us take the optimal surface $S_{opt}(\underline{r})$ and perturb it slightly. We obtain a new surface, $S(\underline{r}) = S_{opt}(\underline{r}) + dS(\underline{r})$. This perturbation $dS$ of the surface induced a perturbation $dn$ of the density $n(\underline{r}) = n_0\Theta[S(\underline{r}) \geq 0]$, given by, cf. (C.1.14):

$$n(\underline{r}) = n_{opt}(\underline{r}) + dn(\underline{r}) = n_0\Theta[S_{opt}(\underline{r}) \geq 0] + n_0 dS(\underline{r})\delta(S_{opt}(\underline{r})) \quad (F.4)$$

This perturbation of $n(\underline{r})$ modifies the operator B by a quantity $dB$, and the flux $\varphi$ by a quantity $d\varphi$. We therefore have, neglecting the second-order term:

$$\begin{aligned}&\left(B_{\underline{r}E\underline{\Omega}}(S_{opt}) + dB_{\underline{r}E\underline{\Omega}}\right)(\varphi + d\varphi) \\&= B_{\underline{r}E\underline{\Omega}}(S_{opt})\varphi(\underline{r},E,\underline{\Omega},S_{opt}) + B_{\underline{r}E\underline{\Omega}}(S_{opt})d\varphi(\underline{r},E,\underline{\Omega}) + dB_{\underline{r}E\underline{\Omega}}\varphi(\underline{r},E,\underline{\Omega},S_{opt}) \quad (F.5)\\&= Q(\underline{r},E,\underline{\Omega})\end{aligned}$$

with:

$$dB_{\underline{r}E\underline{\Omega}}\varphi(\underline{r},E,\underline{\Omega},S_{opt}) = n_0 dS(\underline{r})\delta(S_{opt}(\underline{r}))\left[\begin{array}{l}\sigma_t(E)\varphi(\underline{r},E,\underline{\Omega},S_{opt}) \\ -\int\limits_{E''=0}^{+\infty}\int\limits_{\underline{\Omega}''\in 4\pi}\sigma_s(E'')f(E''\to E,\underline{\Omega}''\to\underline{\Omega})\varphi(\underline{r},E'',\underline{\Omega}'',S_{opt})dE''d\underline{\Omega}''\end{array}\right] \quad (F.6)$$

Since $B\varphi = Q$, we obtain:

$$B_{\underline{r}E\underline{\Omega}}(S_{opt})d\varphi(\underline{r},E,\underline{\Omega}) = -dB_{\underline{r}E\underline{\Omega}}\varphi(\underline{r},E,\underline{\Omega},S_{opt}) \quad (F.7)$$

Using the definition of the Green's function, this equation has the following solution:

$$d\varphi(\underline{r},E,\underline{\Omega}) = -\int\limits_{\underline{r}'\in\Re^3}\int\limits_{\underline{\Omega}'\in 4\pi}\int\limits_{E'=0}^{+\infty} G(\underline{r},E,\underline{\Omega};\underline{r}',E',\underline{\Omega}';S_{opt})dB_{\underline{r}'E'\underline{\Omega}'}\varphi(\underline{r}',E',\underline{\Omega}',S_{opt})d\underline{r}'dE'd\underline{\Omega}' \quad (F.8)$$

For a linear functional *O*, the perturbation $dO\varphi$ of the objective $O\varphi$ induced by the perturbation of the surface is hence given by:

$$\begin{aligned}dO\varphi &= O_{\underline{r}E\underline{\Omega}}d\varphi(\underline{r},E,\underline{\Omega},S) \\&= -\int\limits_{\underline{r}'\in\Re^3}\int\limits_{\underline{\Omega}'\in 4\pi}\int\limits_{E'=0}^{+\infty} O_{\underline{r}E\underline{\Omega}}G(\underline{r},E,\underline{\Omega};\underline{r}',E',\underline{\Omega}';S_{opt})dB_{\underline{r}'E'\underline{\Omega}'}\varphi(\underline{r}',E',\underline{\Omega}',S_{opt})d\underline{r}'dE'd\underline{\Omega}' \quad (F.9)\\&= -\int\limits_{\underline{r}'\in\Re^3} dS(\underline{r}')n_0\delta(S_{opt}(\underline{r}'))\int\limits_{\underline{\Omega}'\in 4\pi}\int\limits_{E'=0}^{+\infty}\left[\begin{array}{l}O_{\underline{r}E\underline{\Omega}}G(\underline{r},E,\underline{\Omega};\underline{r}',E',\underline{\Omega}';S_{opt})\times \\ \sigma_t(E')\varphi(\underline{r}',E',\underline{\Omega}',S_{opt}) \\ -\int\limits_{E''=0}^{+\infty}\int\limits_{\underline{\Omega}''\in 4\pi}\sigma_s(E'')f(E''\to E',\underline{\Omega}''\to\underline{\Omega}') \\ \times\varphi(\underline{r}',E'',\underline{\Omega}'',S_{opt})dE''d\underline{\Omega}''\end{array}\right]d\underline{r}'dE'd\underline{\Omega}'\end{aligned}$$

The perturbation $dL$ of the Lagrangian (F.3) induced by the perturbation $dS$ of the surface is thus, reusing the result (C.1.14):

$$dL = \frac{\partial L}{\partial S} dS = dO\varphi - \lambda \int_{\underline{r}' \in \mathfrak{R}^3} dS(\underline{r}')\delta(S_{opt}(\underline{r}')) d\underline{r}' \quad \text{(F.10)}$$

According to (C.2), we must have $dL = 0 \; \forall dS$, which finally imposes that:

$$\lambda = -n_0 \int_{\underline{\Omega}' \in 4\pi} \int_{E'=0}^{+\infty} \begin{bmatrix} O_{\underline{r}E\underline{\Omega}} G(\underline{r}, E, \underline{\Omega}; \underline{r}', E', \underline{\Omega}', S_{opt}) \times \\ \sigma_t(E')\varphi(\underline{r}', E', \underline{\Omega}', S_{opt}) \\ - \int_{E''=0}^{+\infty} \int_{\underline{\Omega}'' \in 4\pi} \sigma_s(E')f(E'' \to E', \underline{\Omega}'' \to \underline{\Omega}') \\ \times \varphi(\underline{r}', E', \underline{\Omega}', S_{opt}) dE'' d\underline{\Omega}'' \end{bmatrix} dE' d\underline{\Omega}', \quad \text{(F.11)}$$

$\forall \underline{r}'/S_{opt}(\underline{r}') = 0$, i.e. $\forall \underline{r}' \in S_{opt}$

We obtain here a general equation of the surface $S_{opt}$, valid in the framework of the Boltzmann equation, with no restriction on the properties of the propagator or on the neutron energies. Let us hence extract as much information as possible from this calculation method. Let us wonder what result we would have obtained if, instead of optimizing the surface $S(\underline{r})$ of the material, we had optimized its volume density $\rho(\underline{r})$. If we optimize the density $\rho(\underline{r})$, the Lagrangian (F.3) and the derivatives (C.2) become, cf. (1.2):

$$L(\rho, \lambda') = O\varphi - \lambda' \left( \int_{\underline{r} \in \mathfrak{R}^3} \rho(\underline{r}) d\underline{r} - P_{max} \right)$$

$$\left. \frac{\partial L}{\partial \rho} \right|_{\rho=\rho_{opt}} = 0, \quad \left. \frac{\partial L}{\partial \lambda'} \right|_{\rho=\rho_{opt}} = 0 \quad \text{(F.12)}$$

The calculation of the derivative $\partial L/\partial \rho$ then follows the same calculation steps, (F.4) to (F.11), as before. As the atomic density $n(\underline{r})$ of the material is proportional to its volume density $\rho(\underline{r})$, we directly take $n(\underline{r}) = n_{opt}(\underline{r}) + dn(\underline{r})$ in (F.4). We have then to replace $n_0 dS(\underline{r})\delta(S_{opt}(\underline{r}))$ by $dn(\underline{r})$ in equations (F.6), (F.9) and (F.10). Finally, taking (F.10) and noticing that $dL = (\partial L/\partial n)dn = (\partial L/\partial \rho)d\rho$ must be equal to 0 $\forall dn$, we obtain the equation of the optimal density, $\rho_{opt}(\underline{r})$, which is written:

$$\lambda' = - \int_{\underline{\Omega}' \in 4\pi} \int_{E'=0}^{+\infty} \begin{bmatrix} O_{\underline{r}E\underline{\Omega}} G(\underline{r}, E, \underline{\Omega}; \underline{r}', E', \underline{\Omega}', \rho_{opt}) dE' d\underline{\Omega}' \times \\ \sigma_t(E')\varphi(\underline{r}', E', \underline{\Omega}', \rho_{opt}) \\ - \int_{E''=0}^{+\infty} \int_{\underline{\Omega}'' \in 4\pi} \sigma_s(E')f(E'' \to E', \underline{\Omega}'' \to \underline{\Omega}') \\ \times \varphi(\underline{r}', E', \underline{\Omega}', \rho_{opt}) dE'' d\underline{\Omega}'' \end{bmatrix}, \quad \forall \underline{r}' \in \mathfrak{R}^3 \quad \text{(F.13)}$$

This equation is identical to that of the surface $S_{opt}$, (F.11), with the difference that it is valid now inside the material, instead of being valid at its surface. A simpler but not rigorous way to obtain this result is to rewrite the optimization conditions for the density $\rho(\underline{r})$ and for the surface $S(\underline{r})$. If one wishes to optimize the density $\rho(\underline{r})$, one must solve:

$$dL = \frac{\partial L}{\partial \rho} d\rho = 0, \forall d\rho \quad \text{(F.14)}$$

The optimal density is hence given by:

$$\frac{\partial L}{\partial \rho} = 0, \forall \underline{r} \in \Re^3 \quad \text{(F.15)}$$

If one wishes to optimize the shape, $S(\underline{r})$, of the material, one must solve:

$$dL = \frac{\partial L}{\partial \rho} d\rho = \frac{\partial L}{\partial \rho} \rho_0 dS \delta(S_{opt}(\underline{r})) = 0, \forall dS \quad \text{(F.16)}$$

The optimal surface is hence given by:

$$\frac{\partial L}{\partial \rho} = 0, \forall \underline{r} / S_{opt}(\underline{r}) = 0 \text{ i.e. } \forall \underline{r} \in S_{opt} \quad \text{(F.17)}$$

These results are simple but they suggest an efficient way to implement a topology optimization procedure.

## Section G. Construction of the topology optimization A1

In this section, we describe the reasoning steps that have led to the formulation of the topology optimization algorithm A1, used throughout the study.

G.1. First attempt of a topology optimization algorithm

In sections C and D, we have analytically solved the problem (2.1.4), which is a special subcase of (2.1) for a monoenergetic diffusive neutron transport in a non-absorbing material. Let us now go back to the general problem (2.1). We want to solve it with little or no approximations, using tools that are already available and widespread in the community. To achieve this double goal, we propose to use the results obtained at the end of the section F, which suggest that, in the framework of the linear Boltzmann equation, the surface $S_{opt}$ should be the interface between two regions of space, $\partial L/\partial \rho > 0$ and $\partial L/\partial \rho < 0$. If one uses the parametrization (1.6) of the density $\rho(\underline{r})$, tiling the available space with a union of cells $\Theta_i$ of volumes $V_i$ and of densities $\rho_i$, the equations of these regions can be rewritten, cf. (1.10):

$$\frac{\partial L}{\partial \rho} > 0 \text{ (resp} < 0) \Leftrightarrow C_i = \frac{1}{V_i} \frac{\partial O\varphi}{\partial \rho_i} > \lambda \text{ (resp} < \lambda) \quad \text{(G.1)}$$

In (G.1), the derivatives $\partial O\varphi/\partial \rho_i$ can be computed using the method described in section 1. To complete the resolution of the general optimization problem (2.1), all that remains is hence to position the available matter on the correct side of the interface $S_{opt}$, in one of the two regions $C_i > \lambda$ or $C_i < \lambda$. To determine this side, suppose one wishes to maximize the objective $O\varphi$. To achieve this, the available matter must be positioned where its addition increases $O\varphi$, i.e. where $\partial O\varphi/\partial \rho_i > 0$. More, since there is a volume constraint, the available matter should be positioned first in the cells where $\partial O\varphi/\partial \rho_i$ is as large as possible. Consequently, if one wishes to maximize $O\varphi$, the material should be placed in the area $C_i > \lambda$, by choosing the threshold $\lambda$ so that the total volume $V$ of matter positioned be equal to $V_{max}$. Conversely, if one wishes to minimize $O\varphi$, the material should be placed in the area $C_i < \lambda$, choosing anew $\lambda$ so that $V = V_{max}$.

From these simple considerations, we can write a first attempt of an algorithm capable to solve the topology optimization problem (2.1). Starting from an arbitrary vector of initial densities, $\underline{\rho}(0) = (\rho_1(0), \ldots, \rho_N(0))$, where $N$ is the number of cells $\Theta_i$ usable in the optimization procedure, we proposed the following iterative procedure:

Algorithm A0
For $n$ from 1 to $n_{max}$
1) Write the MCNP input by taking for volume densities $\rho_i$ of the cells $\Theta_i$ the densities $\rho_i(n-1)$ obtained at the previous iteration of the algorithm. Add the PERT$i$ cards described section 1. Modify the seed used for the random draws.
2) Run MCNP, read its output file, extract the derivatives $\partial O\varphi/\partial \rho_i$, calculate the coefficients $C_i$ given in (G.1).
3) Create $N$ vectors $\underline{F}_i = (C_i, V_i, \rho_i(n-1))$, one per cell $\Theta_i$ usable in the optimization procedure. Reorder them in descending (resp. increasing) order of $C_i$ if one wants to maximize (resp. minimize) $O\varphi$. Renumber the vectors $\underline{F}_j$ thus classified from $j = 1$ to $j = N$.
4) Procedure for calculating the new densities $\rho_j(n)$:
$V = 0$. For $j$ from 1 to $N$
    $V = V + V_j$ (volume of cell $\Theta_j$)
    If one wants to maximize $O\varphi$: if ($V \leq V_{max}$ & $C_j > 0$) then $\rho_j(n) = \rho_{max}$ else $\rho_j(n) = \rho_{min}$
    If one wants to minimize $O\varphi$: if ($V \leq V_{max}$ & $C_j < 0$) then $\rho_j(n) = \rho_{max}$ else $\rho_j(n) = \rho_{min}$

In A0, the predecessor of A1, the positioning of the material was therefore binary: each cell $\Theta_i$ contained either matter at density $\rho_i = \rho_{max}$ (natural density of the material), or matter at density $\rho_i = \rho_{min} \ll \rho_{max}$ to mimic an empty cell while avoiding crashing MCNP, cf. discussion of equation (1.12) in section 1.

For testing A0, we used the tiling of the available space described in (2.2.3), with $NX = 20$ and $NR = 10$, which gives a fairly coarse pixelisation, cf. fig. G.1-G.2. Indeed, for this first test, the priority was not to make a precise calculation of the shape of the propagator, rather to test the feasibility of a topology optimization algorithm in the framework of the linear Boltzmann equation. We have thus aimed to reduce as much as possible the computing time consumed to process the $(NX+1)(NR+1)-2$ PERT cards of the MCNP calculation, in order to increase as much as possible the number $NPS$ of sampled source neutrons, thereby of simulated trajectories. The objective was to minimize the statistical noise on the $C_i$ values, susceptible to bias the shape of the calculated optimal structure if it is too strong. A bias would indeed have reduced the reach of this feasibility test. In this section, we took $NPS = 10^9$, and we mobilized for this task all our computing resources, cutting the MCNP input into 7 separated inputs, each containing one-seventh of the PERT cards, and we launched each of them on a 12-16 CPU machine. As discussed in section 2.2, we took a material made of 100% of $^{207}$Pb atoms, with a density $\rho_{max} = 11.34$ g/cm$^3$, a distance between the source and the detector $H = 20$ cm, a source energy $E_0 = kT = 2.53\ 10^{-8}$ MeV, and a maximal volume $V_{max} = 10^4$ cm$^3$. We took $\rho_{min} = \rho_{max}/100$ [5], and started the calculations from an uniform initial density configuration $\rho_i(0) = \rho_{max}\ \forall i$ (iteration $n = 0$ of A0 hence does not respect the constraint $V \leq V_{max}$ discussed section 2.2).

The results obtained with this procedure are shown in fig. G.1-G.3. In fig. G.1-G.2, the density profiles $\rho(\underline{r},n) = \Sigma\rho_i(n)\Theta_i(\underline{r})$ obtained with A0 at iterations $n = 0$ to 9 are plotted in a plane containing the symmetry axis SD. The corresponding 3D structures are obtained by revolving these profiles around the axis SD. In fig. G.1-G.2, the squares (which are cross-sections of the cylindrical rings $\Theta_i$) in black are cells in which $\rho_i = \rho_{max}$, i.e. cells filled with lead, while the squares in white are cells in which $\rho_i = \rho_{min}$, i.e. empty cells. The red squares are the source (left) and the detector (right) cells. The red line is the solution obtained section 2.1 equations (2.1.6)-(2.1.8) and fig. 1 in the approximation of a diffusive transport, for $H = 20.0$ cm and $V_{max} = 10^4$ cm$^3$, i.e. for $\beta = 1.496$. In fig. G.1, iteration 0 shows the initial configuration of densities used for the calculations, $\rho_i(0) = \rho_{max}\ \forall i$. A video allowing a better visualization of the sequence of iterations performed by A0, phimax_therm_A0.mp4, is given in Supplemental Material. The evolution of the total neutron flux, $\phi(n)$, in the detector cell, whose maximization is the goal of these calculations, is given in fig. G.3 on the left side as a function of the iteration number $n$. These fluxes $\phi(n)$ are compared to the flux $\phi_{diff}$ calculated with MCNP for the theoretical shape (2.1.6). The MCNP statistical errors on the fluxes $\phi(n)$ being lower than one percent, their error bars are not plotted. Finally, the right part of fig. G.3 gives the evolution of the weight $P(n)$ of the structure in kg with the iteration number $n$, given equation (2.2.4). The maximum permissible weight, $P_{max} = \rho_{max}V_{max} = 113.4$ kg is indicated by the red line.

---

[5] The calculations performed with A0 in this section were our first test of a topology optimization algorithm, so at that moment we did not know well to what level we could reduce $\rho_{min}$ to avoid crashing MCNP.

In Fig. G.1-G.2, we note that the shape of the neutron propagator obtained with the procedure A0 does converge towards the reference solution, in a few iterations. Given the strong non-linearity of the problem (2.1) and the simplicity of algorithm A0, this was an unexpected, motivating success [6]. However, although these results look promising, we note that they are not perfect. We observe indeed, during the first iterations of A0, in fig. G.1-G.2 or in the associated video, that the calculated shape violently oscillates around the theoretical surface, elongating alternatively in the direction $z$ then in the radial direction $y = \pm r$. These oscillations quickly dampen, but do not vanish completely. The structure ends stabilizing in a slightly suboptimal configuration, then oscillates from a slightly asymmetrical shape on the left to a slightly asymmetrical shape on the right, both symmetrical with respect to the plane $z = 0$ containing the middle of the segment SD. These oscillations of the structure remind the response of a mechanical system to a violent shock. It is hence not difficult to intuit that this shock is induced by the sudden transitions from $\rho_{min}$ to $\rho_{max}$ (and vice versa) in some of the cells. To avoid this shock, a simple solution is to vary the densities $\rho_i$ in small steps at each iteration of the algorithm. To do this, we impose a quantization of the densities $\rho_i$, linear or logarithmic, described in section 2.2. This idea led to the writing of algorithm A1, which was then applied with success to solve the various optimization problems addressed in sections 2-5.

---

[6] It is because of this initial success that we decided to investigate in deeper extends the capabilities of a topology optimization algorithm to solve the various problems addressed sections 2 to 5.

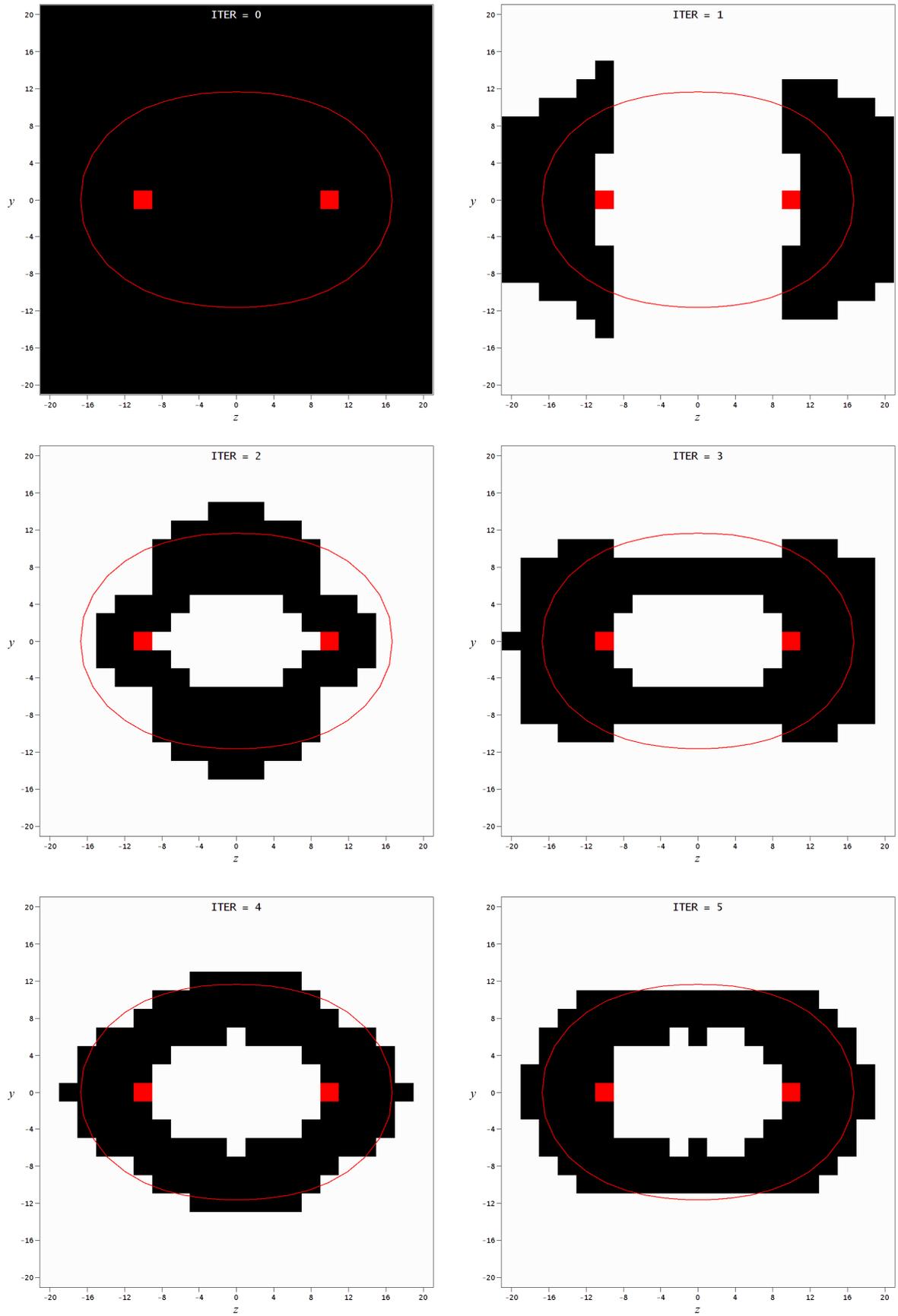

**Figure G.1.** Iterations 0 to 5 of algorithm A0. The units of the axes $y = \pm r$ and $z$ are cm.

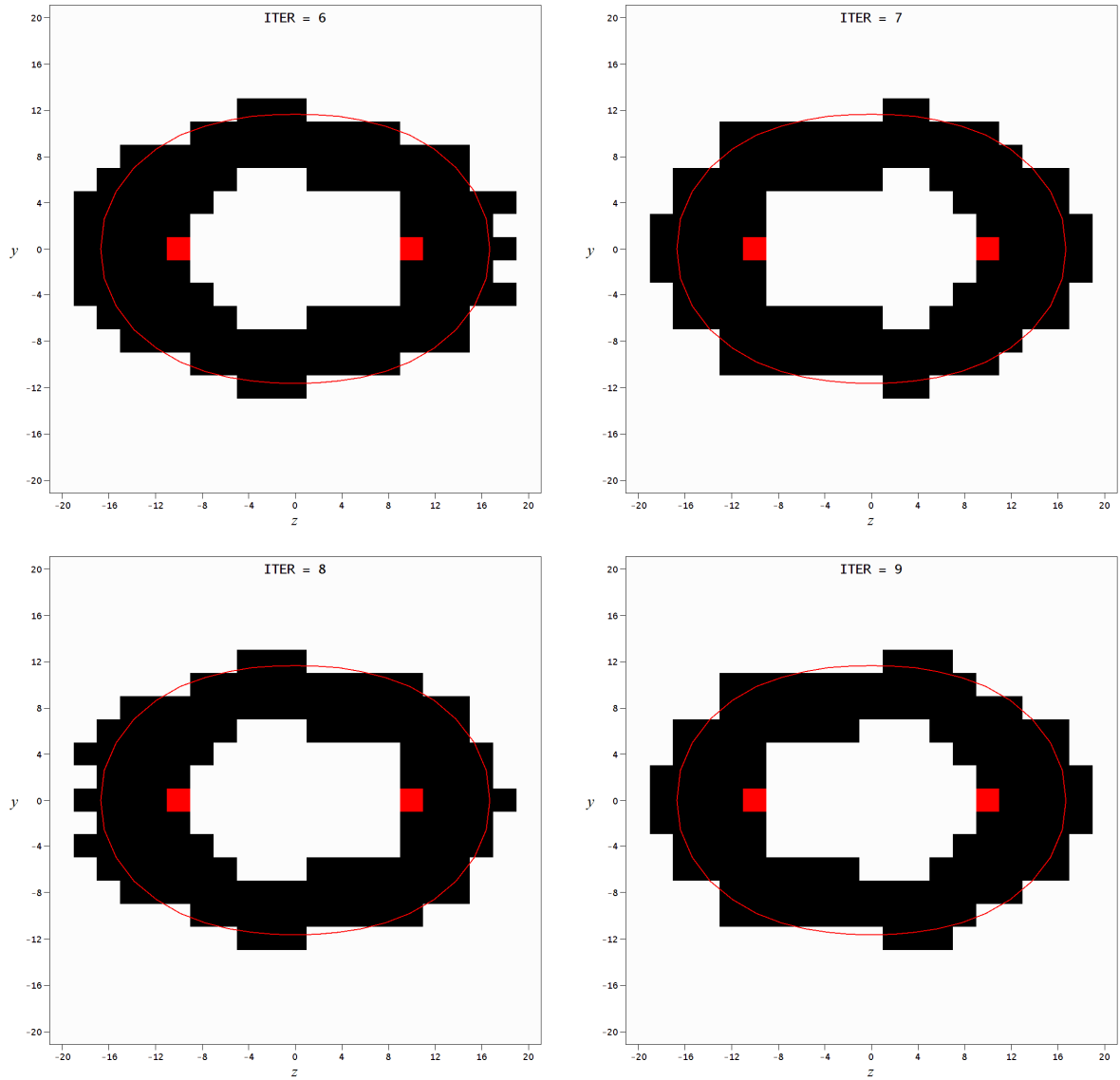

**Figure G.2.** Iterations 6 to 9 of algorithm A0. The units of the axes $y = \pm r$ and $z$ are cm.

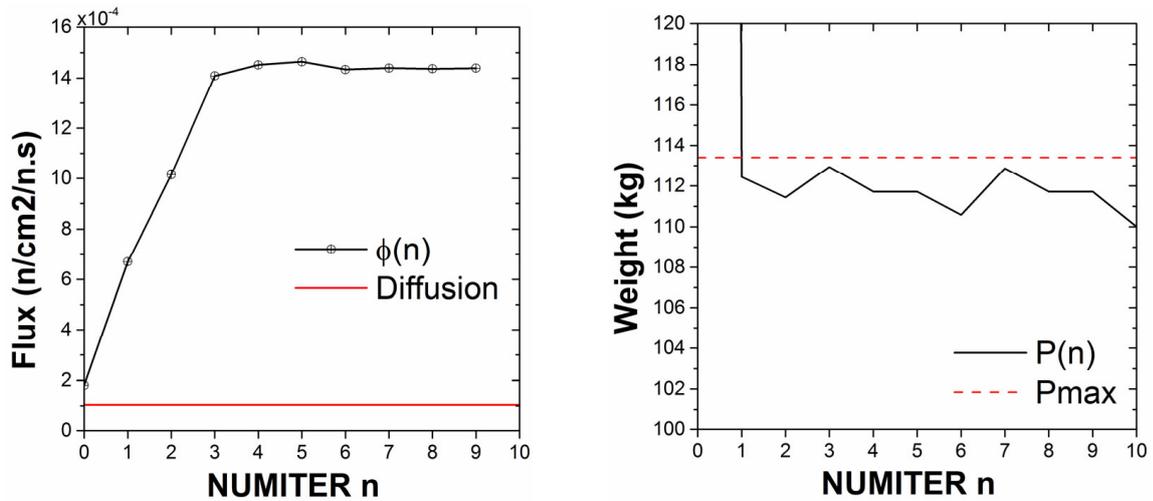

**Figure G.3.** (Left) evolution of the neutron flux, $\phi(n)$, in the detector cell with $n$ (line + circles), compared to the flux $\phi_{diff}$ calculated with MCNP for the shape obtained eq. (2.1.6)-(2.1.8) (red line); (right) evolution of the weight of the structure, $P(n)$, in kg with $n$. The maximum permissible weight, $P_{max}$, is indicated by the red dashed line.

G.2. Demonstration of formulae (2.2.5)-(2.2.6)

*Linear quantization.* Let us take a linear quantization of the density, (2.2.1). In this framework, the initial density $\rho(0)$ obeys to:

$$\rho(0) = \rho_{min} + k\delta\rho, \quad \delta\rho = (\rho_{max} - \rho_{min})/M \quad (G.2.1)$$

where $k$ is an integer. The weight constraint $P \leq P_{max}$ then imposes that:

$$P(0) = \sum_{\substack{i \neq 1+NX/4 \\ i \neq 1+3NX/4}} \rho_i(0)V_i = \rho(0)V_a \leq P_{max} \quad (G.2.2)$$

We seek the density $\rho(0)$ for which the weight of the initial configuration, $P(0)$, is as close as possible to $P_{max}$, while satisfying the constraint (G.2.2). This problem is equivalent to finding the largest integer $k_0$ satisfying the conditions (G.2.1) and (G.2.2). By injecting (G.2.1) in (G.2.2), we have:

$$k \leq \frac{1}{\delta\rho}\left(\frac{P_{max}}{V_a} - \rho_{min}\right) \quad (G.2.3)$$

Introducing the function $floor(x)$, which returns the largest integer smaller than $x$, we obtain:

$$k_0 = floor\left(\frac{1}{\delta\rho}\left(\frac{P_{max}}{V_a} - \rho_{min}\right)\right) \quad (G.2.4)$$

Finally, reinjecting (G.2.4) in (G.2.1), we obtain the formula (2.2.5).

*Logarithmic quantization.* Let us take a logarithmic quantization, (2.2.2), of the density. In this framework, the initial density $\rho(0)$ obeys to:

$$\rho(0) = \rho_{min}\varepsilon^k, \quad \varepsilon = (\rho_{max}/\rho_{min})^{1/M} \quad (G.2.5)$$

By injecting (G.2.5) in (G.2.2), we have:

$$k \leq \frac{1}{\ln(\varepsilon)}\ln\left(\frac{P_{max}}{V_a\rho_{min}}\right) \quad (G.2.6)$$

Using the function $floor(x)$, we then obtain:

$$k_0 = floor\left(\frac{1}{\ln(\varepsilon)}\ln\left(\frac{P_{max}}{V_a\rho_{min}}\right)\right) \quad (G.2.7)$$

Finally, reinjecting (G.2.7) in (G.2.5), we obtain the formula (2.2.6).

G.3. Test of a gradient-type procedure in algorithm A1

Suppose one wishes to improve algorithm A1 by using a gradient-type method rather than the quantization (2.2.1) or (2.2.2) of the density. For a gradient method, one should take:

$$\rho_i(n) = \rho_i(n-1) + \beta C_i \quad (G.3.1)$$

where $\beta$ is a constant. This equation is however incompatible with the constraint $P(n) = P_{max}$. Indeed, by multiplying (G.3.1) by $V_i$ then by summing it over $i$, one obtains:

$$\sum_i \rho_i(n)V_i = P_{max} = \sum_i \rho_i(n-1)V_i + \beta \sum_i C_i V_i = P_{max} + \beta \sum_i C_i V_i \quad (G.3.2)$$

which implies that $\beta = 0$. One may try to alleviate this contradiction by renormalizing the mass of the material, by taking e.g.:

$$\rho_i(n) = \alpha(\rho_i(n-1) + \beta C_i) \quad (G.3.3)$$

then by adjusting the constant $\alpha$ so that $P(n) = P_{max}$. However, this solution is now incompatible with the constraint $\rho_{min} \leq \rho_i(n) \leq \rho_{max}$. A possible gradient-type solution that simultaneously satisfies the constraints $P(n) = P_{max}$ and $\rho_{min} \leq \rho_i(n) \leq \rho_{max}$ is the step 5 of A1,

$$\rho_i(n) = \rho_i(n-1) \pm \delta\rho(C_i) \quad (G.3.4)$$

with $\delta\rho(C_i)$ given by:

$$\delta\rho(C_i) = \beta |C_i| \quad (G.3.5)$$

We would then replace the constant increment $\delta\rho$ at step 5 of algorithm A1 by $\delta\rho(C_i)$.
To test this solution, we have redone the calculations of fig. 2-3 using the procedure (G.3.4)-(G.3.5) instead of the quantization (2.2.1), starting from an uniform initial density configuration $\rho_i(0) = P_{max}/V_a \, \forall i$. We have also taken:

$$\beta = 2\frac{\rho_{max} - \rho_{min}}{\langle |C_j| \rangle}, \quad \langle |C_j| \rangle = \frac{1}{N}\sum_{j=1}^{N} |C_j| \quad (G.3.6)$$

For these new calculations, we have reused the same parameters, $NX$, $NR$, $NPS$, $E_0$, $\rho_{max}$, $\rho_{min}$, $H$ and $P_{max}$, that those used for obtaining fig. 2-3. The results obtained with this gradient-type procedure are given in fig. G.4 for the density configurations $\rho_i(n)$, and in fig. G.5 for the neutron fluxes $\phi(n)$ and the weights $P(n)$ of the structures. The fluxes are compared with those obtained fig. 3. Finally, a video, phimax_grad.mp4, given in Supplemental Material allows to visualize the convergence of the calculated structures.

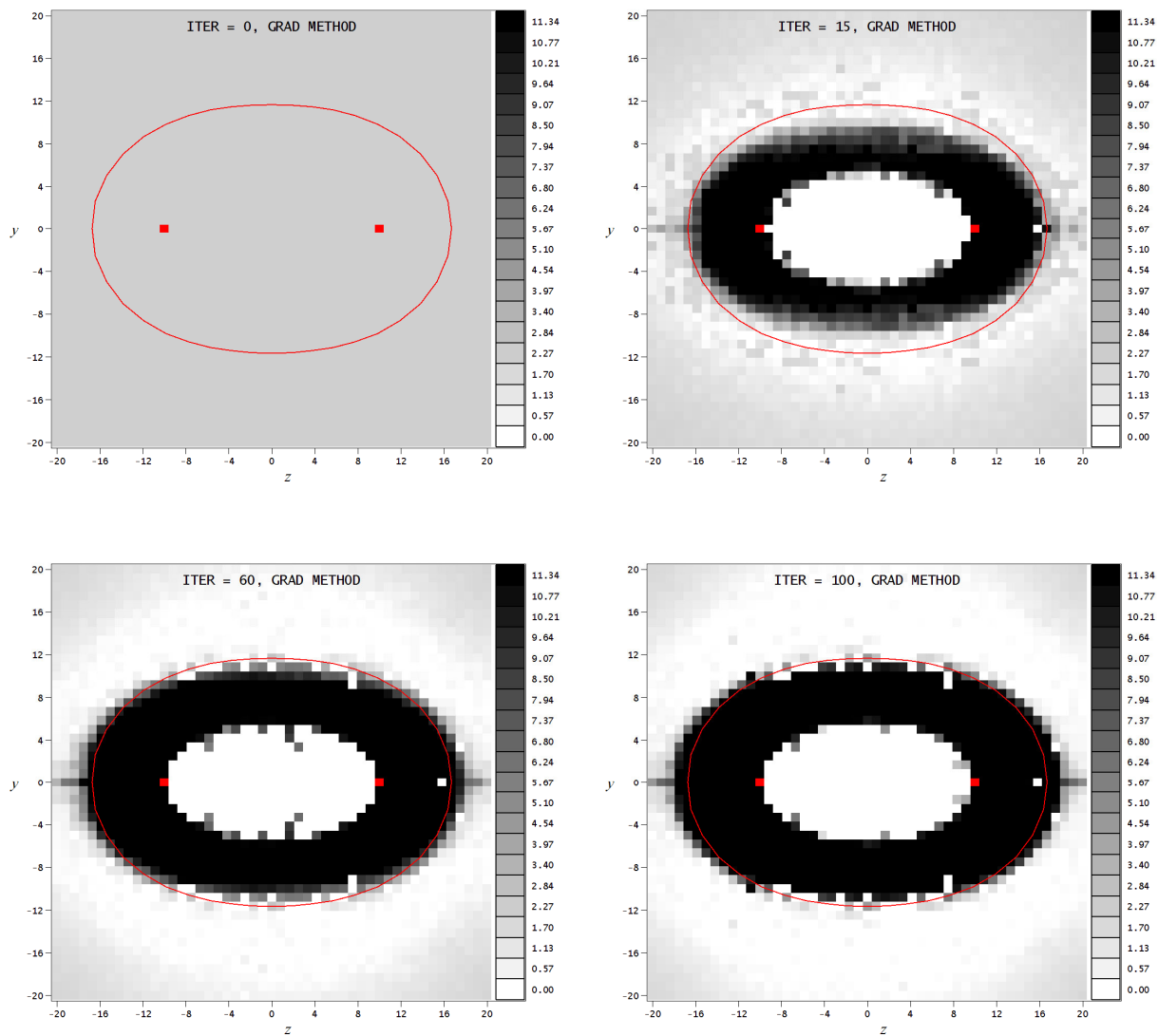

**Figure G.4.** Density maps obtained at iterations 0 to 100 of algorithm A1, using the gradient-type procedure (G.3.4)-(G.3.6), with the parameters used to obtain fig. 2. The units of axes $y = \pm r$ and $z$ are cm. The gray scale gives the values of the densities $\rho_i(n)$ in g/cm$^3$.

We note that the results obtained with the gradient-type procedure (G.3.4)-(G.3.6) are actually rather poor:

(i) Because $\delta\rho(C_i) \to 0$ when $C_i \to 0$, i.e. when the contribution of the cell $\Theta_i$ to the objective $O\varphi$ is small, the structure does not empty correctly on the edges. Worse, as the cells at the periphery have large volumes, and the total volume is constrained, this defect disrupts the construction of the central area of the structure, the most important. The algorithm therefore converges badly, with a resulting shape that is suboptimal and polluted on the edges. To solve this problem, one could modify the function $\delta\rho(C_i)$ in order to have an hybrid between the quantization procedure of the density and the gradient method, for example $\delta\rho(C_i) = \delta\rho + \beta|C_i|$.

(ii) The very strong variability of the values of $|C_i|$, over 5 orders of magnitude, coupled to the statistical fluctuations, induces abrupt modifications of the structure, with cells whose density reaches directly $\rho_{min}$ or $\rho_{max}$ in a single iteration. The video illustrates this problem well. We thus fall back on the main problem of algorithm A0, i.e. the appearance of fluctuations of the structure, for example visible in fig. G.5 on the evolution of the flux with $n$. These oscillations hinder the convergence of the algorithm, and lead to a suboptimal structure. To solve this problem, one could move even more away from the gradient method, by taking a function $\delta\rho(C_i)$ that is non-linear in $C_i$, e.g. $\delta\rho(C_i) = \delta\rho + \beta|C_i|^{1/n}$ with $n \gg 1$. Note that this bad behavior of the gradient-type procedure (G.3.4) underlines by contrast an unnoticed advantage of the quantization procedure: by taking $\delta\rho$ independent on $C_i$, it reduces the impact of the statistical fluctuations of $C_i$ on the structure.

(iii) The asymmetry between the minimum and maximum values of $C_i$, with a minimum at $\sim -10^{-5}$ vs a maximum at $\sim +10^{-6}$, makes the algorithm more efficient at destroying poorly positioned matter rather than at building in important areas. As a result, the formation of the central area, which is already handicapped by the slow evolution of the density in the peripheral cells, is even more slown down. To solve this problem, one could possibly weight the values $C_i$ according to their sign. However, such weighting would introduce a new prior in the optimization procedure. The calculated structure could depend on this choice, which would be better to avoid, considering the little intuition that we generally have of the optimal shape before its calculation in the majority of cases.

In the rest of this study, hence, we will use only the quantization (2.2.1)-(2.2.2) of the density at step 5 of algorithm A1. Although it is probably possible to do better, cf. the above mentioned leads, this procedure has the advantage of simplicity, and will prove to be robust and universal, since we will show in the following that it gives correct results for all the categories of problems addressed in this study.

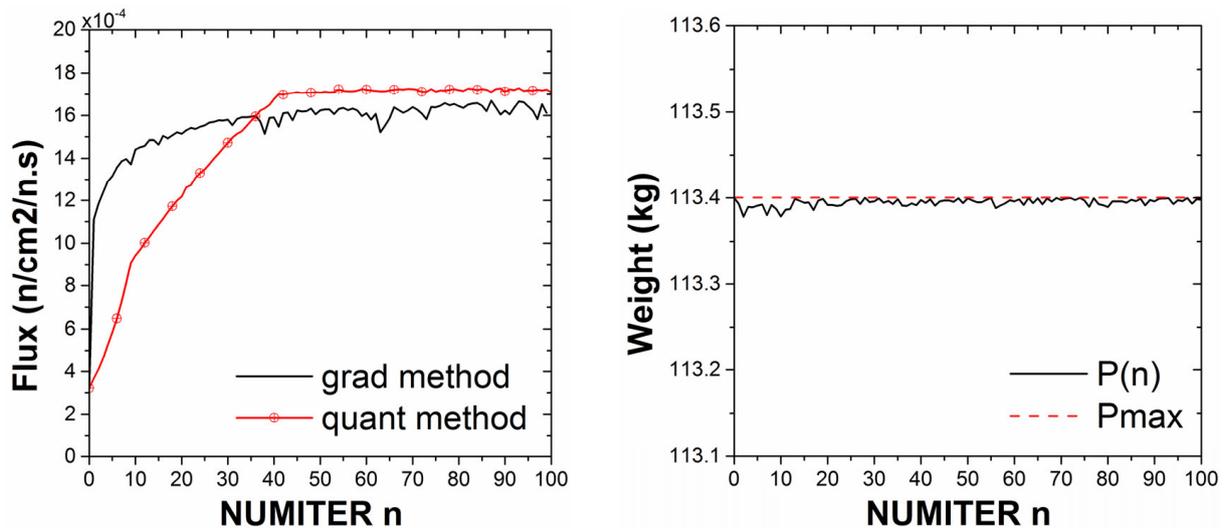

**Figure G.5.** (Left) evolution of the neutron flux, $\phi(n)$, in the detector cell with the iteration number $n$, obtained with the gradient-type procedure (G.3.4)-(G.3.6) (black line). The fluxes are given in neutrons/cm² per source neutron. They are compared to the fluxes obtained fig. 3 with the quantization procedure (2.2.1) (red line + circles); (right) evolution of the weight of the structure, $P(n)$, in kg with $n$, obtained with the procedure (G.3.4)-(G.3.6). The maximum permissible weight, $P_{max}$, is indicated by the red dashed line.

## Section H. Composition of the heavy concrete used in section 2.3

In this section we give the isotopic composition of the heavy concrete used in the calculations of section 2.3, in MCNP format. The values indicated for each isotope are weight fractions, which must be preceded by a sign – in an MCNP input file, see [26] section 5.3.1. The isotope names are in ZA format, first giving the Z than the A number of the isotope.

| Isotope | Weight fraction | Isotope | Weight fraction |
|---------|-----------------|---------|-----------------|
| 56130 | -4.58499E-04 | 20046 | -2.16481E-06 |
| 56132 | -4.43595E-04 | 20048 | -1.05607E-04 |
| 56134 | -1.07898E-02 | 14028 | -1.62613E-02 |
| 56135 | -2.96154E-02 | 14029 | -8.52785E-04 |
| 56136 | -3.55228E-02 | 14030 | -5.86084E-04 |
| 56137 | -5.11925E-02 | 26054 | -2.50033E-03 |
| 56138 | -3.29232E-01 | 26056 | -4.06651E-02 |
| 8016 | -3.12703E-01 | 26057 | -9.56426E-04 |
| 16032 | -1.00622E-01 | 26058 | -1.28534E-04 |
| 16033 | -8.19032E-04 | 13027 | -5.90005E-03 |
| 16034 | -4.73652E-03 | 1001 | -6.49006E-03 |
| 16036 | -2.38254E-05 | 12024 | -1.60968E-03 |
| 20040 | -4.56248E-02 | 12025 | -2.12286E-04 |
| 20042 | -3.19717E-04 | 12026 | -2.43051E-04 |
| 20043 | -6.83007E-05 | 5010 | -4.34975E-05 |
| 20044 | -1.07986E-03 | 5011 | -1.92505E-04 |

**Table H.1**. Isotopic composition of the heavy concrete used in the calculations of section 2.3, in MCNP format.

## Section I. Two examples illustrating the limitations of algorithm A1

We herein propose two counterexamples that illustrate some of the limitations of algorithm A1.

I.1. Counterexample 1

We could redo the calculations of fig. 5-6, starting this time no longer from an uniform initial density configuration $\rho_i(0) = \rho(0)$ $\forall i$, but from a configuration closer to that of the intuitive solution $\phi_{int}$. This new initial configuration, shown in fig. I.1 iteration 0, is obtained by condensing in the cells surrounding the axis SD as much polyethylene (PE) as possible at density $\rho_{max}$, while conforming to the weight constraint $P \leq P_{max}$. Then, conforming to the quantization of $\rho$, the structure is completed with a ring of PE at density $\rho_{max} - \delta\rho$, in order to approach as best as possible the limit $P_{max}$. The results obtained with A1 with this new initial configuration are given in fig. I.1 for the density maps $\rho_i$, and in fig. I.2 for the fluxes $\phi(n)$ and the weights $P(n)$. It can be seen in these figures that A1 recreates a neutron shield almost identical to that obtained in fig. 5. However, we also note that A1 does not succeed in thinning sufficiently the central body of the shield to complete the antenna $a_2$. As a result, at iteration 400, the algorithm has still not managed to reduce the flux $\phi(n)$ under its reference $\phi_{int}$. Worse, in his repeated attempts to close the antenna $a_2$, A1 eventually builds structures that are less optimal than its starting configuration, which is a counter-performance. Here we see appear what we consider to be one of the main weaknesses of A1: it has trouble with disassembling massive structures. This observation explains why we always started, and will always start in the rest of this study, with an uniform initial configuration, as diluted as possible, with no matter already condensed in a suboptimal configuration.

Another problem of algorithm A1, visible in fig. 5 or I.1, is the generation of cells of densities $\rho_{min} + \delta\rho$ (and more rarely $\rho_{min} + 2\delta\rho$) in areas that should be empty. These noised cells, whose positions fluctuate from one iteration to another, are induced by statistical fluctuations on the values $C_i$, inherent to the Monte-Carlo method used for the computation of the derivatives of the objective, and amplified by the choice of the linear quantization of the density. These statistical fluctuations on the values of $C_i$ are enhanced: (i) by the large size of the structure, which dilutes the neutron trajectories; (ii) by the very objective sought: to design a neutron shield. As neutrons have more and more difficulty to cross the shield, as it becomes more efficient, the neutron density in the areas behind it decreases sharply, so the statistical fluctuations on the associated $C_i$ values strongly increase. We can e.g. verify in fig. 5 that the noise in these areas increases with $n$. The optimization of a particle shield with a Monte-Carlo method thus leads to a small paradox: the more efficient the calculated structure, the noisier it will be around. This problem can be solved by increasing the number *NPS* of source particles or, when the available computing resources are insufficient, by using at the last resort convergence acceleration methods, cf. conclusion of the study.

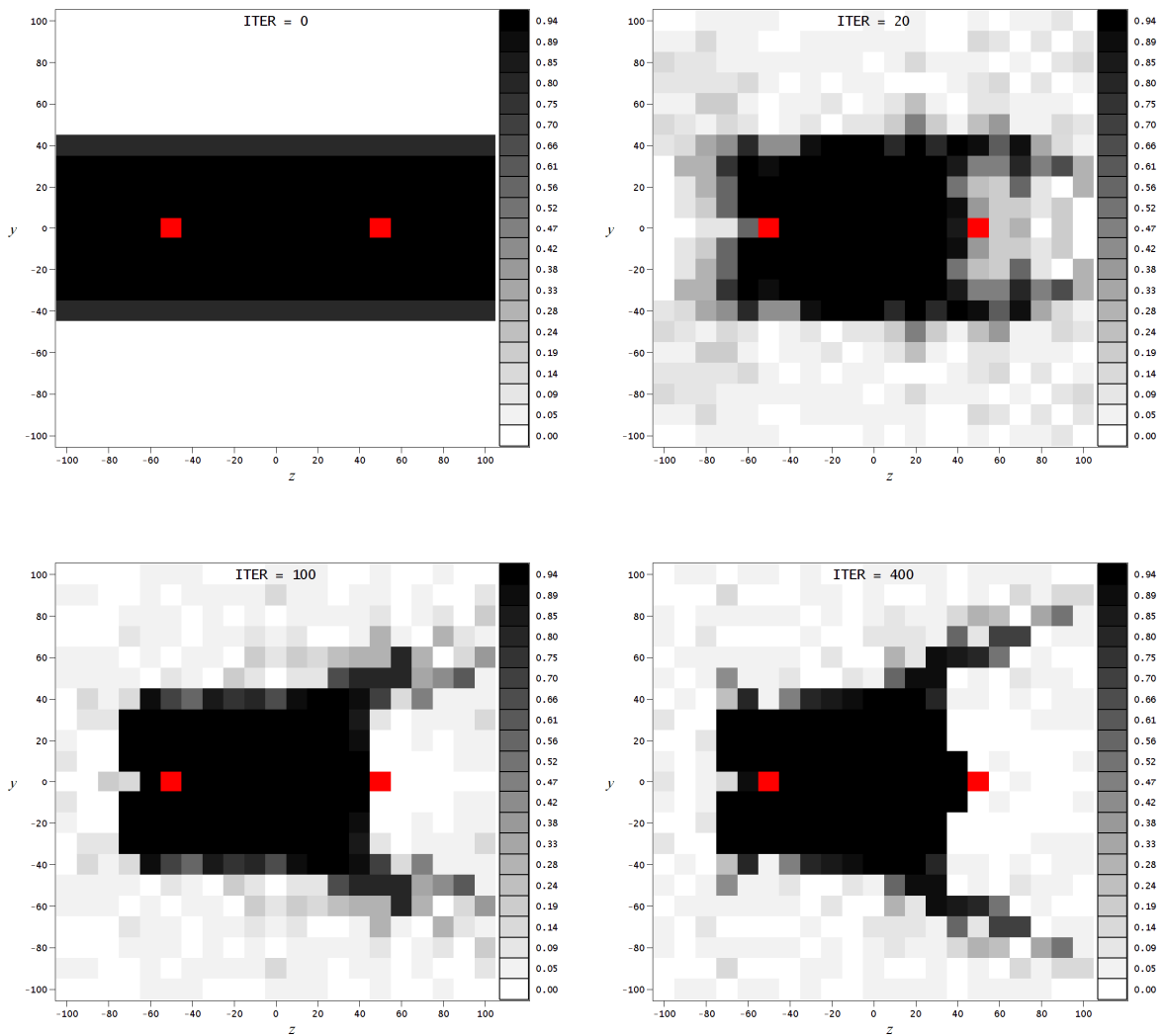

**Figure I.1.** Density maps obtained at iterations 0 to 400 of A1. The initial density configuration is close to that of the intuitive configuration $\phi_{int}$, and is shown at iteration 0. The units of the axes $y = \pm r$ and $z$ are cm. The gray scale gives the values of the PE densities $\rho_i(n)$ in g/cm³.

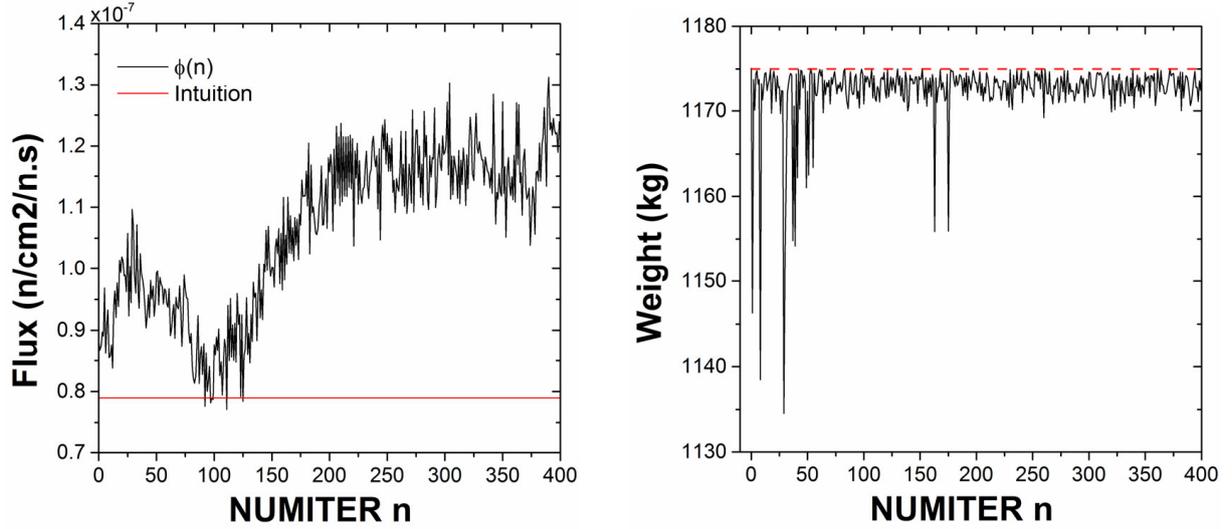

**Figure I.2.** (Left) evolution of the neutron flux, $\phi(n)$, in the cell D with the iteration number $n$. The fluxes $\phi(n)$ are given in neutrons/cm$^2$/s.n, and are compared to the flux $\phi_{int}$ of the intuitive solution indicated by the red line; (right) evolution of the weight of the structure, $P(n)$, in kg with $n$. The maximum permissible weight, $P_{max}$, is indicated by a red dashed line. For these calculations, the initial density configuration is shown fig. I.1 iteration 0.

I.2. Counterexample 2

We could redo the calculations of fig. 5-6, this time by removing the concrete walls. Without them, indeed, the problem of this section has an obvious solution: the optimal shield is the union of the cells $\Theta_i$ parametrized in (2.2.3) with $i \in\ ]1+NX/4, 1+3NX/4[$, i.e. of the cells that contain the axis SD and are comprised between the source and the detector. The addition of any other cell makes the structure suboptimal, because it will act as a reflector that will return towards the detector a fraction of the neutrons that would have leaked from the structure otherwise. For this problem, the calculations were carried out: (i) with $NR = 10$ and $NX = 20$ in the parametrization (2.2.3) of the cells; (ii) with the logarithmic quantization (2.2.2) of the density, with $M = 20$; (iii) starting from an uniform initial density configuration, $\rho_i(0) = \rho(0)\ \forall i$; (iv) using a filter on the statistically absurd PERT$_i$ values: if the MCNP statistical error on PERT$_i$ exceeds 95%, we take PERT$_i = 0$. For these calculations, we chose a material made of 100% of $^{207}$Pb atoms, with $\rho_{min} = 10^{-5}$ g/cm$^3$, $\rho_{max} = 11.34$ g/cm$^3$, $H = 20$ cm, $NPS = 10^9$, $V_{max} = 10^4$ cm$^3$ and $E_0 = 2.53\ 10^{-8}$ MeV. The structures $\rho_i(n)$ thus obtained with A1 are plotted in fig. I.3 for several iterations, and are compared to the optimal solution indicated by the hatched area in red. A video, shield_nowall.mp4, given in Supplemental Material gives the sequence of these results until the convergence of the calculations (look at the left-hand figures, those with the mention $NR = 10$). The red cells are the source (left) and detector (right) cells. The neutron fluxes $\phi(n)$ obtained in the detector cell are given in fig. I.5 (line + blue squares), and are compared to the flux $\phi_{opt}$ calculated with the optimal shield, indicated by the red line. We note in fig. I.3 or in the video that A1 quickly converges towards a structure close to the optimal structure. Almost all superfluous cells are properly removed, with the exception of a small ring that surrounds the optimal structure. This ring seems to be of little importance, but we note in fig. I.5 that it actually has a strong reflective power. After convergence, the flux in the cell D, $10^{-6}$ n/cm$^2$/n.s, is indeed 3.3 times greater than the optimal flux, $\phi_{opt} = 3.10^{-7}$ n/cm$^2$/s.n. The structure calculated with A1, although close to the ideal structure, is thus strongly suboptimal. This counter-performance is probably due to the issue observed in the counterexample 1, i.e. that A1 struggles to disassemble massive structures.

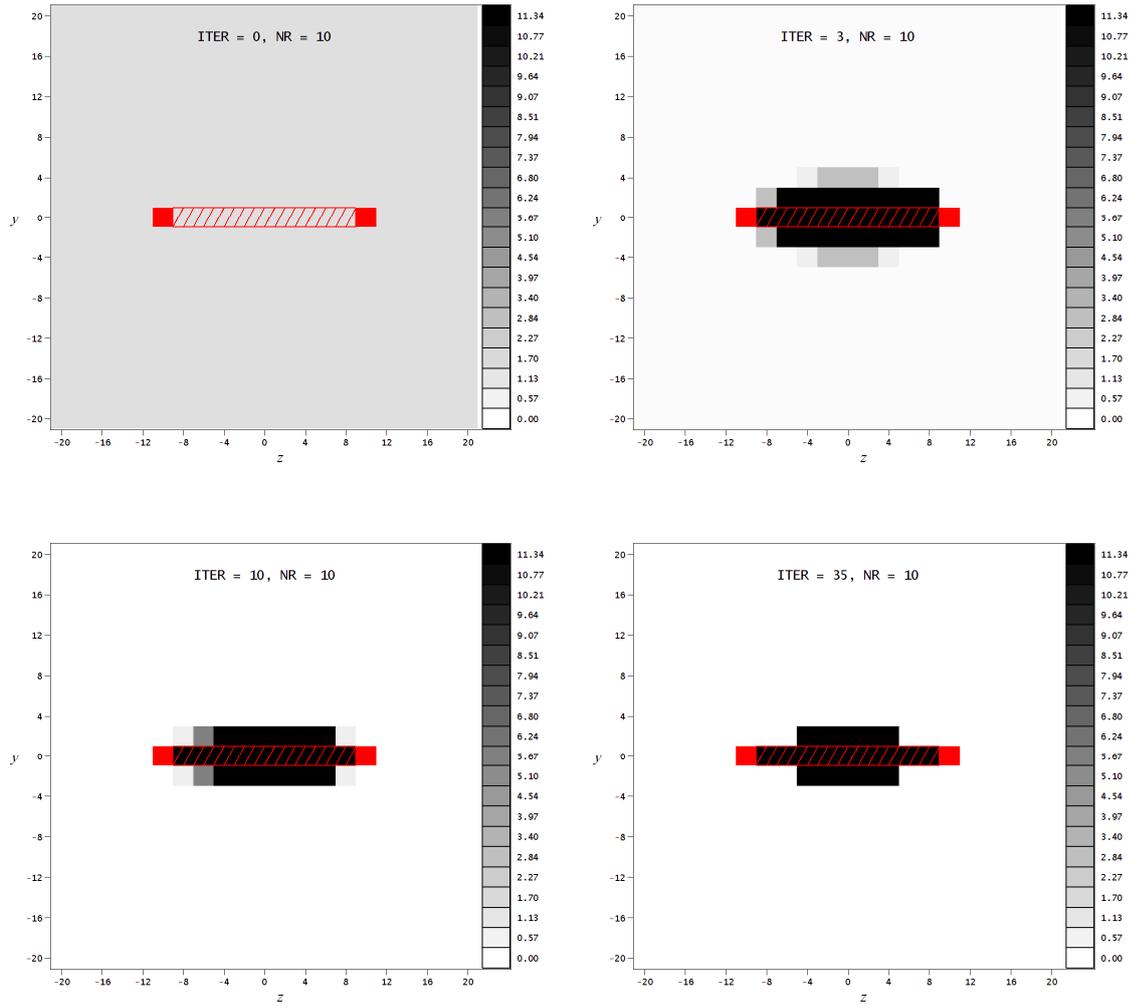

**Figure I.3.** Density maps obtained at iterations 0 to 35 of A1, in the absence of walls, using the parametrization (2.2.3) or (I.2.1) with $NR = 10$. The units of the axes $y = \pm r$ and $z$ are cm. The gray scale gives the values of the densities $\rho_i(n)$ in g/cm³.

Let us now redo these calculations by reducing the size of the cells $\Theta_i$ used to tile the space. To do this, we have chosen the following parametrization:

$$\Theta_i(\underline{r}) = \Theta[R_{j-1} \leq r < R_j] \times \Theta[X_k \leq z < X_{k+1}]$$

$$X_k = H\left(\frac{2k-1}{NX} - 1\right), \quad R_j = \begin{cases} 0 & \text{if } j = 0 \\ \dfrac{H}{20} + H\dfrac{j-1}{NR} & \text{if } j > 0 \end{cases} \quad \text{(I.2.1)}$$

$$i = (NX+1)j + k - NX, \quad k \in [0, NX], \quad j \in [1, NR+1]$$

with $NR = 50$ and $NX = 20$. This parametrization has several advantages: (i) it allows to reduce the thickness of the rings $\Theta_i$ by a factor of 5 compared to that of the previous calculations, with the exception of the cylindrical cells containing the axis SD, left unchanged; (ii) it is equivalent to the previous parametrization, (2.2.3), if one takes $NR = 10$, in order to compare directly the results obtained with (I.2.1) and (2.2.3); (iii) it does not modify the optimal solution $\phi_{opt}$. For this new batch of calculations, in order to compensate for the drop in statistic due to the decrease in cell volume, we have lowered the filter on the PERT$_i$ values down to 50%, instead of the previous 95%.

The density configurations obtained with this procedure are given in fig. I.4 for a few iterations, and are compared anew to the optimal structure indicated by the red dashed area. The video shield_nowall.mp4 gives the sequence of these results until convergence (see the right-hand figures, with the title $NR = 50$). The neutron fluxes calculated with the parametrization (I.2.1) with $NR = 50$ are given in fig. I.5 as a function of $n$, and are compared to the flux obtained for the optimal shield, indicated by the red line. We note in fig. I.4-I.5 or in the video that A1 converges this time correctly towards the optimal solution: the residual ring present for $NR = 10$ is eliminated for $NR = 50$. We hence see appear in this example an important and known issue in topology optimization: an overly coarse pixelation of the available volume, whether it is chosen by default to reduce the computing time or e.g. due to constraints on the machining process of the object, can lead to a suboptimal structure.

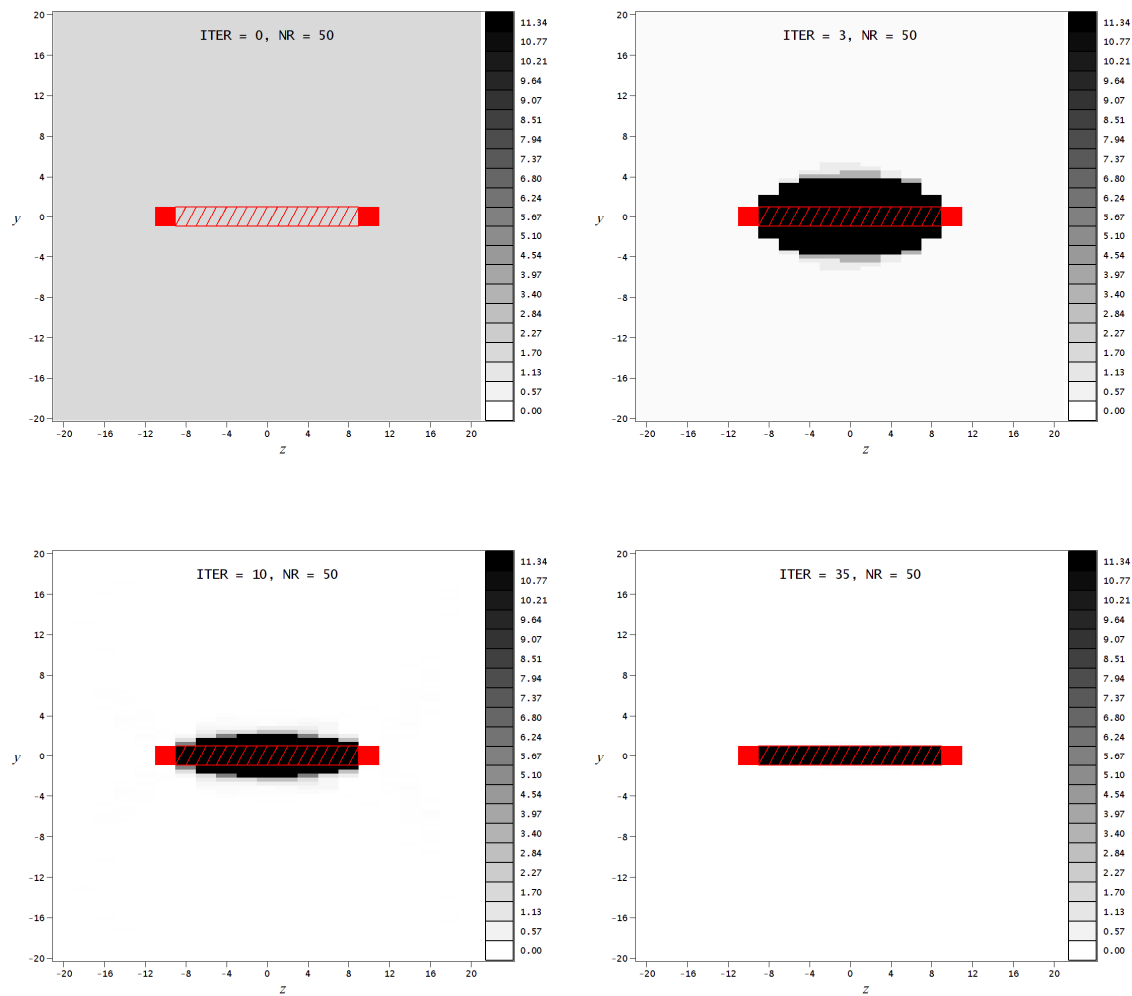

**Figure I.4.** Density maps obtained at iterations 0 to 35 of A1, in the absence of walls, using the parametrization (I.2.1) with $NR = 50$. The units of the axes $y = \pm r$ and $z$ are cm. The gray scale gives the values of the densities $\rho_i(n)$ in g/cm³.

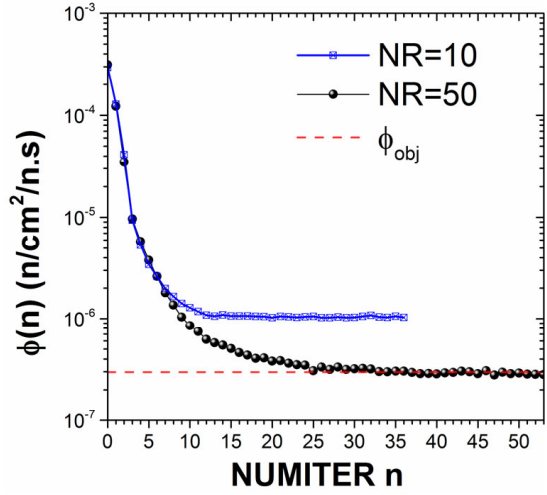

**Figure I.5.** Evolution of the neutron flux, $\phi(n)$, in the detector cell with the iteration number $n$ of algorithm A1, in the absence of walls. The fluxes $\phi(n)$ are given in neutrons/cm² per source neutron, and are compared to the flux $\phi_{opt}$ of the optimal solution, indicated by the red dashed line. The fluxes are calculated by taking $NR = 10$ (line + blue squares) or $NR = 50$ (line + black circles) in the parametrization (I.2.1) of the cells $\Theta_i$.

## Section J. Demonstration of the result (3.2.1)

Consider a small volume of matter $dV$ positioned at a point $\underline{r}$. We want to calculate at the point $\underline{r}_d$ the total flux of neutrons that have passed by $dV$. To perform this calculation, we observe that an integral formulation of the Boltzmann equation is more convenient than its differential version, (0.1)-(0.2). This integral formulation can be found e.g. in [27] pp 95-96, and is:

$$\varphi(\underline{r}, E, \underline{\Omega}) = \varphi_0(\underline{r}, E, \underline{\Omega}) +$$

$$\int_{E'=0}^{+\infty} \int_{s=0}^{+\infty} \int_{\underline{\Omega}' \in 4\pi} \left[ \begin{array}{l} \Sigma_s(\underline{r} - s\underline{\Omega}, E' \to E, \underline{\Omega}' \to \underline{\Omega})\varphi(\underline{r} - s\underline{\Omega}, E', \underline{\Omega}') \\ \times \exp\left( -\int_{s'=0}^{s} \Sigma_t(\underline{r} - s'\underline{\Omega}, E)ds' \right) \end{array} \right] dE' \, ds \, d\underline{\Omega}' \quad (J.1)$$

$$\varphi_0(\underline{r}, E, \underline{\Omega}) = \int_{s=0}^{+\infty} Q(\underline{r} - s\underline{\Omega}, E, \underline{\Omega}) \exp\left( -\int_{s'=0}^{s} \Sigma_t(\underline{r} - s'\underline{\Omega}, E)ds' \right) ds$$

In this integral equation, $\varphi(\underline{r},E,\underline{\Omega})$ is the angular flux of neutrons, $Q(\underline{r},E,\underline{\Omega})$ the neutron source, $\Sigma_t(\underline{r},E)$ the total macroscopic cross-section at the point $\underline{r}$ and at the energy $E$, and $\Sigma_s(\underline{r}, E' \to E, \underline{\Omega}' \to \underline{\Omega})$ the macroscopic cross-section of transition from a state $(E',\underline{\Omega}')$ to the state $(E,\underline{\Omega})$ after a scattering at the point $\underline{r}$.

In (J.1), one can note that $\varphi_0(\underline{r},E,\underline{\Omega})$ is the flux of neutrons having done 0 scattering. One can then note that the flux $\varphi_{n+1}(\underline{r},E,\underline{\Omega})$ of neutrons having done $n+1$ scatterings can be expressed as a function of the flux $\varphi_n(\underline{r},E,\underline{\Omega})$ of neutrons having done $n$ scatterings, as follows:

$$\varphi_{n+1}(\underline{r}, E, \underline{\Omega}) = \int_{E'=0}^{+\infty} \int_{s=0}^{+\infty} \int_{\underline{\Omega}' \in 4\pi} \left[ \begin{array}{l} \Sigma_s(\underline{r} - s\underline{\Omega}, E' \to E, \underline{\Omega}' \to \underline{\Omega}) \times \\ \varphi_n(\underline{r} - s\underline{\Omega}, E', \underline{\Omega}') \exp\left( -\int_{s'=0}^{s} \Sigma_t(\underline{r} - s'\underline{\Omega}, E)ds' \right) \end{array} \right] dE' \, ds \, d\underline{\Omega}' \quad (J.2)$$

The angular flux $\varphi(\underline{r},E,\underline{\Omega})$ is then given by:

$$\varphi(\underline{r},E,\underline{\Omega}) = \sum_{n=0}^{+\infty} \varphi_n(\underline{r},E,\underline{\Omega}) \quad (J.3)$$

Suppose now: (i) that the material used is little absorbing at 14 MeV (which is the case for most materials); (ii) that the volume $dV$ is small enough so that a neutron that makes a collision in it has an almost zero probability of making another collision in it; (iii) that $dV$ does not cross the line connecting the source $\underline{r}_s$ and the detector $\underline{r}_d$; (iv) finally, that there is, outside $dV$, no other matter in space. In this framework, restricted but useful, the angular flux (J.3) can be easily formulated: it reduced to the sum $\varphi_0+\varphi_1$, where the exponentials can be approximated by 1:

$$\varphi(\underline{r},E,\underline{\Omega}) \approx \varphi_0(\underline{r},E,\underline{\Omega}) + \varphi_1(\underline{r},E,\underline{\Omega}) \quad (J.4)$$

with:

$$\varphi_0(\underline{r},E,\underline{\Omega}) = \int_{s=0}^{+\infty} Q(\underline{r}-s\underline{\Omega},E,\underline{\Omega})ds$$

$$\varphi_1(\underline{r},E,\underline{\Omega}) = \int_{E'=0}^{+\infty}\int_{s=0}^{+\infty}\int_{\underline{\Omega}'\in 4\pi} \Sigma_s(\underline{r}-s\underline{\Omega},E'\to E,\underline{\Omega}'\to\underline{\Omega})\varphi_0(\underline{r}-s\underline{\Omega},E',\underline{\Omega}')dE'\,ds\,d\underline{\Omega}' \quad (J.5)$$

$$= \int_{E'=0}^{+\infty}\int_{s=0}^{+\infty}\int_{s'=0}^{+\infty}\int_{\underline{\Omega}'\in 4\pi} \Sigma_s(\underline{r}-s\underline{\Omega},E'\to E,\underline{\Omega}'\to\underline{\Omega})Q(\underline{r}-s\underline{\Omega}-s'\underline{\Omega}',E',\underline{\Omega}')dE'\,ds\,ds'\,d\underline{\Omega}'$$

The total neutron flux $\phi(\underline{r}_s\to\underline{r}_d,E)$ generated at the point $\underline{r}_d$ by a point source placed at $\underline{r}_s$, is thus:

$$\phi(\underline{r}_s\to\underline{r}_d,E) \approx \phi_0(\underline{r}_s\to\underline{r}_d,E) + \phi_1(\underline{r}_s\to\underline{r}_d,E)$$

$$\phi_0(\underline{r}_s\to\underline{r}_d,E) = \int_{\underline{\Omega}\in 4\pi} \varphi_0(\underline{r}_d,E,\underline{\Omega})d\underline{\Omega} = \int_{s=0}^{+\infty}\int_{\underline{\Omega}\in 4\pi} Q(\underline{r}_d-s\underline{\Omega},E,\underline{\Omega})ds\,d\underline{\Omega}$$

$$\phi_1(\underline{r}_s\to\underline{r}_d,E) = \int_{\underline{\Omega}\in 4\pi} \varphi_1(\underline{r}_d,E,\underline{\Omega})d\underline{\Omega} \quad (J.6)$$

$$= \int_{E'=0}^{+\infty}\int_{s=0}^{+\infty}\int_{\underline{\Omega}\in 4\pi}\int_{s'=0}^{+\infty}\int_{\underline{\Omega}'\in 4\pi} \begin{bmatrix} \Sigma_s(\underline{r}_d-s\underline{\Omega},E'\to E,\underline{\Omega}'\to\underline{\Omega}) \\ \times Q(\underline{r}_d-s\underline{\Omega}-s'\underline{\Omega}',E',\underline{\Omega}') \end{bmatrix} dE'\,ds\,d\underline{\Omega}\,ds'\,d\underline{\Omega}'$$

Our objective is the formulation at the point $\underline{r}_d$ of the total flux $\phi(\underline{r}_s\to dV\to\underline{r}_d,E)$ of neutrons that have passed by the volume $dV$. As $\phi_0$ is the contribution of neutrons that have not performed any scattering, the flux $\phi_0(\underline{r}_s\to\underline{r}_d,E)$ in (J.6) is induced by source neutrons having ballistic trajectories from $\underline{r}_s$ to $\underline{r}_d$. So, as $dV$ does not cross the line linking $\underline{r}_s$ to $\underline{r}_d$, none of these neutrons can pass by $dV$. Conversely, the neutrons taken into account by $\phi_1(\underline{r}_s\to\underline{r}_d,E)$ have, by definition, performed one scattering, and since there is no other matter in space than $dV$, this scattering can only have occurred in $dV$. The flux $\phi(\underline{r}_s\to dV\to\underline{r}_d,E)$ sought is therefore the flux $\phi_1(\underline{r}_s\to\underline{r}_d,E)$. Let us calculate it. For a monoenergetic isotropic [7] point source emitting $Q_0$ neutrons of energy $E_0$ at the point $\underline{r}_s$, the source term is written:

---

[7] With a good approximation, a 14 MeV D-T generator is monoenergetic and isotropic.

$$Q(\underline{r}, E, \underline{\Omega}) = \frac{Q_0}{4\pi} \delta(\underline{r} - \underline{r}_s) \delta(E - E_0) \quad (J.7)$$

In this problem, $E_0 = 14$ MeV. Noticing then that $d\underline{r}' = s^2 ds d\underline{\Omega}$ is a volume element in spherical coordinates, we can simplify the integrals (J.6) by making the changes of variables:

$\underline{r}' = s\underline{\Omega}, \quad \underline{r}'' = s'\underline{\Omega}' \quad (J.8)$

As $\|\underline{\Omega}\| = \|\underline{\Omega}'\| = 1$, this implies that:

$$s = \|\underline{r}'\|, \quad s' = \|\underline{r}''\|, \quad \underline{\Omega} = \frac{\underline{r}'}{\|\underline{r}'\|}, \quad \underline{\Omega}' = \frac{\underline{r}''}{\|\underline{r}''\|}, \quad \int_{s=0}^{+\infty} \int_{\underline{\Omega}\in 4\pi} g(\ldots, s\underline{\Omega}, \underline{\Omega}) ds d\underline{\Omega} = \int_{\underline{r}'\in\mathfrak{R}^3} g\left(\ldots, \underline{r}', \frac{\underline{r}'}{\|\underline{r}'\|}\right) \frac{d\underline{r}'}{\|\underline{r}'\|^2}$$
(J.9)

Using these changes of variables and the expression (J.7) of the source, the fluxes $\phi_0$ and $\phi_1$ in (J.6) can be rewritten as follows:

$$\phi_0(\underline{r}_s \to \underline{r}_d, E) = \frac{Q_0 \delta(E - E_0)}{4\pi} \int_{\underline{r}'\in\mathfrak{R}^3} \delta(\underline{r}_d - \underline{r}_s - \underline{r}') \frac{d\underline{r}'}{\|\underline{r}'\|^2} = \frac{Q_0 \delta(E - E_0)}{4\pi \|\underline{r}_d - \underline{r}_s\|^2}$$

$$\phi_1(\underline{r}_s \to \underline{r}_d, E) = \phi(\underline{r}_s \to dV \to \underline{r}_d, E) \quad (J.10)$$

$$= \frac{Q_0}{4\pi} \int_{\underline{r}''\in\mathfrak{R}^3} \Sigma_s\left(\underline{r}_s + \underline{r}'', E_0 \to E, \frac{\underline{r}''}{\|\underline{r}''\|} \to \frac{\underline{r}_d - \underline{r}_s - \underline{r}''}{\|\underline{r}_d - \underline{r}_s - \underline{r}''\|}\right) \frac{d\underline{r}''}{\|\underline{r}_d - \underline{r}_s - \underline{r}''\|^2 \|\underline{r}''\|^2}$$

For an infinitesimal volume $dV$ centered at a point $\underline{r}$, the scattering cross-section $\Sigma_s(\underline{v}, E_0 \to E, \underline{\Omega}' \to \underline{\Omega})$ can be written $dV \delta(\underline{v} - \underline{r}) \Sigma_s(E_0 \to E, \underline{\Omega}' \to \underline{\Omega})$. Observing now that the probability that a neutron of initial direction $\underline{\Omega}'$ ends with a direction $\underline{\Omega}$ after a collision depends only on the scalar product $\underline{\Omega}' \cdot \underline{\Omega}$ by rotational symmetry, one can note that $\Sigma_s(E_0 \to E, \underline{\Omega}' \to \underline{\Omega}) = \Sigma_s(E_0 \to E, \underline{\Omega} \cdot \underline{\Omega}')/(2\pi)$. The flux $\phi_1$ can thus be rewritten:

$$\phi_1(\underline{r}_s \to \underline{r}_d, E) = \frac{Q_0 dV}{8\pi^2 \|\underline{r} - \underline{r}_s\|^2 \|\underline{r} - \underline{r}_d\|^2} \Sigma_s\left(E_0 \to E, \frac{\underline{r} - \underline{r}_s}{\|\underline{r} - \underline{r}_s\|} \cdot \frac{\underline{r}_d - \underline{r}}{\|\underline{r}_d - \underline{r}\|}\right) \quad (J.11)$$

If the volume $dV$ contains several isotopes of atomic fractions $\chi_k(\underline{r})$, the result (J.11) becomes:

$$\phi_1(\underline{r}_s \to \underline{r}_d, E, \underline{\chi}) = \frac{Q_0 dV}{8\pi^2 \|\underline{r} - \underline{r}_s\|^2 \|\underline{r} - \underline{r}_d\|^2} \sum_{k=1}^{N} \frac{\chi_k(\underline{r})}{\sum_{k=1}^{N} \chi_k(\underline{r})} \Sigma_k\left(E_0 \to E, \frac{\underline{r} - \underline{r}_s}{\|\underline{r} - \underline{r}_s\|} \cdot \frac{\underline{r}_d - \underline{r}}{\|\underline{r}_d - \underline{r}\|}\right) \quad (J.12)$$

where $\Sigma_k(E_0 \to E, \underline{\Omega} \cdot \underline{\Omega}')/(2\pi)$ is the macroscopic cross-section of transition from the state $(E_0, \underline{\Omega}')$ to the state $(E, \underline{\Omega})$ after a collision with an isotope $k$.

Finally, by integrating (J.12) over all volumes $dV = d\underline{r}$ of the screen, i.e. over all $\underline{r} \in V$, we obtain the flux $\phi_1(\underline{r}, E)$ of neutrons that have passed through the screen, given in (3.2.1).

## Section K. Demonstration of the result (3.2.3)

We want to find the fractions $\chi_{opt}(\underline{r})$ that minimize the distance between the flux $\phi_1(\underline{r}_d,E,\underline{\chi})$ and the objective spectrum $\phi_{obj}(E)$. We must thus solve the optimization problem (3.1.5), where the vector $\underline{x}$ is here the vector of the fractions $\underline{\chi}(\underline{r}) = (\chi_1(\underline{r}), ..., \chi_N(\underline{r}))$. A clever way to rewrite the Lagrangian associated with this problem is to impose the normalization of the spectrum $\phi_1$ as a constraint (see definitions (3.1.1) to (3.1.3)):

$$L(\underline{\chi}, \lambda) = \left\| \phi_1 - \frac{\phi_{obj}}{\|\phi_{obj}\|} \right\|^2 - \lambda \left( \|\phi_1\|^2 - 1 \right) \quad (K.1)$$

To solve this optimization problem, let us take $\underline{\chi}(\underline{r}) = \underline{\chi}_{opt}(\underline{r})$ then perturb slightly this vector. We have now $\underline{\chi}(\underline{r}) = \underline{\chi}_{opt}(\underline{r}) + d\underline{\chi}(\underline{r})$. The perturbation $d\underline{\chi}$ will induce a perturbation $d\phi$ of the flux, thereby a perturbation $dL$ of the Lagrangian. As in section C.2, the optimal vector $\underline{\chi}_{opt}(\underline{r})$ sought will then be, by definition, the vector that obeys the following equation:

$$dL = 0, \forall d\underline{\chi} \quad (K.2)$$

The perturbation $dL$ can be obtained by reinjecting the vector $\underline{\chi}(\underline{r}) = \underline{\chi}_{opt}(\underline{r}) + d\underline{\chi}(\underline{r})$ into (K.1). Neglecting the second order terms in $\|d\phi\|^2$, we find, cf. definitions (3.1.1)-(3.1.2):

$$L + dL = \left\| \phi_1 + d\phi - \frac{\phi_{obj}}{\|\phi_{obj}\|} \right\|^2 - \lambda \left( \|\phi_1 + d\phi\|^2 - 1 \right) \quad (K.3)$$
$$\Rightarrow dL = 2\langle d\phi, g \rangle$$

where the function $g$ is given by:

$$g(E, \underline{\chi}, \lambda) = (1 - \lambda)\phi_1(\underline{r}_d, E, \underline{\chi}) - \frac{\phi_{obj}(E)}{\|\phi_{obj}\|} \quad (K.4)$$

By reinjecting now the vector $\underline{\chi}(\underline{r}) = \underline{\chi}_{opt}(\underline{r}) + d\underline{\chi}(\underline{r})$ into the formula (3.2.1) of $\phi_1$, we can obtain the perturbation $d\phi$ of the flux, required to formulate the perturbation $dL$ of the Lagrangian in (K.3). We find, neglecting anew the second-order terms:

$$d\phi(E) = \int_{\underline{r} \in V} d\underline{\chi}(\underline{r}) \cdot \underline{\kappa}(\underline{r}, E) d\underline{r} \quad (K.5)$$

where the vector $\underline{\kappa}(\underline{r},E)$ is given by:

$$\underline{\kappa}(\underline{r}, E) = \frac{\underline{\eta}(\underline{r}, E)}{\underline{\chi}(\underline{r}) \cdot \underline{1}} - \frac{\underline{\chi}(\underline{r}) \cdot \underline{\eta}(\underline{r}, E)}{(\underline{\chi}(\underline{r}) \cdot \underline{1})^2} \underline{1} \quad (K.6)$$

Finally, by reinjecting the result (K.5) into (K.3), we find:

$$dL = 2\int_{\underline{r}\in V} d\underline{\chi}(\underline{r}).\langle\underline{\kappa}(\underline{r},E),g(E,\underline{\chi},\lambda)\rangle d\underline{r} \quad (K.7)$$

The equation (K.2) hence shows that the optimal fraction vector $\underline{\chi}_{opt}(\underline{r})$ is solution of a system of $N$ equations, which is written:

$$\langle\underline{\kappa}(\underline{r},E),g(E,\underline{\chi},\lambda)\rangle\big|_{\underline{\chi}=\underline{\chi}_{opt}} = \underline{0}, \forall \underline{r}\in V \quad (K.8)$$

Using the result (K.6), this system can be rewritten:

$$\sum_{j=1}^{N}\frac{\chi_{j_{opt}}(\underline{r})}{\underline{\chi}_{opt}(\underline{r}).\underline{1}}\langle\eta_j(\underline{r},E),g(E,\underline{\chi}_{opt},\lambda)\rangle = \langle\eta_k(\underline{r},E),g(E,\underline{\chi}_{opt},\lambda)\rangle, \forall \underline{r}\in V, \forall k\in[1,N] \quad (K.9)$$

We note that this system is a system of the type:

$$u_k(\underline{r}) = \langle\eta_k(\underline{r},E),g(E,\underline{\chi}_{opt},\lambda)\rangle$$
$$\sum_{j=0}^{N}\alpha_j(\underline{r})u_j(\underline{r}) = u_k(\underline{r}), \forall \underline{r}\in V, \forall k\in[1,N] \quad (K.10)$$

whose solution is a function that can only depend on $\underline{r}$: $u_k(\underline{r}) = f(\underline{r}) \forall \underline{r}, \forall k\in[1,N]$. But, as $\eta_N = 0$, cf. (3.2.2), we have $u_N = 0$, so $f(\underline{r}) = 0$. We thus obtain, using (K.4):

$$u_k(\underline{r}) = 0, \forall \underline{r}\in V, \forall k\in[1,N]$$
$$\Rightarrow \langle\eta_k(\underline{r},E),\phi_1(\underline{r}_d,E,\underline{\chi}_{opt})\rangle = \frac{\langle\eta_k(\underline{r},E),\phi_{obj}(E)\rangle}{(1-\lambda)\|\phi_{obj}\|}, \forall \underline{r}\in V, \forall k\in[1,N-1] \quad (K.11)$$

Finally, by reinjecting the expression (3.2.1) of the flux into (K.11), we show that the optimal vector $\underline{\chi}_{opt}$, thereby the composition of the screen to use so that the flux $\phi_1(\underline{r}_d,E)$ at its output be as close as possible to the objective spectrum $\phi_{obj}(E)$, obeys the following system of equations:

$$\sum_{j=1}^{N-1}\int_{\underline{r}'\in V}\frac{\chi_{j_{opt}}(\underline{r}')}{\sum_{j=1}^{N}\chi_{j_{opt}}(\underline{r}')}\langle\eta_k(\underline{r},E),\eta_j(\underline{r}',E)\rangle d\underline{r}' = \frac{\langle\eta_k(\underline{r},E),\phi_{obj}(E)\rangle}{(1-\lambda)\|\phi_{obj}\|}, \forall \underline{r}\in V, \forall k\in[1,N-1] \quad (K.12)$$

In (K.12), the multiplier $\lambda$ can then be found using the constraint $\|\phi_1\| = 1$, cf. (K.1), by taking $\underline{\chi} = \underline{\chi}_{opt}$ in the expression (3.2.1).

## Section L. Demonstration of the result (3.2.6)-(3.2.7)

Let us take $N = 2$ in the system (3.2.3) or (K.12), with:

$$\chi_1(\underline{r}) + \chi_2(\underline{r}) = 1, \forall \underline{r}\in V \quad (L.1)$$

By injecting the formula (3.2.4) into the equations (3.2.2) then (K.12), we obtain:

$$\int_{\underline{r}'\in V} \frac{\chi_{1opt}(\underline{r}')d\underline{r}'}{\|\underline{r}'-\underline{r}_s\|^2 \|\underline{r}'-\underline{r}_d\|^2} \langle f_1(E_0 \to E, \mu(\underline{r}')), f_1(E_0 \to E, \mu(\underline{r})) \rangle$$
$$= \frac{8\pi^2}{(1-\lambda)\|\phi_{obj}\|Q_0\Sigma_1(E_0)} \langle f_1(E_0 \to E, \mu(\underline{r})), \phi_{obj}(E) \rangle, \forall \underline{r} \in V \quad (L.2)$$

Using the formula (3.2.5) of the density $f_1$, then making the change of variables $x = E/E_0$, we show that the scalar product of the left-hand side of (L.2) is:

$$\langle f_1(E_0 \to E, \mu(\underline{r}')), f_1(E_0 \to E, \mu(\underline{r})) \rangle = \frac{1}{(1-\alpha)^2 E_0} \int_{x=\alpha}^{1} \delta(\mu(\underline{r}')-h(x))\delta(\mu(\underline{r})-h(x))dx \quad (L.3)$$

with:

$$h(x) = \frac{A+1}{2}\sqrt{x} - \frac{A-1}{2}\sqrt{\frac{1}{x}} \quad (L.4)$$

We then note that:

$$\int_{x=\alpha}^{1} \delta(\mu(\underline{r}')-h(x))\delta(\mu(\underline{r})-h(x))dx = \delta(\mu(\underline{r})-\mu(\underline{r}')) \int_{x=\alpha}^{1} \delta(\mu(\underline{r})-h(x))dx \quad (L.5)$$

To calculate the integral (L.5), we can perform the change of variables $y = h(x)$. For an isotope whose number $A$ is greater than 1, i.e. for all elements heavier than hydrogen, we observe that the function $h(x)$ is monotonically increasing over $[\alpha, 1]$. For $A > 1$, the change of variables $y = h(x)$ is therefore well defined over the interval $[\alpha, 1]$. However, for hydrogen, the element that interests us, we have $A < 1$. We then observe that $h(x)$ is no longer monotonic over the interval $[\alpha, 1]$: $h(x)$ decreases from $x = \alpha$ to $x = \alpha^{1/2}$, reaches a minimum at $x = \alpha^{1/2}$, then increases from $x = \alpha^{1/2}$ to $x = 1$. In particular, for $A < 1$ we have:

$$h(\alpha) = h(1) = 1$$
$$h\left(\sqrt{\alpha} = \frac{1-A}{A+1}\right) = \sqrt{1-A^2} \quad \text{for } A < 1 \quad (L.6)$$

To use the change of variables $y = h(x)$ when $A < 1$, we must first subdivide the interval $[\alpha, 1]$ into two intervals, $[\alpha, \alpha^{1/2}]$ and $[\alpha^{1/2}, 1]$. We then obtain:

$$\int_{x=\alpha}^{1} \delta(\mu - h(x))dx = \int_{x=\alpha}^{\sqrt{\alpha}} \delta(\mu - h(x))dx + \int_{x=\sqrt{\alpha}}^{1} \delta(\mu - h(x))dx$$
$$= \int_{y=h(\alpha)=1}^{h(\sqrt{\alpha})=\sqrt{1-A^2}} \delta(\mu - y)\frac{dy}{h'(x)} + \int_{x=\sqrt{1-A^2}}^{h(1)=1} \delta(\mu - y)\frac{dy}{h'(x)} \quad (L.7)$$

where *h'(x)* is the derivative of *h(x)*, given by:

$$h'(x) = \frac{(A+1)x + A - 1}{4x^{3/2}} \quad (L.8)$$

It remains only to express *x* as a function of *y* in the integrals (L.7). To do this, we must solve the equation *y = h(x)*. For *A* < 1, this equation has two solutions, *x = g₊(y)* and *x = g₋(y)*, given by:

$$x = g_\pm(y), \quad g_\pm(y) = \left(\frac{y \pm \sqrt{y^2 + A^2 - 1}}{A+1}\right)^2 \quad (L.9)$$

We then show that: (i) function *g₊(y)* is increasing over *y* ∈ [(1−*A*²)^(1/2), 1], from *g₊*((1−*A*²)^(1/2)) = *α*^(1/2) to *g₊*(1) = 1; (ii) function *g₋(y)* is decreasing over *y* ∈ [(1−*A*²)^(1/2), 1], from *g₋*((1−*A*²)^(1/2)) = *α*^(1/2) to *g₋*(1) = *α*. In the integrals (L.7), we must hence take *x = g₋(y)* over the interval *x* ∈ [*α*, *α*^(1/2)], and *x = g₊(y)* over the interval *x* ∈ [*α*^(1/2), 1]. We thus obtain:

$$\int_{x=\alpha}^{1} \delta(\mu - h(x))dx = -\int_{y=\sqrt{1-A^2}}^{1} \frac{\delta(\mu - y)}{h'(g_-(y))}dy + \int_{x=\sqrt{1-A^2}}^{1} \frac{\delta(\mu - y)}{h'(g_+(y))}dy$$

$$= \frac{\Theta[\mu \in [\sqrt{1-A^2}, 1]]}{h'(g_+(\mu))} - \frac{\Theta[\mu \in [\sqrt{1-A^2}, 1]]}{h'(g_-(\mu))} \quad (L.10)$$

We then show that:

$$h'(g_\pm(\mu)) = \pm \frac{\sqrt{\mu^2 + A^2 - 1}}{2} \frac{1}{g_\pm(\mu)}$$

$$g_+(\mu) + g_-(\mu) = 2\frac{2\mu^2 + A^2 - 1}{(A+1)^2} \quad (L.11)$$

We thus obtain:

$$\int_{x=\alpha}^{1} \delta(\mu - h(x))dx = \frac{4}{(A+1)^2} \frac{2\mu^2 + A^2 - 1}{\sqrt{\mu^2 + A^2 - 1}} \Theta[\mu \in [\sqrt{1-A^2}, 1]] \quad (L.12)$$

Finally, by reinjecting this result in equation (L.5), then in (L.3), we obtain the scalar product in the left-hand term of (L.2):

$$\langle f_1(E_0 \to E, \mu(\underline{r}')), f_1(E_0 \to E, \mu(\underline{r}))\rangle$$

$$= \frac{\delta(\mu(\underline{r}') - \mu(\underline{r}))}{(1-\alpha)^2 E_0} \frac{4}{(A+1)^2} \frac{2\mu(\underline{r})^2 + A^2 - 1}{\sqrt{\mu(\underline{r})^2 + A^2 - 1}} \Theta[\mu(\underline{r}) \in [\sqrt{1-A^2}, 1]] \quad (L.13)$$

The calculation of the right-hand term of (L.2) uses similar steps. Using the formula (3.2.5) of $f_1$ then making the change of variables $x = E/E_0$, we obtain:

$$\langle f_1(E_0 \to E, \mu(\underline{r})), \phi_{obj}(E) \rangle = \frac{1}{1-\alpha} \int_{x=\alpha}^{1} \delta(\mu(\underline{r}) - h(x)) \phi_{obj}(E_0 x) dx \quad \text{(L.14)}$$

Then performing the change of variables $y = h(x)$ using the functions $g_\pm(y)$, we find:

$$\int_{x=\alpha}^{1} \delta(\mu - h(x)) \phi_{obj}(E_0 x) dx = \int_{x=\alpha}^{\sqrt{\alpha}} \ldots + \int_{x=\sqrt{\alpha}}^{1} \ldots$$

$$= -\int_{y=\sqrt{1-A^2}}^{1} \frac{\delta(\mu - y) \phi_{obj}(E_0 g_-(y))}{h'(g_-(y))} dy + \int_{x=\sqrt{1-A^2}}^{1} \frac{\delta(\mu - y) \phi_{obj}(E_0 g_+(y))}{h'(g_+(y))} dy \quad \text{(L.15)}$$

$$= \left[ \frac{\phi_{obj}(E_0 g_+(\mu))}{h'(g_+(\mu))} - \frac{\phi_{obj}(E_0 g_-(\mu))}{h'(g_-(\mu))} \right] \Theta\!\left[ \mu \in \left[\sqrt{1-A^2}, 1\right] \right]$$

Finally, using the result (L.11), then noticing that:

$$g_-(\mu) = g_+(-\mu) \quad \text{(L.16)}$$

we obtain the scalar product of the right-hand term of (L.2):

$$\langle f_1(E_0 \to E, \mu(\underline{r})), \phi_{obj}(E) \rangle = \frac{2\Theta\!\left[ \mu(\underline{r}) \in \left[\sqrt{1-A^2}, 1\right] \right]}{(1-\alpha)\sqrt{\mu(\underline{r})^2 + A^2 - 1}} Z(\mu(\underline{r})) \quad \text{(L.17)}$$

with:

$$Z(x) = \phi_{obj}(E_0 g(x)) g(x) + \phi_{obj}(E_0 g(-x)) g(-x)$$

$$g(x) = \left( \frac{x + \sqrt{x^2 + A^2 - 1}}{A + 1} \right)^2 \quad \text{(L.18)}$$

Now let us take cylindrical coordinates $\underline{r} = (r, \theta, z)$, whose axis Oz is the line SD passing through $\underline{r}_s = (0, 0, 0)$ and $\underline{r}_d = (0, 0, z_d)$, where $r$ is the distance from $\underline{r}$ to SD. By using the axial symmetry of the problem, we note that $\chi_1(\underline{r})$ does not depend on the angle $\theta$. The left-hand term of equation (L.2), which will be noted *LHT*, is thus rewritten:

$$LHT = \int_{\underline{r}' \in V} \frac{\chi_{1opt}(\underline{r}') d\underline{r}'}{\|\underline{r}' - \underline{r}_s\|^2 \|\underline{r}' - \underline{r}_d\|^2} \langle f_1(E_0 \to E, \mu(\underline{r}')), f_1(E_0 \to E, \mu(\underline{r})) \rangle$$

$$= \int_{r'=0}^{+\infty} \int_{\theta'=0}^{2\pi} \int_{z'=-\infty}^{+\infty} \Theta[\underline{r}' \in V] \frac{\chi_{1opt}(r', z') r' dr' d\theta' dz'}{(r'^2 + z'^2)(r'^2 + (z' - z_d)^2)} \langle f_1(E_0 \to E, \mu(r', z')), f_1(E_0 \to E, \mu(r, z)) \rangle \quad \text{(L.19)}$$

with, cf. (3.2.2):

$$\mu(r,z) = \frac{-r^2 + z(z_d - z)}{\sqrt{r^2 + z^2}\sqrt{r^2 + (z_d - z)^2}} \quad \text{(L.20)}$$

Let us now consider the case where the volume $V$ of the material is a screen of thickness $dz$, positioned between the planes $z = z_e - dz/2$ and $z = z_e + dz/2$. We thus have:

$$\Theta[\underline{r}' \in V] = \Theta[z_e - dz/2 \leq z' \leq z_e + dz/2] \quad \text{(L.21)}$$

For a thin screen, i.e. for $dz \to 0$, we note that:

$$\Theta[\underline{r}' \in V] = dz\, \delta(z' - z_e) \quad \text{(L.22)}$$

For $\underline{r} \in V$, cf. (L.2), we thus obtain, using the result (L.13):

$$LHT = \frac{8\pi dz}{(1-\alpha)^2 (A+1)^2 E_0} \frac{2\mu(r,z_e)^2 + A^2 - 1}{\sqrt{\mu(r,z_e)^2 + A^2 - 1}} \Theta\!\left[\mu(r,z_e) \in \left[\sqrt{1-A^2},1\right]\right]$$

$$\times \int_{r'=0}^{+\infty} \frac{\chi_{1opt}(r')\, r'\, dr'}{(r'^2 + z_e^2)(r'^2 + (z_d - z_e)^2)} \delta(\mu(r',z_e) - \mu(r,z_e)) \quad \text{(L.23)}$$

Then, using the property (O.7) of the Dirac function, cf. section O, we observe for a thin screen that:

$$\delta(\mu(r',z_e) - \mu(r,z_e)) = \delta(r - r') \left|\frac{d\mu(r,z_e)}{dr}\right|^{-1} \quad \text{(L.24)}$$

The left-hand term of (L.2) is thus:

$$LHT = \frac{8\pi dz}{(1-\alpha)^2 (A+1)^2 E_0} \frac{2\mu(r,z_e)^2 + A^2 - 1}{\sqrt{\mu(r,z_e)^2 + A^2 - 1}} \Theta\!\left[\mu(r,z_e) \in \left[\sqrt{1-A^2},1\right]\right] \left|\frac{d\mu(r,z_e)}{dr}\right|^{-1}$$

$$\times \frac{r}{(r^2 + z_e^2)(r^2 + (z_d - z_e)^2)} \chi_{1opt}(r) \quad \text{(L.25)}$$

We then show that the derivative of $\mu(r,z_e)$ is:

$$\frac{d\mu(r,z_e)}{dr} = -r\, \frac{z_d^2 (r^2 + z_e(z_d - z_e))}{(r^2 + z_e^2)^{3/2} (r^2 + (z_d - z_e)^2)^{3/2}} \quad \text{(L.26)}$$

Finally, using the results (L.17), (L.24), (L.25), then noticing that $(1-\alpha)(A+1)^2 = 4A$, we obtain the formula of $\chi_{1opt}(r)$, optimal concentration of hydrogen in the screen, which is written:

$$\chi_{1opt}(r) = C \frac{\Theta\left[\mu(r,z_e) \in \left[\sqrt{1-A^2},1\right]\right]}{2\mu(r,z_e)^2 + A^2 - 1} \frac{r^2 + z_e(z_d - z_e)}{\sqrt{r^2 + z_e^2}\sqrt{r^2 + (z_d - z_e)^2}} Z(\mu(r,z_e))$$

$$C = \frac{8\pi E_0 A z_d^2}{(1-\lambda)\|\phi_{obj}\| Q_0 \Sigma_1(E_0) dz}$$

(L.27)

Functions $Z(x)$ and $\mu(r,z)$ are given in equations (L.18) and (L.20). As for the multiplier $\lambda$ and thus the constant $C$, they are theoretically fixed by the constraint $\|\phi_1\| = 1$ in (K.1).

## Section M. Sensitivity to the mass ratio $A$

In this section, we show that the optimal density profile $\rho_{opt}(r)$ (3.2.6)-(3.2.8) and the spectrum $\phi_1(E,\underline{r}_d)$ are very sensitive to the value taken for $A$.

To illustrate this point, we plot in fig. M.1 on the left the densities $\rho_{opt}(r)$ obtained with equation (3.2.6)-(3.2.8) for: (i) two very close values of $A$, 0.99917 and 1; (ii) a point isotropic source of 14 MeV neutrons placed at the point $r = z = 0$; (iii) a distance between the source and the detector $z_d = 100$ cm; (iv) a screen placed halfway between the source and the detector, at $z_e = 50$ cm; (v) a Gaussian objective spectrum, $\phi_{obj}(E) = \exp(-(E-E_{obj})^2/(2\sigma^2))$, centered on $E_{obj} = 100$ keV, with a width $\sigma$ equal to 20 keV. The product $\beta C$ of the constants appearing in (3.2.6)-(3.2.8) is adjusted so that the maximum of these two densities $\rho_{opt}(r)$ be equal to $\rho_{max} = 1.0$ g/cm$^3$. Then, in fig. M.1 on the right, we show the two energy spectra $\phi_1(E)$ calculated with MCNP in the detector D, modeled by a sphere of radius 0.5 cm and of center $r = 0$, $z = z_d$, by interposing between the source and the detector two screens of thickness $dz = 0.1$ cm, comprised between the planes $z = 49.95$ cm and 50.05 cm, which contain the densities $\rho_{opt}(r)$ of hydrogen plotted fig. M.1 on the left.

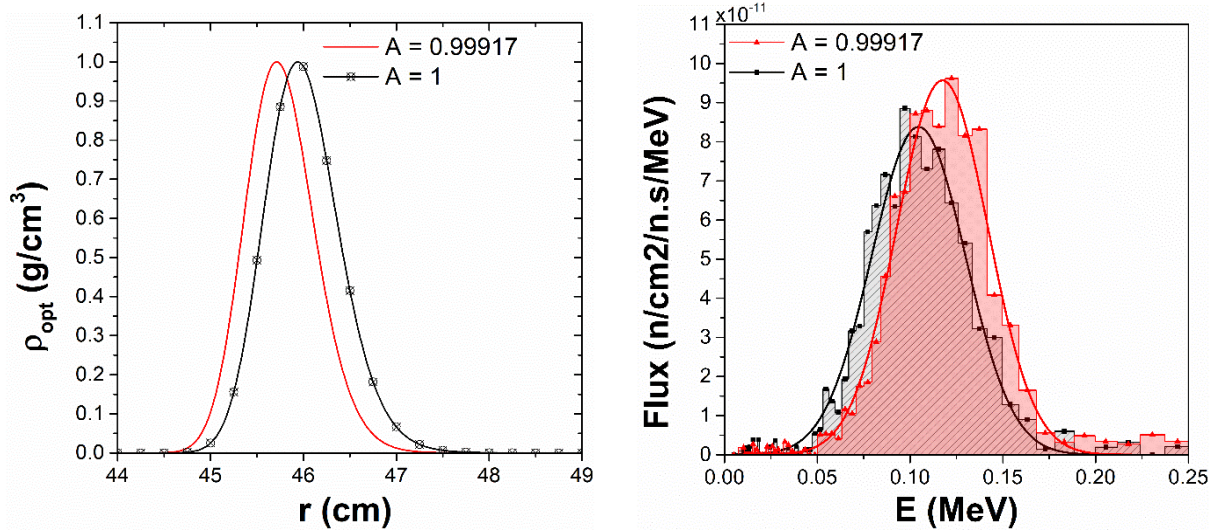

**Figure M.1.** (Left) densities $\rho_{opt}(r)$ obtained for $A = 0.99917$ (red line) and $A = 1$ (black line + circles); (right) spectra $\phi_1(E)$ calculated with MCNP in the detector cell D for the theoretical densities $\rho_{opt}(r)$ drawn on the left side, for $A = 0.99917$ (red triangles) and $A = 1$ (black squares). These spectra are fitted by Gaussians of formula $\exp(-(E-E_c)^2/(2w^2))$. The fitted parameters $E_c$ and $w$ obtained are: (i) $E_c = 117.4 \pm 0.7$ keV and $w = 24.6 \pm 0.6$ keV for $A = 0.99917$; (ii) $E_c = 104.6 \pm 0.8$ keV and $w = 25.3 \pm 0.8$ keV for $A = 1$.

In fig. M.1, we verify that the spectra $\phi_1(E)$ calculated with MCNP using the densities $\rho_{opt}(r)$ plotted on the left are correctly fitted by Gaussians. The widths of these Gaussians, about 25 keV for both, are a little larger than the width of the objective spectrum, 20 keV. This widening is due to the non-zero radius of the detector D as well as the non-zero thickness of the screen used in the MCNP calculations. We also observe that the centroids of the Gaussians do not coincide perfectly with that of the objective Gaussian, $E_{obj} = 100$ keV. There must therefore be a small positioning error in the formula (3.2.8), an explanation of which will be given in section 3.2.2. Finally, we note that approaching $A = 0.99917$ by 1, i.e. modifying $A$ by 0.083% only, in the formula of $\rho_{opt}$ results in a 0.23 cm shift of the position (~46 cm) of its maximum. This shift induces in turn a 12.8 keV shift, i.e. a 11.5% modification, of the centroid of the spectrum $\phi_1(E)$ calculated with MCNP. The sensitivity of the spectrum to the value taken for $A$ is thus considerable. In order to be able to finely compare the predictions of algorithm A1 to the theoretical predictions (3.2.8), it is therefore imperative to use the same value of $A$ in the MCNP calculations that in the theoretical calculations. It was thus mandatory to make in section L the effort to calculate $\chi_{1opt}(r)$ for the difficult case $A < 1$.

## Section N. Design of a fast neutron monochromator

In this section, we apply the resolution procedure described in section 3.1 to solve a challenging problem. Suppose that the neutron source operated is anew a standard 14 MeV D-T generator, installed at a location $r_s$ of a work area. The users of this source have in stock a quantity $P_{max}$ of material, and would like to machine it so that the neutrons that pass through it all have the same, unique energy, $E_{obj} = 100$ keV, at a point $r_d$. Is such a monochromator of fast neutrons feasible, and if so, how to design it?

By taking the volume densities $\rho$ of the material in its cells $\Theta$ of volumes $V$ as the parameters $x$ to be optimized, the problem to be solved is a type (3.1.5) problem, with:

$$Q(r,E,\Omega) \propto \delta(r-r_s)\delta(E-E_0)$$
$$P(x=\rho) = \rho.V, \quad \phi_{obj}(E) \propto \delta(E-E_{obj}) \qquad (N.1)$$
$$V_D \to 0 \Rightarrow \phi(E, x=\rho) = \int_{\Omega \in 4\pi} \varphi(r_d, E, \Omega, x=\rho) d\Omega$$

with $E_0 = 14$ MeV and $E_{obj} = 100$ keV. At first glance, the problem (3.1.5)+(N.1) seems difficult, but it has the interesting property of possessing a simple analytical solution. We will find this solution in the following section, N.1, then we will use it as a reference to test in section N.2 the topology optimization procedure proposed in section 3.1.

N.1. Reference solution of the problem (3.1.5)+(N.1)

To solve the problem (3.1.5)+(N.1), let us position a small volume of matter $dV$ at a point $r$. To simplify the calculations, we will suppose that the material contained in $dV$ is monoatomic, and outside $dV$, we will suppose that the space is devoid of any matter. Neutrons emitted by the source will have a probability of passing by $dV$, of being scattered in it, and of passing by the point $r_d$. In section J, we showed that the total flux at the point $r_d$ of neutrons that have passed by $dV$ is equal to, cf. (J.11)+(3.2.4):

$$\phi_1(E) = \frac{Q_0 \Sigma_s(E_0) dV}{8\pi^2 \|\underline{r}_d - \underline{r}\|^2 \|\underline{r} - \underline{r}_s\|^2} f\left(E_0 \to E, \frac{\underline{r} - \underline{r}_s}{\|\underline{r} - \underline{r}_s\|} \cdot \frac{\underline{r}_d - \underline{r}}{\|\underline{r}_d - \underline{r}\|}\right) \quad (N.2)$$

where $\Sigma_s(E)$ is the macroscopic scattering cross-section at energy $E_0$, and $f(E' \to E, \underline{\Omega}.\underline{\Omega}')/(2\pi)$ is the probability density of transition from the state $(E',\underline{\Omega}')$ to the state $(E,\underline{\Omega})$. The result (N.2) is valid for a monoenergetic isotropic point source emitting $Q_0$ neutrons of energy $E_0$ at the point $\underline{r}_s$, provided: (1) that the material contained in $dV$ is little absorbing at 14 MeV; (2) that $dV$ is small enough so that a neutron that performs a collision in it has an almost zero probability of making another collision in it; (3) that $dV$ does not cross the line passing by $\underline{r}_s$ and $\underline{r}_d$. Suppose now that the scatterings performed by the neutrons in $dV$ are elastic. In this case, for a monoatomic material, the distribution $f$ can be approximated using (3.2.5), and we have:

$$\phi_1(E) = \frac{Q_0 \Sigma_s(E_0) dV \Theta[\alpha \leq E/E_0 \leq 1]}{8\pi^2 (1-\alpha) E_0 \|\underline{r}_d - \underline{r}\|^2 \|\underline{r} - \underline{r}_s\|^2} \delta\left(\frac{\underline{r}_d - \underline{r}}{\|\underline{r}_d - \underline{r}\|} \cdot \frac{\underline{r} - \underline{r}_s}{\|\underline{r} - \underline{r}_s\|} - \frac{A+1}{2}\sqrt{\frac{E}{E_0}} + \frac{A-1}{2}\sqrt{\frac{E_0}{E}}\right) \quad (N.3)$$

where the parameters $A$ and $\alpha$ is given in (3.2.5). We thus observe that the flux $\phi_1(E)$ is always zero, except: (i) if $\alpha E_0 \leq E \leq E_0$; (ii) and if the term in the Dirac function in (N.3) is equal to zero. The problem (3.1.5)+(N.1) therefore possesses a very simple solution: in order to have only neutrons with an energy $E_{obj}$ at the point $\underline{r}_d$ after their passing through a volume element $dV$, one must position $dV$ at a point $\underline{r}$ satisfying the following equation:

$$\underline{\Omega}.\underline{\Omega}' = \frac{\underline{r}_d - \underline{r}}{\|\underline{r}_d - \underline{r}\|} \cdot \frac{\underline{r} - \underline{r}_s}{\|\underline{r} - \underline{r}_s\|} = \frac{A+1}{2}\sqrt{\frac{E_{obj}}{E_0}} - \frac{A-1}{2}\sqrt{\frac{E_0}{E_{obj}}} \quad (N.4)$$

As for condition (i), it imposes a minimum and a maximum bound on the energies $E_{obj}$ that can be obtained with this solution, which are:

$$\alpha E_0 \leq E_{obj} \leq E_0 \quad (N.5)$$

A more rigorous demonstration of the result (N.4)-(N.5) is given in Supplemental Material O. The equation (N.4) can be easily solved using the law of sines. By applying this law to the triangle of vertices S, M and D, of resp. coordinates $\underline{r}_s$, $\underline{r}$ and $\underline{r}_d$, we note that the set of points $\underline{r}$ satisfying the equation (N.4), i.e. the shape of the fast neutron monochromator sought, is a surface $S$ generated by revolving around the axis SD a circle $C$ of radius $R$ passing by the points S and D, cf. fig. N.1. The radius $R$ of this circle is:

$$R = \frac{H}{2}\frac{1}{\sin(\underline{\Omega},\underline{\Omega}')} = \frac{H}{2}\left[1 - \left(\frac{A+1}{2}\sqrt{\frac{E_{obj}}{E_0}} - \frac{A-1}{2}\sqrt{\frac{E_0}{E_{obj}}}\right)^2\right]^{-1/2} \quad (N.6)$$

By taking cylindrical coordinates $\underline{r} = (r, \theta, z)$, whose axis Oz is the axis SD, whose coordinate $r$ is the distance from $\underline{r}$ to the axis SD, and whose origin $r = z = 0$ is the point S, the equation of this circle is:

$$r = \left|\pm\sqrt{R^2 - (z - H/2)^2} - \sqrt{R^2 - (H/2)^2}\right|, \quad -R + H/2 \leq z \leq R + H/2 \quad (N.7)$$

The surface $S$ can be divided into two surfaces, $S_1$ and $S_2$. The surface $S_1$ is the surface generated by the rotation of the arc of circle $C_1$, delimited by the points S and D and having the shortest perimeter, see fig. N.1. The surface $S_2$ is the surface generated by the rotation of the arc of circle $C_2$, delimited by the points S and D and having the longest perimeter, see fig. N.1.

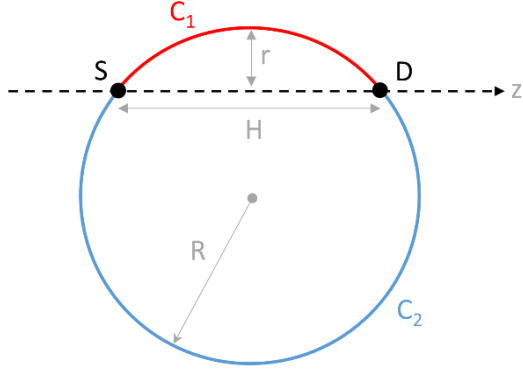

**Figure N.1.** Arcs of circle $C_1$ and $C_2$ whose rotation around the axis SD generates the surface $S$ of the fast neutron monochromator.

In the problem (3.1.5)+(N.1), we took $E_0$ = 14 MeV and $E_{obj}$ = 100 keV. For these energies, we note that the only material satisfying the condition (N.5), i.e. $E_{obj}/E_0 = 7.1 \cdot 10^{-3} \geq \alpha$, is hydrogen. Indeed, for $^1$H, we have $A = 0.99917$ so $\alpha \approx 1.7 \cdot 10^{-7}$. Whereas for the next isotope in order of mass, deuterium $^2$H, we have $A = 1.9968$ so $\alpha \approx 0.11 > E_{obj}/E_0$. For $^1$H, however, we observe that the solution surface $S$ has to be truncated. Indeed, because (see discussion from result (L.4) to result (L.6) in section L)

$$\sqrt{1-A^2} \leq \frac{A+1}{2}\sqrt{\frac{E_{obj}}{E_0}} - \frac{A-1}{2}\sqrt{\frac{E_0}{E_{obj}}} \leq 1, \forall E_{obj} \in [\alpha E_0, E_0] \quad (N.8)$$

when $A \leq 1$, we note that $\underline{\Omega}.\underline{\Omega}'$ is always positive for a neutron-proton collision, so a neutron hitting a proton cannot be backscattered. As a result, none of the neutrons colliding in surface $S_2$ will be able to reach the point $\underline{r}_d$, so this surface must be removed. In conclusion, for $E_0$ = 14 MeV and $E_{obj}$ = 100 keV, a simple solution to the problem (3.1.5)+(N.1) is a thin layer of hydrogen deposited on the surface $S_1$, of equation:

$$r = \sqrt{R^2 - (z-H/2)^2} - \sqrt{R^2 - (H/2)^2}, \quad 0 \leq z \leq H \quad (N.9)$$

*Comment.* Giving the values of the coefficients $A$ of $^1$H or $^2$H with a large number of digits after the decimal point is not a luxury. Indeed, cf. section M, the sensitivity to $A$ of the neutron energy at the point $\underline{r}_d$ after a collision in the monochromator is very large. Hence, if an experimental test or an application of the result (N.9) were to be conducted, the required accuracy on $A$ would probably be such that the contribution of the mass of the electron of $^1$H to $A$ would have to be taken into account. By default, in the data libraries distributed with MCNP, the coefficient $A$ includes this mass, perhaps because neutrons used in reactor physics are often thermal. However, for neutron-proton collisions involving fast neutrons, this mass should not be added.

N.2. Resolution of the problem (3.1.5)+(N.1) with algorithm A1

We propose now to calculate the shape of the fast neutron monochromator using the topology optimization procedure described in section 3.1, then to compare the results thus obtained with the solution (N.9). As in section 2.2, to implement this procedure, we start by tiling the space using a union of cells $\Theta_i(\underline{r})$, parametrized in cylindrical coordinates as follows:

$$\Theta_i(\underline{r}) = \Theta[R_{j-1} \leq r < R_j] \times \Theta[X_k \leq z < X_{k+1}]$$
$$X_k = -\frac{H}{3} + \frac{2H}{3NX}k, \quad R_j = jR_{min}, \quad R_{min} = \frac{2H}{3NR} \quad \text{(N.10)}$$
$$i = NXj + k - NX + 1, \quad k \in [0, NX], \quad j \in [1, NR]$$

In the MCNP input, the 14 MeV neutron source is positioned at the point $\underline{r}_s$ of coordinates $r = 0$, $z = -H/2$. The point detector $\underline{r}_d$ is modeled by a sphere D of radius $R' = 0.5$ cm, whose center's coordinates are $r = 0$, $z = H/2$. The value taken for $R'$ is a compromise: (i) we must take a value of $R'$ as small as possible, as the result (N.9) is valid for a point detector $\underline{r}_d$ i.e. for $R' \to 0$; (ii) but, in the MCNP calculations, we cannot use an infinitely small radius $R'$ neither, because this will induce large statistical errors on the energy spectrum in D, as the trajectories of neutrons passing by D would be too few. In (N.10), the cells obtained for $j = 1$, contained in the area $r < R_{min}$, are left empty, in order to ease the subtraction of the contribution $\phi_0$ (of the source neutrons that reach $\underline{r}_d$ in ballistic flight without performing a collision), i.e. the peak of source neutrons at 14 MeV, from the energy spectrum. Using this trick, we impose in the MCNP input that the source emits $NPS = 10^{10}$ neutrons, whose directions are isotropically sampled in the portion of solid angle delimited by the angles $\theta_{min}$ and $\theta_{max}$ defined by $\tan\theta_{min} = 6/5NR$ [8] and $\tan\theta_{max} = 4$. This procedure allows to save computing time: (i) by forcing the neutrons to pass through the structure (N.10); (ii) by eliminating the need for computing the difference $\phi - \phi_0 = \phi_1$ of the spectra with and without the structure. For this study, we take $H = 20$ cm and $NX = NR = 10$. In accordance with the conclusions of the previous section, we fill the cells $\Theta_i$ with a material made of 100% of $^1$H in the MCNP input, with volume densities $\rho_i$ that can range from $\rho_{min} = 10^{-5}$ g/cm$^3$ to $\rho_{max} = 0.01$ g/cm$^3$. The value taken here for $\rho_{max}$ is also a compromise: (i) it must be small enough to minimize the probability of neutrons undertaking multiple collisions in the structure. Indeed, according to the main hypothesis of section N.1, at the origin of the solution (N.9), the neutrons must perform at most one collision in the material; (ii) but it cannot be too small neither, otherwise the number of collisions in the structure will be small, which will induce large statistical fluctuations on the energy spectrum. The other parameters of the modeling are identical to those used in section 3.2.2: (i) we reuse the energy binning of the histograms of $\phi$ and $\phi_{obj}$ given in (3.2.10), with $NBIN = 41$, $E_{min} = 10^{-6}$ MeV and $E_{max} = 20$ MeV; (ii) we reuse the histogram of the objective spectrum given in equation (3.2.11),; (iii) we retake $\omega_j = 1 \ \forall j$ in the expression of the scalar product $<\phi, \phi_{obj}>$, cf. (3.1.8); (iv) we retake $P_{max} = +\infty$ in algorithm A1; (iv) we reuse the threshold at 90% in the filter on the PERT$ij$ values. The density maps $\rho_i(n)$ obtained with algorithm A1 are shown in fig. N.2 on the left for some of its iterations. The gray scale of the figures gives the density values in g/cm$^3$, in logarithmic scale. As before, the 3D structures are generated by revolving these maps around axis SD. These results were obtained using: (i) the logarithmic quantization (2.2.2) of the density, with $M = 50$; (ii) an uniform initial density profile, $\rho_i(0) = \rho(0)\ \forall i$, with $\rho(0)$ given in (2.2.6). In the figures, the sphere of radius $R' = 0.5$ cm modeling the point detector $\underline{r_d}$ is indicated by a circle to the right of the structures, and the analytical solution (N.9) is indicated by the red line. On the right side of fig. N.2, we show the histograms of the spectra $\phi_1(E)$ obtained in D for each structure $\rho_i(n)$ plotted on the left. These spectra are normalized to 1 and are compared to the histogram (3.2.11) of the objective spectrum $\phi_{obj}(E)$. For information, the total neutron flux obtained in D is equal to $4.0\ 10^{-8}$ ($\pm 3.6\%$) n/cm² per source neutron at iteration $n = 100$. The mutual evolution of the structure and of the spectrum $\phi_1(E)$ throughout the iterations $n$ of A1 is shown in a video, mimic_gauss_bulk.mp4, given in Supplemental Material. Finally, we plot fig. N.3 the evolution of the objective $O\phi = 1/2 - 1/2 <\phi_1,\phi_{obj}>/\|\phi_1\|\|\phi_{obj}\|$, cf. (3.1.5), with the iteration number $n$. This figure helps visualize the convergence of $\phi_1$ towards $\phi_{obj}$. As a reminder, $O\phi$ is by construction comprised between 0 and 1, and the closer it is to 0, the closer $\phi_1$ is to $\phi_{obj}$.

---

[8] Logically, we should have taken $\tan\theta_{min} = 4/5NR$. However, because of a typo in one of our programs, identified tardily, we used the value $\tan\theta_{min} = 6/5NR$ instead. This error has no impact on the results shown in fig. N.2-N.3, because the cells $\Theta_i$ activated by the algorithm are located very far from the edge $r = R_{min}$ of the structure. In fact, this error is even beneficial, because it reduces a little the noise on the spectrum induced by grazing collisions undertaken by some neutrons in the cells of density $\rho_{min}$ close to the edge $r = R_{min}$.

The results shown in fig. N.2 are encouraging. We observe that the structure calculated with A1 converges towards the reference solution (N.9): the cells with densities $\rho_i > \rho_{min}$ are those crossed by the theoretical surface indicated by the red line. However, we note fig. N.3 that the convergence of $\phi_1$ towards $\phi_{obj}$ is middling, with $O\phi$ reaching 0.5 at best. Indeed, we observe in fig. N.2 that the histograms of $\phi_1(E)$ are noised at high energy, around 14 MeV, and at low energy, around and below $E_{obj} = 100$ keV. The high-energy noise is due to neutrons that perform grazing angle collisions in the cells with density $\rho_{min} = 10^{-5}$ g/cm³ that border the axis SD. The probability of these collisions appears to be of the same order of magnitude as the probability that a neutron collides in the distant $\rho_{max}$ area, scatters at large angle, then crosses the structure to reach the detector. This noise at high energy is thus an artifact of our topology optimization procedure, due to the fact that we cannot take $\rho_{min} = 0$ strictly without causing an MCNP error. The low-energy noise, on the other hand, is induced by neutrons performing multiple collisions in the structure. According to the calculations in section N.1, the objective $\phi_{obj}(E) = \delta(E-E_{obj})$ can only be achieved if the neutrons perform at most one collision in the structure, in the surface (N.9). We note fig. N.2 that algorithm A1 intelligently disadvantages these multiple collisions by reducing the density of the eligible cells in the central part of the structure (caution, the density gray scale is logarithmic). However, A1 cannot succeed in eliminating completely the noise at low energy, because the latter is also an artifact of the optimization procedure. This noise results from the fact we try to approach the solution of the problem, which is an infinitely thin surface, using a rather coarse assembly of cylindrical rings of thickness $dr = 2H/3NR = 1.3$ cm and of width $dz = 2H/3NX = 1.3$ cm, which are both not completely negligible in front of the characteristic size of the problem, ~20 cm. This non-negligible size of the rings $\Theta_i$ induces another perturbation: as it can be seen in the histograms of the spectra, it widens the objective dirac, $\delta(E-E_{obj})$, into a (deformed) Gaussian. This widening is also due to the fact that we model the detector $\underline{r_d}$, which is theoretically point-like, using a sphere of radius $R' = 0.5$ cm in order to reduce the computing time.

In this section, we have obviously tried to limit these disturbances by reducing the density $\rho_{max}$, the radius $R'$, and/or the size of the cylindrical rings by increasing $NR$ and $NX$. However, the computing time with the parameters used here, $NR = NX = 10$, $NPS = 10^{10}$, $\rho_{max} = 0.01$ g/cm³, is already quite long, ~11 to 24 hours per iteration on one of our 16-CPU machines. Since we could not go much further with the computing power at our disposal, we have elaborated a second test of the algorithm A1, more precise, presented in section 3.2.

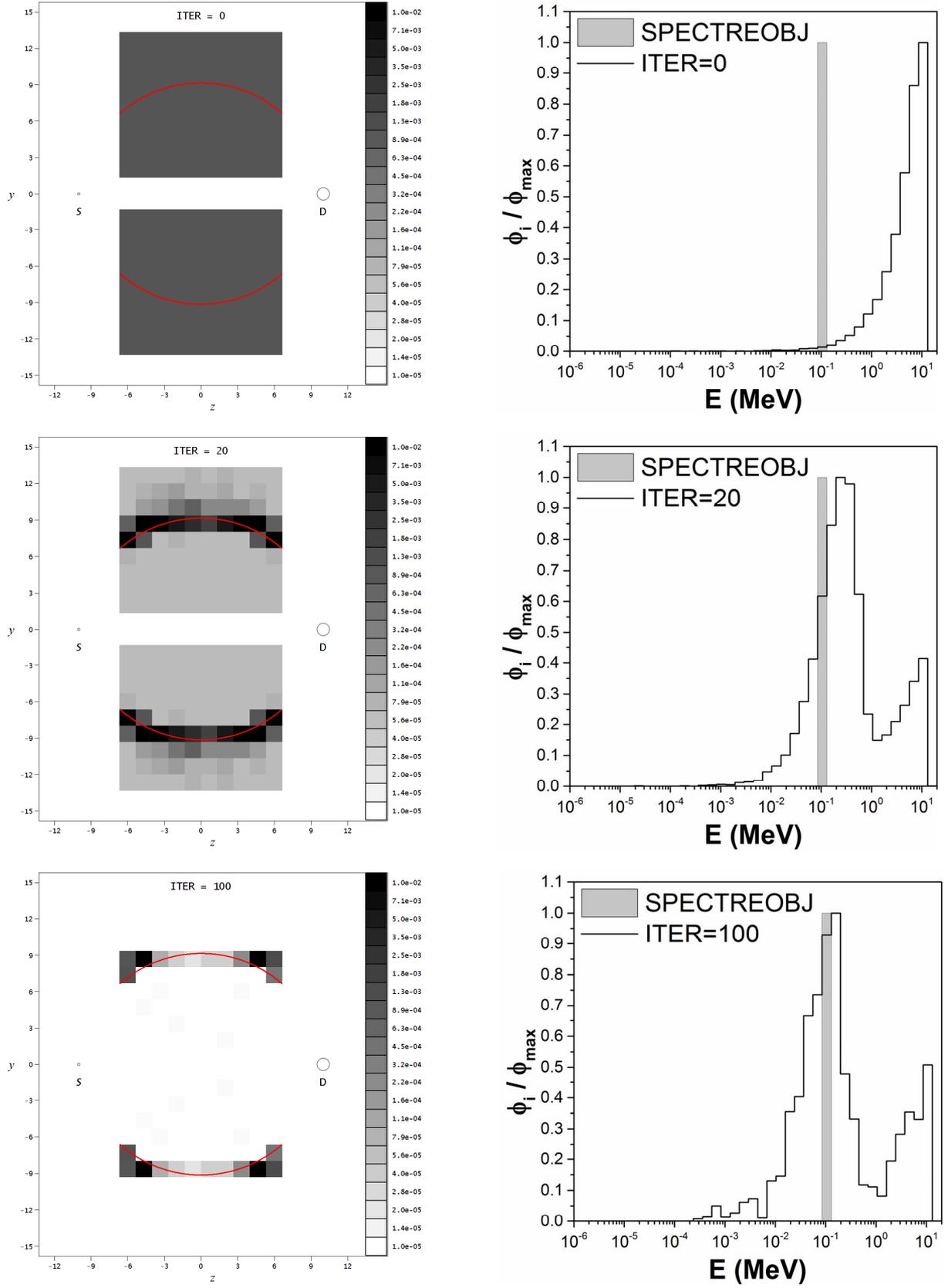

**Figure N.2.** (Left) density maps obtained at iterations 0, 20, 100 of algorithm A1. The units of axes $y = \pm r$ and $z$ are cm. The gray scale gives the values of the densities $\rho_i(n)$ in g/cm$^3$. The line in red is the reference solution (N.9); (right) histograms of the spectrum $\phi_i(E)$, normalized to 1 and compared to the histogram (N.12) of the objective spectrum $\phi_{obj}$ with $E_{obj}$ = 100 keV (gray area).

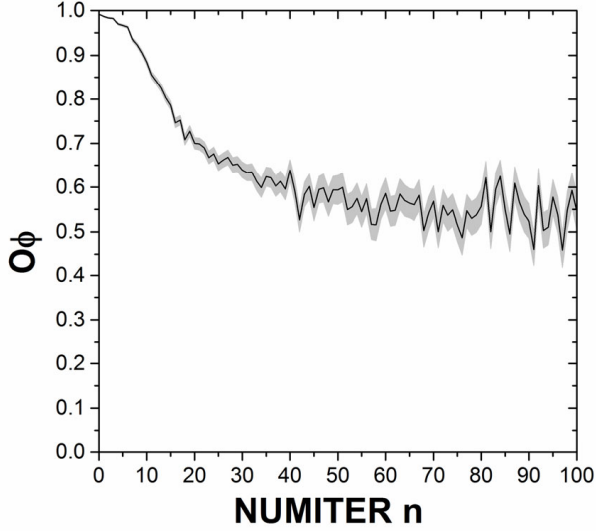

**Figure N.3.** Evolution of $O\phi$ with the iteration number $n$. The gray area gives the statistical error on $O\phi$.

## Section O. Demonstration of the result (N.4)-(N.5)

The flux $\phi_1(E)$ of neutrons: (i) generated at the point $\underline{r}_d$ by a monoenergetic source of energy $E_0$, isotropic, point-like, positioned at $\underline{r}_s$; (ii) that pass through an infinitesimal volume of matter $dV$ positioned at a point $\underline{r}$, of macroscopic scattering cross-section $\Sigma_s(E)$; (iii) that perform in $dV$ at most one elastic scattering, supposedly isotropic in the center of mass frame, is given in (N.3). We want to find the set of points $\underline{r}$ for which $\phi_1(E)$ be as close as possible to the objective spectrum $\phi_{obj}(E) = \delta(E-E_{obj})$. In section 3.1, we have defined a distance, $d(\phi,\phi_{obj})$, between two spectra $\phi$ and $\phi_{obj}$, cf. (3.1.3). However, in this section, we propose a particular, more elegant solution, based on the properties of the Dirac function. Let us take here the following objective:

$$O\phi = (\langle E \rangle - E_{obj})^2 + \sigma^2 \quad (O.1)$$

where $\langle E \rangle$ is the average energy of the spectrum $\phi$, and $\sigma^2$ its variance, defined by:

$$\langle E \rangle = \frac{\int_{E=0}^{+\infty} E\phi(E)dE}{\int_{E=0}^{+\infty} \phi(E)dE}, \quad \sigma^2 = \langle E^2 \rangle - \langle E \rangle^2 = \frac{\int_{E=0}^{+\infty} E^2\phi(E)dE}{\int_{E=0}^{+\infty} \phi(E)dE} - \langle E \rangle^2 \quad (O.2)$$

We note that $O\phi \geq 0 \; \forall \phi$, the equality $O\phi = 0$ being reached iff $\phi(E) \propto \delta(E-E_{obj})$. Finding the set of points $\underline{r}$ that minimizes the distance between $\phi_1(E)$ and $\phi_{obj}(E) = \delta(E-E_{obj})$ hence amounts to solving the following optimization problem:

$$\min_{\underline{r}} O\phi \quad (O.3)$$

To solve (O.3), we first note that the objective $O\phi$ can be rewritten in a simpler way, by expanding the term $(\langle E \rangle - E_{obj})^2$:

$$O\phi = \left(\langle E\rangle^2 - 2E_{obj}\langle E\rangle + E_{obj}^2\right) + \langle E^2\rangle - \langle E\rangle^2 = \left\langle (E - E_{obj})^2\right\rangle \quad (O.4)$$

Now let us inject the expression (N.3) of $\phi_1(E)$ in the objective (O.4). We obtain:

$$O\phi = \frac{\int_{E=\alpha E_0}^{E_0}(E - E_{obj})^2 \delta(f(E))dE}{\int_{E=\alpha E_0}^{E_0} \delta(f(E))dE} \quad (O.5)$$

where function $f(E)$ is given by:

$$f(E) = \frac{r_d - r}{\|r_d - r\|} \cdot \frac{r - r_s}{\|r - r_s\|} - \frac{A+1}{2}\sqrt{\frac{E}{E_0}} + \frac{A-1}{2}\sqrt{\frac{E_0}{E}} \quad (O.6)$$

But the Dirac function has the following property:

$$\delta(f(E)) = \sum_i \delta(E - E_i)|f'(E_i)|^{-1} \quad (O.7)$$

where $f'$ is the derivative of $f$, and where the energies $E_i$ are (simple) roots of $f(E)$, defined by:

$$f(E_i) = 0 \quad (O.8)$$

By injecting (O.7) into (O.5), we obtain:

$$O\phi = \frac{\sum_i (E_i - E_{obj})^2 |f'(E_i)|^{-1} \Theta[\alpha E_0 \leq E_i < E_0]}{\sum_i |f'(E_i)|^{-1} \Theta[\alpha E_0 \leq E_i < E_0]} \quad (O.9)$$

We note that the absolute minimum of $O\phi$, $O\phi = 0$, is reached: (i) if $f(E)$ has a single root $E_i$ in the interval $[\alpha E_0, E_0]$; (ii) if this root $E_i$ is equal to $E_{obj}$. Since the absolute minimum $O\phi = 0$ is reached iff $\phi_1(E) \propto \delta(E - E_{obj})$, we thus show that the set of points $r$, i.e. the shape of the fast neutron monochromator sought, obeys the equation $E_i = E_{obj}$, provided $E_{obj} \in [\alpha E_0, E_0]$, i.e. is given by, cf. (O.8):

$$f(E_{obj}) = 0 \Leftrightarrow \frac{r_d - r}{\|r_d - r\|} \cdot \frac{r - r_s}{\|r - r_s\|} - \frac{A+1}{2}\sqrt{\frac{E_{obj}}{E_0}} + \frac{A-1}{2}\sqrt{\frac{E_0}{E_{obj}}} = 0 \quad (O.10)$$

and $\alpha E_0 \leq E_{obj} \leq E_0$

We retrieve the result (N.4)-(N.5).

## Section P. Resolution of the problem (4.3)

The Lagrangian of the problem (4.3) can be written:

$$L(\underline{\chi}, \lambda) = \underline{\chi}.\underline{V} - \lambda(k_{eff} - 1) \quad (P.1)$$

The optimal distribution of fuel $\chi_{opt}$ thus obeys the following system:

$$\left.\frac{\partial L}{\partial \underline{\chi}}\right|_{\underline{\chi}=\underline{\chi}_{opt}} = \underline{0}, \quad \left.\frac{\partial L}{\partial \lambda}\right|_{\underline{\chi}=\underline{\chi}_{opt}} = 0 \quad (P.2)$$

The equations (P.2) impose that:

$$\frac{\partial L}{\partial \chi_i} = V_i - \lambda \frac{\partial k_{eff}}{\partial \chi_i} = 0, \forall i \implies C_i = \frac{1}{V_i}\frac{\partial k_{eff}}{\partial \chi_i} = \frac{1}{\lambda} = \lambda', \forall i \quad (P.3)$$

As before, the solution of the optimization problem, here the optimal distribution of fuel $\chi_{opt}$, is given by an isovalue of coefficients $C_i$, here $(1/V_i)\partial k_{eff}/\partial \chi_i$, which verifies the constraint of the problem, here $k_{eff} = 1$. To solve the problem (4.3), we hence have to calculate these derivatives.

### P.1. Calculation of the derivatives $\partial k_{eff}/\partial \chi_i$ using the MCNP code

As for the previous problems in sections 2-3, the $N$ derivatives $\partial k_{eff}/\partial \chi_i$ can all be obtained in a single MCNP calculation. The procedure to follow is summarized here. The MCNP calculation must be performed in N mode and in KCODE mode, see ref. [26], in particular its section 5.5.2. As we are studying a reactor, a TOTNU card must be used, so that the contribution of the delayed neutrons be taken into account. In KCODE mode, the MCNP output file o gives the estimated $k_{eff}$ of the reactor. For this study, we will use the estimator of the $k_{eff}$ constructed by combining the collision, absorption and tracking estimators, see [25] section VIII.B. The thermalization of neutrons in water is modeled using a MT card. Finally, the source neutrons are distributed over the volume of the reactor using a KSRC card, see [26] section 5.5.3.

Suppose now that, after an iteration $n$ of the optimization algorithm to be written, a vector $\underline{\chi}(n)$ has been obtained, which is still suboptimal. A new iteration has to be performed. To do this, we will write a new MCNP input, in which we will define for each tube $\Theta_i$ two materials, using two cards M (see [26] section 5.3.1), numbered $M_i$ and $M_{N+i}$ and given by:

| | | |
|---|---|---|
| M$i$ | 92235.70c | $\varepsilon \chi_i(n)$ |
| | 92238.70c | $(1-\varepsilon)\chi_i(n)$ |
| | 1001.70c | $(2/3)(1-\chi_i(n))$ |
| | 8016.70c | $(1/3)(1-\chi_i(n))$ |
| M$i$+N | 92235.70c | $\varepsilon(\chi_i(n)+\Delta\chi)$ |
| | 92238.70c | $(1-\varepsilon)(\chi_i(n)+\Delta\chi)$ |
| | 1001.70c | $(2/3)(1-(\chi_i(n)+\Delta\chi))$ |
| | 8016.70c | $(1/3)(1-(\chi_i(n)+\Delta\chi))$ |

where $\varepsilon$ is the enrichment degree of the fuel, here 0.03, and $\Delta\chi$ a fraction increment, for example 0.05. The card M$_i$ recalls the isotopic composition of the material obtained at the previous iteration *n* of the algorithm for the tube n°*i*. It is to be used in the writing of the MCNP cell n°*i* that defines the tube $\Theta_i$. The card $M_{i+N}$ gives a perturbed composition of this material, which is needed to calculate the corresponding derivative $\partial k_{eff}/\partial \chi_i$. The index of this card is shifted by *N*, as there are already *N* cards $M_i$, numbered from 1 to *N*, one per tube *i*. At each pair of cards $M_i$ and $M_{i+N}$ then corresponds a pair of volume densities $\rho_i$ and $\rho_{i+N}$, given by:

$$\begin{aligned}\rho_i(n) &= \chi_i(n)\rho_U + (1-\chi_i(n))\rho_{H_2O} \\ \rho_{i+N}(n) &= (\chi_i(n)+\Delta\chi)\rho_U + (1-(\chi_i(n)+\Delta\chi))\rho_{H_2O}\end{aligned} \quad (P.1.1)$$

where $\rho_U$ is the density of enriched uranium, and $\rho_{H2O}$ that of water. The density $\rho_i$ is to be used in the writing of the MCNP cell n°*i*, and the density $\rho_{i+N}$ is needed to calculate the corresponding derivative $\partial k_{eff}/\partial \chi_i$. For this task, we must add in the MCNP input file one PERT card per tube *i*, given here by:

PERT*i*:n  CELL=*i*  MAT=*i*+N  RHO=−$\rho_{i+N}$  METHOD=2

After calculation, the MCNP output file o contains a table of three columns, which gives at each line the PERT number *i*, the associated perturbation $dk_i = (\partial k_{eff}/\partial \chi_i)\Delta\chi$, and the statistical error on $dk_i$. We have thereby access to the *N* derivatives $\partial k_{eff}/\partial \chi_i = dk_i/\Delta\chi$ sought. We can now write an optimization algorithm then test it.

P.2. Algorithm for optimizing the fuel distribution

The optimal fuel distribution $\chi_{opt}$ is given by the isovalue $\lambda$' of coefficients $C_i$ that satisfies the condition $k_{eff} = 1$, cf. (P.3). Hence, the optimization algorithm to be written to solve the problem (4.3) should be similar to the algorithm A1 described in section 2.2, simply replacing $\rho$ by $\chi$, $\delta\rho$ by $\delta\chi$ and *P* by $k_{eff}$. There are, however, two additional issues to address.

*Issue* 1. A card $M_i$ that contains a zero atomic fraction causes an error in the MCNP calculation, then its termination. So, cf. the upper description of this card, we cannot have $\chi_i(n) = 0$ nor 1 strictly. Neither can we have $\chi_i(n) \geq 1-\Delta\chi$, otherwise the atomic fractions of $^1$H and $^{16}$O in the card $M_{i+N}$ can become null or negative. All the more since, in a MCNP input, a negative atomic fraction is interpreted as a weight fraction, and the mixing of atomic and weight fractions in a card M causes another error, and stops the MCNP run. These software limitations impose the addition of a constraint on the atomic fractions in the problem (4.3), similar to the constraint (1.12) used in the previous sections on the densities $\rho_i$, given by:

$$\chi_i \in [\chi_{min} > 0, \chi_{max} = 1-\Delta\chi[, \forall i \quad (P.2.1)$$

*Issue* 2. Suppose we performed a MCNP calculation starting from a suboptimal configuration $\chi(n)$. Using the procedure described in section P.1, we obtain a list of values $C_i$. In this section, we want to minimize the fuel mass at constant $k_{eff}$, so we should position the fuel where it is the most efficient, by adding e.g. a quantity $\delta\chi$ to the starting fractions $\chi_i(n)$ in the tubes with the highest $C_i > 0$, and by removing fuel where it is the least efficient. With a little organization, this phase can be reduced to the test of *N*+1 fuel configurations. Among these configurations, we must then find the one that reaches $k_{eff} \approx 1$ with the smallest fuel mass. To follow this optimization procedure, we must thus calculate the $k_{eff}$ of *N*+1 fuel configurations, so perform *N*+1 additional MCNP calculations per iteration of the algorithm. If the number *N* of tubes is large (which will be the case here), the associated computing time can be sizable. Fortunately, this issue can be overcome using the coefficients $C_i$ calculated upper. Indeed, when a fraction $\chi_i(n)$ is increased by a quantity $\delta\chi$, the resulting modification of the $k_{eff}$ is directly $\delta k_{eff} = C_i V_i \delta\chi$. With this observation, we can estimate the $N+1$ $k_{eff}$ coefficients sought without redoing a MCNP calculation per possible configuration.

With these comments, by taking inspiration from the algorithm A1 described in section 2.2, a possible writing of an optimization algorithm usable to solve the problem (4.3)+(P.2.1) is:

Algorithm B0
For $n$ from 1 to NUMITER
  1) Write the MCNP input taking as fractions $\chi_i$ in the cells $\Theta_i$ the fractions $\chi_i(n-1)$ obtained at the previous iteration of the algorithm, see card $M_i$ in section P.1. Add the cards $M_{i+N}$ and PERT$i$ described in section P.1. Modify the seed used for the random draws.
  2) Run MCNP, read its output file, extract the $\partial k_{eff}/\partial\chi_i$ derivatives, compute the $C_i$ coefficients.
  3) Create $N$ vectors $\underline{F}_i = (C_i, V_i, \chi_i(n-1))$, one per cell $\Theta_i$. Reorder them in descending order of $C_i$. Renumber the vectors $\underline{F}_j$ thus classified from $j = 1$ to $j = N$.
  4) Increment the fractions $\chi_i$ by assuring they verify the constraint (P.2.1) and compute the $k_{eff}$ of the associated reactor configurations:
For $m$ from 0 to $N$
  $k(m) = k_{eff}(n-1)$, $k_{eff}$ of the previous configuration $\chi(n-1)$, given in the MCNP output file o
  If $m = 0$  For $j$ from 1 to $N$
        $temp = \chi_j(n-1) - \delta\chi$
        If $temp \geq \chi_{min}$ then $c(j,0) = temp$ else $c(j,0) = \chi_{min}$ endif
        $k(0) = k(0) + (c(j,0) - \chi_j(n-1)) \times V_j C_j$
  If $m > 0$  For $j$ from 1 to $N$
        If $j \leq m$  $temp = \chi_j(n-1) + \delta\chi$
          If $temp < \chi_{max}$ then $c(j,m) = temp$ else $c(j,m) = \chi_{max} - 10^{-5}$ endif
          $k(m) = k(m) + (c(j,m) - \chi_j(n-1)) \times V_j C_j$
        If $j > m$  $temp = \chi_j(n-1) - \delta\chi$
          If $temp \geq \chi_{min}$ then $c(j,m) = temp$ else $c(j,m) = \chi_{min}$ endif
          $k(m) = k(m) + (c(j,m) - \chi_j(n-1)) \times V_j C_j$
  5) Determine the configuration with the least uranium approaching $k_{eff} = 1$, giving priority to the subcritical configuration, i.e. for $k_{eff} \leq 1$.
If $k(m) > 1$ $\forall m \in [0, N]$ then $m_{opt} = 0$ (always supercritical: remove some fuel in all cells)
Elseif $k(m) < 1$ $\forall m \in [0, N]$ then $m_{opt} = N$ (always subcritical: add some fuel in all cells)
Else  For $m$ from 0 to $N-1$
      If $(k(m) > 1$ and $k(m+1) \leq 1)$ then $(m_{opt} = m+1$ and break) endif
      If $(k(m) \leq 1$ and $k(m+1) > 1)$ then $(m_{opt} = m$ and break) endif
Endif. The new fractions $\chi_i(n)$ are then given by $c(i, m_{opt})$.

The results shown in fig. 12-13 were obtained for $\Delta\chi = 0.05$, $\delta\chi = 10^{-3}$ and $\chi_{min} = 10^{-5}$.
In algorithm B0, the fractions $\chi$ are modified by quantized steps, of the type (2.2.1). However, we have observed that a gradient procedure, $\underline{\chi}(n+1) = \underline{\chi}(n) + \alpha\underline{C}$, can work seemingly correctly for this kind of problem. Indeed, as MCNP automatically renormalizes the atomic fractions in the $M_i$ cards so that their sum be 1, the constraints $\chi_i < \chi_{max}$ and $\Sigma\chi_i = 1$ can be removed. Adapting the constraint $\chi_i > 0$ is then feasible using the procedure given at step 4 of B0. We tested this hybrid procedure on a problem with a simple geometry, but whose challenge was to simultaneously optimize the fractions of ~200 different isotopes, with plausible results.

P.3. Additional discussion of the results shown in fig. 12-13

In the video okeff.mp4, we observe that the calculated structure still evolves after convergence, rotating and slowly drifting towards the upper left corner of the reactor. This rotational-translational movement is not really an artifact. Indeed, for the problem of this section, there are no source nor detector cells to anchor the structure. Therefore, as the water surrounding the fuel acts as an infinite medium, the structure can drift freely herein. Adding anchor points, e.g. control bars, or using the planar symmetries of the reactor should suppress this effect.

Fig. 13, we note that the initial configuration of the reactor, $\chi_i(0) = 0.02\ \forall i$, is supercritical. Then, from $n = 1$ to $3$, we verify that B0 behaves correctly, decreasing the amount of uranium in the reactor until $k_{eff}$ falls below 1. But afterwards, we note that $k_{eff}$ rises anew above 1, notably from $n = 4$ to $22$, which should normally be forbidden at step 5 of B0. The explanation of this behavior is interesting. At step 4 of B0, the $k_{eff}$ coefficients, $k(m)$, of the configurations $c(j,m)$ are estimated using the derivatives $\partial k_{eff}/\partial \chi_i$, in accordance with the solution of the issue 2 given section P.2. However, this estimate is not perfectly accurate. First, it can be biased by statistical fluctuations. Secondly, the perturbations $dk_i = (\partial k_{eff}/\partial \chi_i)\Delta\chi$ are printed in a MCNP6.1 output file o in the decimal form with only five significant digits. Therefore, when a perturbation $dk_i$ is lower than 1 pcm, which sometimes happens in our calculations involving many small localized perturbations, it is indicated as zero. The accumulation of these small but numerous numerical deviations eventually results in a noticeable difference between the $k_{eff}$ value obtained with a KCODE calculation and the $k(m_{opt})$ value estimated with the derivatives. To reduce this difference, the source code of MCNP could be modified so that the perturbations $dk_i$ be written with a better precision, in scientific form.

## Section Q. Resolution of the problem (5.1.3)-(5.1.4)

The Lagrangian $L$ of the problem (5.1.3)-(5.1.4) is given equation (1.9). The optimal density configuration $\rho_{opt}$ sought hence obeys the following system of equations:

$$\frac{\partial L}{\partial \lambda} = 0,\quad \frac{\partial L}{\partial \rho_i} = 0 \Leftrightarrow C_i = \frac{1}{V_i}\left(\frac{1}{\phi_1^\beta}\frac{\partial \phi_0}{\partial \rho_i} - \beta\frac{\phi_0}{\phi_1^{\beta+1}}\frac{\partial \phi_1}{\partial \rho_i}\right) = \lambda, \forall i \quad (Q.1)$$

As before, we note that the optimal configuration is given by an isovalue of coefficients $C_i$ that verifies the weight constraint. We can thus solve the problem (5.1.3)-(5.1.4) using algorithm A1, by simply replacing the coefficients $C_i$ of section 2.2 by those given in (Q.1). The use of A1 adds the constraint (1.12) to the problem. As before, the coefficients $C_i$ can all be obtained in a single MCNP calculation, by using one PERT card per coefficient. Using a tally F4:p to calculate the gamma fluxes $\phi_0$ and $\phi_1$, one must add to the MCNP input file the following lines:

F4:p *Index_of_cell_*$\Theta_{in}$ *Index_of_cell_*$\Theta_{out}$
PERT1:p  CELL=1  RHO=−1.05×$\rho_1$  METHOD=2
…
PERT*N*:p  CELL=*N*  RHO=−1.05×$\rho_N$  METHOD=2

To randomly draw the energies of the gamma-rays emitted by the $^{60}$Co source, the following lines must also be added to the MCNP input file:
SDEF −20 0 0 ERG=d1 PAR=2
SI1 L 1.1732 1.3325
SP1 0.4997 0.5003

From the MCNP output files o or m, we can then extract the values of the derivatives $\partial\phi_0/\partial\rho_i$ and $\partial\phi_1/\partial\rho_i$ appearing in (Q.1), given by their corresponding PERT*i*/(0.05$\rho_i$) ratios.